\documentclass{amsart}

\usepackage[lite]{amsrefs}
\usepackage{amssymb}
\usepackage[all,cmtip]{xy}
\usepackage{amsthm}
\usepackage{amsmath}
\usepackage{amsfonts}
\usepackage[hidelinks]{hyperref} 
\usepackage{enumerate}
\usepackage[titletoc]{appendix}
\usepackage{fancyhdr}
\usepackage{xcolor}
\usepackage{comment}
\usepackage{tikz-cd}
\usepackage{tikzpagenodes}
\usepackage{caption}
\usepackage{float}
\usetikzlibrary[patterns]
\usepackage{mathtools}

\usepackage[margin = 1in]{geometry}
\usepackage{setspace}

\spacing{1.2}

\allowdisplaybreaks

\numberwithin{equation}{section}
\newtheorem{definition}{Definition}[section]
\newtheorem{example}[definition]{Example}
\newtheorem{theorem}[definition]{Theorem}
\newtheorem{lemma}[definition]{Lemma}
\newtheorem{corollary}[definition]{Corollary}
\newtheorem{proposition}[definition]{Proposition}
\newtheorem{remark}[definition]{Remark}

\newcommand{\CC}{\mathbb{C}}

\newcommand{\ZZ}{\mathbb{Z}}
\newcommand{\g}{\mathfrak{g}}
\newcommand{\h}{\mathfrak{h}}
\newcommand{\p}{\mathfrak{p}}
\newcommand{\n}{\mathfrak{n}}

\newcommand{\sll}{\mathfrak{sl}}
\newcommand{\vac}{ \left| 0 \right>}
\newcommand{\ad}{\text{ad}}

\newcommand{\Mod}[1]{\ (\mathrm{mod}\ #1)}

\def\cA{{\mathcal A}}

\def\cD{{\mathcal D}}
\def\cE{{\mathcal E}}
\def\cF{{\mathcal F}}

\def\cH{{\mathcal H}}

\def\cS{{\mathcal S}}

\def\cW{{\mathcal W}}

\def\go{{\mathfrak o}}
\def\gp{{\mathfrak p}}

\def\gs{{\mathfrak s}}

\begin{document}
\title[]{
Supersymmetries in the theory of W-algebras
}

\author[A.Linshaw]{Andrew Linshaw}
\address[A.Linshaw]{Department of Mathematics, University of Denver, 2199 S University Blvd, Denver, CO 80210, USA}
\email{andrew.linshaw@du.edu}
\thanks{A. L. is partially supported by National Science Foundation Grant DMS-2401382, and Simons Foundation Grant MPS-TSM 00007694}

\author[A. Song]{Arim Song}
\address[A. Song]{Department of Mathematics, University of Denver, 2199 S University Blvd, Denver, CO 80210, USA}
\email{Arim.Song@du.edu}
\thanks{A. S. is supported by Basic Science Research Program through the National Research Foundation of Korea(NRF) funded by the Ministry of Education(RS-2024-00409689)}

\author[U.R.Suh]{Uhi Rinn Suh}
\address[U.R.Suh]{Department of Mathematical Sciences and Research institute of Mathematics, Seoul National University, Gwanak-ro 1, Gwanak-gu, Seoul 08826, Korea}
\email{uhrisu1@snu.ac.kr}
\thanks{U.R. S. is supported by NRF Grant \#2022R1C1C1008698 and  Creative-Pioneering Researchers Program by Seoul National University}

\thanks{ }

\begin{abstract}
 Let $\g$ be a basic Lie superalgebra and $f$ be an odd nilpotent element in an $\mathfrak{osp}(1|2)$ subalgebra of $\g$. We provide a mathematical proof of the statement that the W-algebra $W^k(\g,F)$ for $F=-\frac{1}{2}[f,f]$ is a vertex subalgebra of the SUSY W-algebra $W_{N=1}^k(\g,f)$, and that it commutes with all weight $\frac{1}{2}$ fields in $W_{N=1}^k(\g,f)$. Note that it has been long believed by physicists \cite{MadRag94}. In particular, when $f$ is a minimal nilpotent, we explicitly describe superfields which generate $W^k_{N=1}(\g,f)$ as a SUSY vertex algebra and their OPE relations in terms of the $N=1$ $\Lambda$-bracket introduced in \cite{HK07}. In the last part of this paper, we define $N=2,3$, and small or big $N=4$ SUSY vertex operator algebras as conformal extensions of $W^k_{N=1}(\mathfrak{sl}(2|1),f_{\text{min}})$, $W^k_{N=1}(\mathfrak{osp}(3|2),f_{\text{min}})$, $W^k_{N=1}(\mathfrak{psl}(2|2),f_{\text{min}})$, and $W^k_{N=1}(D(2,1;\alpha)\oplus \mathbb{C},f_{\text{min}})$, respectively, for the minimal odd nilpotent $f_{\text{min}}$, and examine some examples. 
\end{abstract}

\maketitle

\vskip 6mm

\setcounter{equation}{0}

\section{Introduction}\label{section:Intro}

A supersymmetry in a vertex algebra $V$ is defined as an odd derivation $D$ satisfying $D^2=\partial$, where $\partial$ denotes the translation operator on $V$, and a vertex algebra with a supersymmetry is called a supersymmetric (SUSY) vertex algebra \cite{HK07}. 
This concept first appeared in the physics literature (see, e.g., \cite{EH91, Ito91, Ito92}) in the context of superconformal field theories via the superfield formalism. On the mathematical side, Barron investigated the geometric foundations of the theory in a series of papers \cite{B1996, B2000, B2003, B2004}, while Heluani and Kac took a more algebraic approach in their work \cite{HK07}, developing a solid framework for the structure theory of SUSY vertex algebras.

% A vertex algebra equipped with such a derivation is called a supersymmetric (SUSY) vertex algebra. The formal notion of SUSY vertex algebras was introduced in \cite{HK07}, where the authors also developed foundational aspects of their structural theory.

Understanding the supersymmetry of a given SUSY vertex algebra $V$ provides a significant advantage in analyzing its vertex algebra structure. In particular, the behavior of $D(a)$ is strongly influenced by that of its superpartner $a\in V$. Thus, the question of whether supersymmetry exists is of considerable interest. When a vertex algebra $V$ admits a superconformal vector $G$, the associated odd endomorphism $G_{(0)}$ naturally induces a supersymmetry on $V$.
For example, the Kac-Todorov vector \cite{KT85} in the SUSY affine vertex algebra 
$V^k_{N=1}(\g)$ for a non-critical level 
$k$ is a superconformal vector, making it a SUSY vertex algebra that contains the affine vertex algebra $V^k(\g).$  Not all vertex algebras naturally admit supersymmetries. However, in certain cases, one can construct a supersymmetric (SUSY) extension \cite{SY25}. A SUSY vertex algebra $V_{N=1}$ is referred to as a SUSY extension of a vertex algebra $V$
if it is embedded in $V_{N=1}$ as a subalgebra, even though $V_{N=1}$ is not a conformal extension. For instance, a SUSY affine vertex algebra is a SUSY extension  of the corresponding affine vertex algebra. In such situation, the supersymmetry of the SUSY extension can be used to analyze the structure and properties of the original one.

The main object of this paper is a SUSY W-algebra which was first introduced in the physics paper \cite{MadRag94} and reinterpreted with the language of SUSY vertex algebra in \cite{MRS21}. For the representation theory of SUSY W-algebra, 
When $\g$ is a basic Lie superalgebra and $f$ is an odd nilpotent in an $\mathfrak{osp}(1|2)$ subalgebra of $\g,$ the SUSY W-algebra $W^k_{N=1}(\g,f)$ of level $k$ is defined as the cohomology of the SUSY BRST complex. This complex is a SUSY extension of the BRST complex for the usual W-algebra $W^k(\g,F)$ when $F=-\frac{1}{2}[f,f]$. Thus it is natural to ask whether $W^k_{N=1}(\g,f)$ is a SUSY extension of $W^k(\g,F)$. Indeed, in \cite{MadRag94}, the authors presented evidence suggesting that  $W^k_{N=1}(\g,f)$ is an extension of $W^k(\g,F)$ by weight $1/2$ free fields. The primary goal of the first part of this paper is to provide a rigorous proof of this assertion.

One piece of evidence for the above claim can be observed by comparing the generating types of $W^k_{N=1}(\g,f)$ and $W^k(\g,F)$.
The structure of SUSY W-algebra is governed by the $\mathfrak{osp}(1|2)$ representation theory. With respect to the adjoint action of a subalgebra $\mathfrak{s}\simeq\mathfrak{osp}(1|2)$ of $\g$, $\g$ decomposes as a direct sum of irreducible modules which are all odd-dimensional.
Suppose 
\begin{equation} \label{eq:g_osp}
    \g= \bigoplus_{i\in \mathbb{N}} \Big( \ V_i^{\oplus m_i} \oplus W_i^{\oplus n_i}\ \Big)
\end{equation}
where $V_i$ (resp. $W_i$) is an irreducible $\mathfrak{osp}(1|2)$-module of dimension $2i-1$ with  super-dimension $1$ (resp. $-1$).
Each component $V_i$ or $W_i$ is spanned by a basis which consists of one lowest weight vector $v_i\in \g^f_{\frac{1}{2}(1-i)}= \text{ker}(\ad\, f)\cap \g_{\frac{1}{2}(1-i)}$ and $(\text{ad} \, e)^m v_i$ for $m=1,\cdots, 2i-2.$ Then by the structure theory in \cite{MRS21},
$W^k_{N=1}(\g,f)$ is strongly generated by $m_i$ and $n_i$ odd and even weight $(\frac{i}{2}, \frac{i+1}{2})$ superfields, respectively. We denote this SUSY generating type by 
\begin{equation}\label{eq:type of W_{N=1}- 1}
    W_{N=1}\Big(\Big(\frac{1}{2}\Big)^{n_1}, 1^{n_2}, \Big(\frac{3}{2}\Big)^{n_3}, \cdots ;\Big(\frac{1}{2}\Big)^{m_1}, 1^{m_2}, \Big(\frac{3}{2}\Big)^{m_3}, \cdots \Big).
\end{equation}
In other words, a minimal strong generating set of $W^k_{N=1}(\g,f)$ consists of $m_1,\, n_1+m_2, \, n_2+m_3, \cdots$  weight $\frac{1}{2}, 1, \frac{3}{2}, \cdots$ odd fields and $n_1,\, m_1+n_2,\, m_2+n_3, \, \cdots$  weight $\frac{1}{2}, 1, \frac{3}{2} \cdots$ even fields, which means it is strongly generated vertex algebra of type
\begin{equation} \label{eq:type of W_{N=1}-2}
    W \Big( \Big(\frac{1}{2}\Big)^{n_1}, 1^{m_1+n_2}, \Big(\frac{3}{2}\Big)^{m_2+n_3}, 2^{m_3+n_4},\  \cdots \  ;  \Big(\frac{1}{2}\Big)^{m_1}, 1^{n_1+m_2}, \Big(\frac{3}{2}\Big)^{n_2+m_3}, 2^{n_3+m_4}, \ \cdots \  \Big).
\end{equation}
On the other hand, the dimension $(2i-1)$ irreducible $\mathfrak{osp}(1|2)$ representation is decomposed into two irreducible $\sll(2)$ modules with dimensions $i$ and $i-1$ and each irreducible component consists of elements with the same parity. 
Hence if we see $\g$ in \eqref{eq:g_osp} as an $\sll(2)$ module via the $\ad \, \mathfrak{s}_{\bar{0}}$ action then $\g$ is decomposed as 
\begin{equation}\label{eq:g_sl}
        \g= \bigoplus_{i\in \mathbb{N}}\Big( \ (V_i)_{\bar{0}}^{\oplus m_i}\oplus (V_i)_{\bar{1}}^{\oplus m_i} \oplus (W_i)_{\bar{0}}^{\oplus n_i}\oplus (W_i)_{\bar{1}}^{\oplus n_i}\ \Big),
\end{equation}
 and this implies $W^k(\g,F)$ is a vertex algebra of type
\begin{equation}\label{eq:type of W_{N=0}}
       W \Big( 1^{m_1+n_2}, \Big(\frac{3}{2}\Big)^{m_2+n_3}, 2^{m_3+n_4},\  \cdots \  ;   1^{n_1+m_2}, \Big(\frac{3}{2}\Big)^{n_2+m_3}, 2^{n_3+m_4}, \ \cdots \  \Big).
\end{equation}
From \eqref{eq:type of W_{N=1}-2} and \eqref{eq:type of W_{N=0}}, it is clear that the generating types of $W^k(\g,F)$ and $W^k_{N=1}(\g,f)$ coincide up to the weight $1/2$ part.

When $f$ is principal, the SUSY W-algebra $W^k_{N=1}(\g,f)$ does not have the weight $1/2$ part. In \cite{GSS25}, the third and fourth authors with their collaborator, proved that the SUSY W-algebra is isomorphic to the original W-algebra $W^k(\g,F)$ for all $k\neq -h^\vee$. Their proof relies on the injectivity of two Miura maps: one from $W^k(\g,F)$ to $V^{\tau_k}(\g_0)\otimes \mathcal{F}^{ne}$, and the other from  $W^k_{N=1}(\g,f)$ to $V^{\psi_k}_{N=1}$ \cite{Nak23, GSS25}, where $\mathcal{F}^{ne}$ denotes the neutral free fermion vertex algebra generated by $\g_{1/2}$. The target spaces of these Miura maps are isomorphic, and this correspondence facilitates a comparison between the principal ordinary and SUSY W-algebras. In fact, the image of each Miura map coincides with the kernel of the corresponding screening operators \cite{Genra17, Song24free}. By comparing these screening operators, one can establish the isomorphism between $W^k_{N=1}(\g,f)$ and  $W^k(\g,F)$. The supersymmetry of the SUSY W-algebra, naturally inherited from that of the SUSY BRST complex, is well-understood. Consequently, the vertex algebra structure of the original W-algebra  can be analyzed via the supersymmetry induced from the corresponding SUSY W-algebra.

In this paper, we consider an odd nilpotent $f$ in an $\mathfrak{osp}(1|2)$ subalgebra $\mathfrak{s}$ in $\g$. For non-principal $f$ and $F=-\frac{1}{2}[f,f]$, $W^k(\g,F)$ does not have a supersymmetry. However we could show $W^k_{N=1}(\g,f)$ is a minimal SUSY extension of $W^k(\g,F)$ meaning that no SUSY extension exists properly included in $W^k_{N=1}(\g,f)$.
To establish this, we compare the images of the Miura maps associated with both algebras. Unlike the principal case, the target spaces of these Miura maps are not isomorphic. However, we constructed the injective map
\begin{equation} \label{eq:map_miura}
    V^{\tau_k}(\g_0)\otimes \mathcal{F}^{ne} \hookrightarrow V^{\psi_k}_{N=1}(\g_0)
\end{equation}
which induces the embedding $W^k(\g,F) \hookrightarrow W^k_{N=1}(\g,f)$. Using the map \eqref{eq:map_miura}, we rigorously prove the following theorem, which was anticipated in \cite{MadRag94}. Moreover, since the isomorphism in Theorem \ref{main theorem} arises explicitly from the map \eqref{eq:map_miura}, our result shows a precise element-wise correspondence between the two algebras.

\begin{theorem}[Theorem \ref{thm:nonSUSY vs SUSY}] \label{main theorem}
Let $\mathcal{F}(\g^f_0)$ be the vertex subalgebra of $W_{N=1}^k(\g,f)$ generated by weight $1/2$ fields. For $F=-\frac{1}{2}[f,f]$ and $k\neq -h^\vee$, we have  
\begin{equation} \label{eq:main}
        W^k(\g,F) \simeq \textup{Com}(\mathcal{F}(\g^f_0), W_{N=1}^{k}(\g,f)), \quad   W_{N=1}^{k}(\g,f)\simeq  W^k(\g,F) \otimes \mathcal{F}(\g^f_0). 
\end{equation}
Moreover, the embedding of $W^k(\g,F)$ into $W^k_{N=1}(\g,F)$ is directly induced from the map \eqref{eq:map_miura}.
\end{theorem}

    Since $W^k(\g,F)$ is a subalgebra of $W^k_{N=1}(\g,f),$ elements of weight $\Delta$ in $W^k(\g,F)$ can be interpreted as elements of $W^k_{N=1}(\g,f)$ that are expressed as normally ordered products of elements of weight less than $\Delta$, together with the odd derivation $D$. We refer to such elements as {\it Type 1}, and classify all others as {\it Type 2}. Recall the generating types described in equation  \eqref{eq:type of W_{N=0}} for $W^k(\g,F).$ Its minimal strong generating set consists of:
    \begin{itemize}
        \item $m_{i}$ (resp. $n_{i}$)  Type 1 even (resp. odd) weight  $\frac{i+1}{2}$ elements,
        \item $n_{i+1}$ (resp. $m_{i+1}$)  Type 2 even (resp. odd) weight  $\frac{i+1}{2}$ elements
    \end{itemize}   for $i\geq 1$, where $
        m_i= \text{dim}\, \big(\g^f_{\frac{1}{2}(1-i)}\big)_{\bar{0}}, \text{ and }n_i= \text{dim}\, \big(\g^f_{\frac{1}{2}(1-i)}\big)_{\bar{1}}$,
In addition, by combining the theorem with known results on W-algebras, we can derive essential features of SUSY W-algebras and their cosets. 

\begin{enumerate}
     \item For generic $k\in \mathbb{C}$, the SUSY W-algebra $W_{N=1}^k(\g,f)$ and coset vertex algebras by its SUSY affine vertex subalgebras are simple. (Corollary \ref{cor:simplicity} and \ref{cor:simplicityofcosets}) 
    \item For any $k\neq -h^\vee$, the categories of ordinary modules of $W^k(\g,F)$ and $W^k_{N=1}(\g,f)$ are equivalent. (Corollary \ref{cor:ordinary module})
    \item Orbifolds and SUSY affine cosets of a SUSY W-algebra $W_{N=1}^k(\g,f)$ for generic level $k$ are strongly finitely generated. (Corollary \ref{cor:SFGproperty})
\end{enumerate}

\noindent 
Note that the authors of \cite{CCS24+} investigated positive energy modules of SUSY W-algebras using the finite SUSY W-algebras introduced in \cite{CCS24+, GSS25} and showed the equivalence between the categories of nonSUSY and SUSY principal W-algebras.
The result (2) provides a generalization of this  theorem.

   \vskip 2mm
   
In the second part of this paper, we focus on the case where 
$f$ and $F$ are minimal odd and even nilpotent elements, respectively. Minimal 
W-algebras form one of the most well-known families of vertex algebras \cite{KW04,KPP23,ACKL17}, and are particularly notable for their deep connections to superconformal algebras, which lie at the core of supersymmetry theory. In Section 5, we examine the SUSY vertex algebra structure of SUSY minimal 
W-algebras, and in the subsequent sections, we explore their relationships with the 
$N=2,3,4$ superconformal algebras.

Recall that the structure of minimal W-algebra $W^k(\g,F)$ crucially depends on the action of its weight-1 subspace \cite{ACKL17}. To uncover the underlying supersymmetry of a minimal W-algebra, it is essential to view it as a subalgebra of the corresponding minimal SUSY 
W-algebra. This naturally leads us to examine the weight 1 subspace of $\textup{Com}(\mathcal{F}(\g^f_0), W^k_{N=1}(\g,f))$ which is spanned by $\text{dim}\ \g^f_0$  Type 1 fields and $\text{dim}\ \g^f_{-1/2}=\text{dim}\ \g^F_0-\text{dim}\ \g^f_0$ Type 2 fields. Note that the Type 1 fields generate affine vertex algebra $V^{\kappa}(\g_0^f)$ and by including the Type 2 fields as additional generators, we obtain $V^{\kappa}(\g_0^F)$.  In Section \ref{subsec:list of minimal}
, we classify all odd minimal nilpotents for classical Lie superalgebras
based on the classification of minimal nilpotent orbits in \cite{KW04}. We also describe $\g^F_0$ as a $\g^f_0$-module (see Table \ref{table1} and \ref{table2}).

 Moreover, we find a set of superfields that strongly generates $W^k_{N=1}(\g,f)$ and $N=1$ $\Lambda$-brackets between them. Here the $\Lambda$-bracket introduced in \cite{HK07} is a way of expressing OPEs between two superfields. More precisely, in a SUSY vertex algebra, the $\Lambda$-bracket $[a{}_\Lambda b]$ for the couple $\Lambda=(\chi,\lambda)$ of odd and even indeterminates $\chi$ and $\lambda$ is defined by
 \begin{equation}\label{eq:Lambda a b}
     [a{}_\Lambda b]=\chi [a{}_\lambda b]+ [D(a){}_\lambda b],
 \end{equation}
 where $[a{}_\lambda b]=\sum_{i\in \mathbb{Z}_+}\frac{\lambda^n}{n!}a_{(n)}b$ and the sesquilinearity of $\Lambda$-bracket \eqref{eq:Lambda a b} also determines the OPEs between $(a,D(b))$ and $(D(a),D(b)).$
 
 The minimal strong generating set of a SUSY minimal W-algebra consists of $\text{dim}\ \g^f_{0}$ weight $(\frac{1}{2},1)$ superfields,  $\text{dim}\g^f_{-1/2}$ weight $(1,\frac{3}{2})$ superfields and one additional weight $(\frac{3}{2},2)$ superfield. The superfield of weight $(\frac{3}{2},2)$ superfield can be chosen as the superconformal vector, while the $(\frac{1}{2},1)$ superfields generate the SUSY affine vertex algebra associated to $\g^f_0.$
The only remaining case is weight $(1,\frac{3}{2})$ superfields which we explicitly computed in Proposition \ref{prop:weight 1 elt} using the SUSY BRST complex. Moreover, the weight 1 components of the superfields we described lie in the centralizer of $\mathcal{F}(\g^f_0)$ and by Theorem \ref{main theorem}, this implies that these weight 1 fields can be regarded as elements of the ordinary minimal W-algebra. Finally, by combining the main theorem with the properties that stated below, we obtain the complete set of $\Lambda$-brackets among the superfields (Proposition \ref{prop:minimal Lambda braket 1}, \ref{prop:minimal Lambda braket 2} and \ref{prop:minimal Lambda braket 3}): 
\begin{itemize}
    \item $\lambda$-bracket relations on $W^k(\g,F)$ in \cite{KW04,KPP23},
    \item the supersymmetry $D$ in $W^k_{N=1}(\g,f)$ is $G_{(0)}$ for the superconformal vector $G$ in \cite{Song24free},
    \item in terms of the isomorphism \ref{eq:main}, $G-\tau_{KT}$ is in the usual minimal W-algebra, where $\tau_{KT}$ is the Kac-Todorov superconformal vector of the SUSY affine vertex algebra of $\g^f_0.$
\end{itemize}

In the final part of the paper, we explore the concept of 
$N=2,3,4$ SUSY vertex operator algebras (VOAs). Recall that a vertex algebra possessing a superconformal vector necessarily exhibits supersymmetry. Equivalently, any vertex algebra that is a conformal extension of $W^k(\mathfrak{osp}(1|2),f_{\text{min}})$ admits a supersymmetry. Similarly, $N=2$ and $N=3$ superconformal vector introduced in \cite{HK07} along with their derivatives, generate the corresponding $N=2$ and $N=3$ superconformal algebras as described in \cite{KW04}. These works also demonstrate that the associated universal enveloping vertex algebras can be constructed by adding free fields to the minimal W-algebras for $\mathfrak{sl}(2|1)$ and $\mathfrak{sp0}(2|3)$, respectively. Our main theorem shows that these vertex algebras are precisely the SUSY minimal W-algebras for $\mathfrak{sl}(2|1)$ and $\mathfrak{spo}(2|3)$. Furthermore, the small and big $N=4$ superconformal algebras also are shown to be related to minimal W-algebras. The small  $N=4$ superconformal algebra generates the minimal W-algebra for $\mathfrak{psl}(2|2)$ which is isomorphic to the corresponding SUSY W-algebra. The big $N=4$ superconformal algebra yields a vertex algebra obtained by extending the minimal W-algebra  $W^k(D(2,1;\alpha),F)$ with five free fields. In terms of SUSY W-algebras, this extended vertex algebra is isomorphic to the SUSY W-algebra for $D(2,1;\alpha)\oplus \mathbb{C}.$ According to the observations, we introduce the following notions:

\begin{definition}\, \label{def:introN=2,3,4}
\begin{enumerate}
    \item An $N=2$ VOA is a conformal extension of $W^k_{N=1}(\mathfrak{sl}(2|1), f_{\text{min}})$.
    \item An $N=3$ VOA is a conformal extension of $W^k_{N=1}(\mathfrak{spo}(2|3), f_{\text{min}})$.
    \item A small $N=4$ VOA is a conformal extension of $W^k_{N=1}(\mathfrak{psl}(2|2),f_{\text{min}})$.
    \item A big $N=4$ VOA is a conformal extension of $W^k_{N=1}(D(2,1;\alpha)\oplus \mathbb{C}, f_{\text{min}})$.
\end{enumerate}
\end{definition}
We note that neither the small nor big $N=4$ superconformal algebra possesses an $N=4$ superconformal vector in the sense of \cite{HK07}. 
In Section \ref{subsec:N=2superconformal}, we provide an explanation for why the definition based on the $N=4$ superconformal vector is considered less appropriate. In particular, $W^k_{N=1}(D(2,1;\alpha)\oplus \mathbb{C}, f_{\text{min}})$ itself is a conformal extension of $W^k_{N=1}(\mathfrak{sl}(2|1), f_{\text{min}})$,  $W^k_{N=1}(\mathfrak{spo}(2|3), f_{\text{min}})$ or $W^k_{N=1}(\mathfrak{psl}(2|2),f_{\text{min}})$. Hence a big $N=4$ VOA can be also regarded as a $N=2$, $N=3$ or small $N=4$ VOA.

Section 7, we provide a variety of examples of $N=2,3,4$ VOAs introduced in Definition \ref{def:introN=2,3,4}. First, there are many examples $N=2$ SUSY $W$-algebras, including the principal $W$-algebras $W^k_{N=1}(\mathfrak{sl}(n+1|n))$, as well as some infinite families of cosets of SUSY W-algebras by SUSY affine subVOAs. In a separate paper \cite{CKLSS25} with two other collaborators, we have shown that all of these arise as $1$-parameter quotients of a universal $2$-parameter $N=2$ vertex algebra $\cW^{N=2}_{\infty}$. Similarly, we will give several infinite families of SUSY W-algebras and cosets of SUSY W-algebras which are big $N=4$ VOAs, and we expect them all to arise as quotients of a similar universal $2$-parameter $N=4$ VOA $\cW^{N=4}_{\infty}$. Since after changing conformal vector, the big $N=4$ VOA can be viewed as a conformal extension of either the small $N=4$ VOA or the $N=3$ VOA, all our examples of big $N=4$ VOAs can also be viewed as small $N=4$ VOAs or $N=3$ VOAs.

There do not seem to be examples of SUSY W-algebras which are small $N=4$ or $N=3$ VOAs, but {\it not} big $N=4$ VOAs, aside from $W^{k}_{N=1}(\mathfrak{osp}(4|2), f_{\text{min}})$ and $W^{k}_{N=1}(\mathfrak{spo}(2|3), f_{\text{min}})$ themselves. One can also ask whether there exist other VOAs depending continuously on $1$ or more parameters which are small $N=4$ or $N=3$ VOAs, but not big $N=4$ VOAs. We will give an example of a $1$-parameter small $N=4$ VOA with this property, which has infinitely many strong generators that close linearly under OPE. In addition, we mention a more interesting example that has appeared recently in physics. Starting with work of Beem, Meneghetti and Rastelli \cite{BMR19}, a small $N=4$ VOA $\mathcal{W}_{\Gamma}$ with a fixed central charge has been conjecturally attached to any Coxeter group $\Gamma$. In the case when $\Gamma$ is the symmetric group $S_n$, a rigorous construction of these vertex algebras was given by Arakawa, Kuwabara and M\"oller in \cite{AKM23}. Very recently, it was conjectured in \cite{BM25} that $W_{S_n}$ for all $n\geq 2$ should arise as quotients of a unifying $1$-parameter $N=4$ VOA. We will also give a $2$-parameter $N=3$ VOA which is not a big $N=4$ VOA.

There are numerous examples of VOAs with supersymmetry in the literature beyond the W-algebra theory. For example, the chiral de Rham complex $\Omega^{\text{ch}}_M$, which was introduced by Malikov, Schechtman, and Vaintrob in \cite{MSV99}, is a sheaf of VOAs that exists on any smooth manifold $M$ in the smooth, holomorphic, or algebraic setting. If $M$ is a compact, Calabi-Yau manifold of complex dimension $d$, the space of global sections $H^0(M, \Omega^{\text{ch}}_M)$ is an $N=2$ VOA which is an extension of the (simple) $N=2$ algebra with central charge $3d$ \cite{LS23}. This extended algebra was introduced by Odake \cite{O86}, and studied in detail in the case $d=3$. The total cohomology algebra $H^{\bullet}(M, \Omega^{\text{ch}}_M)$ is also an $N=2$ VOA, since it is a module over $H^{0}(M, \Omega^{\text{ch}}_M)$. Similarly, if $M$ is a compact, hyperkahler manifold of complex dimension $2d$, $H^0(M, \Omega^{\text{ch}}_M)$ is isomorphic to the (simple) small $N=4$ algebra with central charge $c = 6d$, so that $H^{\bullet}(M, \Omega^{\text{ch}}_M)$ is a small $N=4$ VOA \cite{LS23}. Finally, if $M$ is a $7$-dimensional manifold with $G_2$ holonomy, or an $8$-dimensional manifold with $\text{Spin}_7$ holonomy,  $H^{0}(M, \Omega^{\text{ch}}_M)$ is a conformal extension of (a homomorphic image of) $W^{1/3}(\mathfrak{osp}(4|2), F_{\text{min}})$ or $W^{1/3}(\mathfrak{spo}(2|3), F_{\text{min}})$, respectively \cite{H17}. These examples are almost not quite big $N=4$ or $N=3$ VOAs because they do not contain the SUSY VOAs $W^{1/3}_{N=1}(\mathfrak{osp}(4|2), f_{\text{min}})$ or $W^{1/3}_{N=1}(\mathfrak{spo}(2|3), f_{\text{min}})$, respectively.

Finally, in the appendix, we will give the generators for the $N=3$ algebra and big and small $N=4$ algebras, show how they are related to their non-SUSY counterparts, and give the explicit embeddings of the $N=3$ and small $N=4$ algebras in the big $N=4$ algebra. 

\vskip 10mm

\textbf{Acknowledgements}

The third author would like to thank Victor Kac for valuable discussions on the minimal SUSY 
W-algebra. This paper was written while she was staying in Denver, and she appreciates the warm hospitality of the University of Denver.

\section{Vertex algebras and SUSY vertex algebras} 
\subsection{Vertex algebras} Let $V=V_{\bar{0}}\oplus V_{\bar{1}}$ be a vector superspace over $\mathbb{C}$, where $V_{\bar{0}}$ (resp. $V_{\bar{1}}$) is the even (resp. odd) subspace of $V.$ For $i=0,1$ and an element $a\in V_{\bar{i}}$, the parity $i$ is denoted by $p(a).$
A vector superspace $V$ is called a vertex algebra if it is endowed with  an even element $\vac$, binary operation $: \ \ :$ called {\it normally ordered product}, even endomorphism $\partial$ and $\lambda$-bracket $[ \ {}_\lambda \ ]: V\otimes V \to \mathbb{C}[\lambda]\otimes V$ for the formal variable $\lambda$ satisfying the following properties introduced in  
\cite{BK03}:
\begin{enumerate}[(i)]
    \item $(V, \partial, [\ {}_\lambda \ ])$ is a Lie conformal algebra (LCA),
    \item $(V, :\ \ :, \partial, \vac )$ is a differential algebra with the quasi-commutativity and the quasi-associativity
    \begin{enumerate}[]
        \item (quasi-commutativity) $:\!ab\!:-(-1)^{p(a)p(b)}:\!ba\!:\, = \int^0_{-\partial}[a{}_\lambda b] d\lambda$
        \item (quasi-associativity) $::\!ab\!:\!c\!:- :\!a\!:\!bc\!::\, =\, :\! (\int^{\partial}_0 d\lambda a)[b{}_\lambda c] \!: + (-1)^{p(a)p(b)} :\! (\int^{\partial}_0 d\lambda b)[a{}_\lambda c] \!:\, , $
    \end{enumerate}
    \item The $\lambda$-bracket and normally ordered product are compatible by the Wick formula
    \begin{enumerate}[]
        \item (Wick formula) $[\, a \, {}_\lambda :\! bc \!:\,] =\,:\! [\, a\, {}_\lambda \, b\, ]c \!: +(-1)^{p(a)p(b)} :\! b[\, a\, {}_\lambda \, c\, ]\! : + \int_0^\lambda [[\, a\, {}_\lambda\,  b\, ]{}_\mu \, c\, ],\,$
    \end{enumerate}
\end{enumerate}
    for $a,b,c\in V$.
In particular, the $\lambda$-bracket on a vertex algebra is given by the series of $(n)$-th product for $n\in \mathbb{Z}_+:$
\begin{equation}
    [a{}_\lambda b]=\sum_{n\in \mathbb{Z}_+}\frac{\lambda^n}{n!}a_{(n)}b
\end{equation}
and the $(n)$-th product $a_{(n)}b$ is also called the $(n+1)$-th order pole of the operator product expansion (OPE). By the condition (i), every vertex algebra is a Lie conformal algebra. Conversely, for a given Lie conformal algebra $R$, one can construct a vertex algebra called universal enveloping vertex algebra $V(R)$ \cite{BK03}.  In addition, if a vertex algebra $V$ has a conformal vector $L$, i.e., $L_{(0)}=\partial$, $L_{(1)}$ is diagonalizable operator on $V$ and $[L{}_\lambda L]= (\partial+2\lambda)+\frac{c}{12}\lambda^3$ for the constant $c$ called the central charge, then $V$ is called a vertex operator algebra (VOA) and the eigenvalue $\Delta_a$ of $a\in V$ for $L_{(1)}$ is called the conformal weight of $a.$

We say a vertex algebra $V$ is {\it strongly generated} by a subset $S$ if 
\begin{equation} \label{eq:strongly gen}
    V=\text{Span}_{\mathbb{C}}\{\, :\partial^{n_s}a_{i_s}\cdots \partial^{n_2}   a_{i_2} \partial^{n_1} a_{i_1}: \, |\ a_{i_1},\cdots, a_{i_s}\in S, \, n_1\cdots n_s\in \mathbb{Z}_+ \,\}.
\end{equation} 
Note that the normally ordered product of $s$ elements is performed from right to left. Suppose $S$ is a totally ordered subset of $V$ and the total order on the set $\cup_{n\in \mathbb{Z}_+}\partial^n S$ is defined by 
$\partial^{n_1}a_{i_1} < \partial^{n_2}a_{i_2}$ (or equivalently, $(a_{i_1},n_1)<(a_{i_2},n_2)$) if and only if $a_{i_1}<a_{i_2}$ or $a_{i_1}=a_{i_2}$ and $n_1<n_2$ for $a_{i_1},a_{i_2}\in S.$ If  $\cup_{n\in \mathbb{Z}_+}\partial^n S$ is a PBW basis of $V$, that is, 
\begin{equation}
    \{\, :\!A_{i_s}\cdots A_{i_2}A_{i_1}\!: | A_{i_p}\in \cup_{n\in \mathbb{Z}_+}\partial S, A_{i_{p+1}}\geq A_{i_{p}} \text{ if } p(A_{i_{p}})=0, \  A_{i_{p+1}}> A_{i_{p}}\text{ if } p(A_{i_{p}})=1\}
\end{equation}
is a basis of $V$ then $V$ is said to be a vertex algebra freely generated by $S$. The universal enveloping vertex algebra $V(R)$ of a Lie conformal algebra $R= \mathbb{C}[\partial]\otimes \g$ is known to be freely generated by any basis of $\g.$ 
When $V$ is a VOA freely generated by $r^{\bar{0}}_i$ (resp. $r^{\bar{1}}_i$) of conformal weight $\Delta_i$  even (resp. odd) elements, we denote 
\begin{equation} \label{eq:generating type}
    V=W(\Delta_1^{r^{\bar{0}}_1}, \Delta_2^{r^{\bar{0}}_2}, \Delta_3^{r^{\bar{0}}_3}, \cdots;\Delta_1^{r^{\bar{1}}_1}, \Delta_2^{r^{\bar{1}}_2}, \Delta_3^{r^{\bar{1}}_3}, \cdots ).
\end{equation}

The most fundamental example of a freely generated vertex algebra is an affine vertex algebra $V^k(\g)$ of level $k\in \mathbb{C}$ where $\g$ is a Lie superalgebra with a supersymmetric invariant bilinear form $(\, | \, )$. It is the quotient vertex algebra of $V(R^{cur}(\g))$ by $K-k$ where $R^{cur}(\g)=\mathbb{C}[\partial]\otimes \g\oplus \mathbb{C}K$ is the current Lie conformal algebra endowed with the $\lambda$-bracket 
\begin{equation}
    [a{}_\lambda b]= [a,b]+K\lambda(a|b), \quad [a{}_\lambda K]=[K{}_\lambda K]=0
\end{equation}
for $a,b\in \g.$ Let $h^\vee$ be the dual Coxeter number of $\g$, that is, the eigenvalue of the adjoint action of $\frac{1}{2}\sum u_i u^i$ for the bases $\{u^i\}$ and $\{u_i\}$ of $\g$ such that $(u^i|u_j)=\delta_{ij}.$ Suppose $k\neq -h^\vee$. Then the affine vertex algebra $V^k(\g)$ has the Sugawara conformal vector $L_{\g}$ with the central charge $c_{Sug}$, where
\begin{equation}
    L_{\g}= \frac{1}{2(k+h^\vee)} \sum_{i} :u_i u^i: \  \text{ and } \ c_{Sug}=\frac{k \, \text{sdim}\ \g}{k+h^\vee}.
\end{equation}
Then it is well known that $\Delta_{a}=1$ for any $a\in \g$ and $V^k(\g)=W\big(1^{\text{dim}\ \g_{\bar{0}}};1^{\text{dim}\ \g_{\bar{1}}} \big)$.

\subsection{SUSY vertex algebras} \label{sec: SUSY VA}
A {\it supersymmetry} of a vertex algebra $V$ is an odd derivation $D=\sqrt{\partial}$ with respect to both $\lambda$-bracket and normally ordered product, i.e. 
\begin{equation} \label{eq:D, derivation}
    D(:\!ab\!:)=\,:\!D(a)b\!:+(-1)^{p(a)}:\!a D(b)\!:\,, \quad D[a{}_\lambda b]= [Da{}_\lambda b]+(-1)^{p(a)}[a{}_\lambda Db].
\end{equation}
If a vertex algebra has such an odd derivation then it is called a {\it supersymmetric (SUSY) vertex algebra.} 
The OPE of a SUSY vertex algebra can be described by so-called $\Lambda=(\lambda,\chi)$-bracket, where $\chi$ is an odd variable and $\lambda=-\chi^2$. A $\Lambda$-bracket $R\otimes R\to \mathbb{C}[\chi]\otimes R$ on a $\mathbb{C}[D]$-module $R$ is a parity reversing linear map satisfying 
\begin{enumerate}[]
    \item (sesquilinearity) $[Da{}_\Lambda b]= \chi[a{}_\Lambda b].$ 
    \quad $[a{}_\Lambda Db]= (-1)^{p(a)+1}(\chi+D)[a{}_\Lambda b],$ 
\end{enumerate}
If the $\Lambda$-bracket satisfies the following two properties 
\begin{enumerate}[]
    \item (skew-symmetry) $[a{}_\Lambda b]= (-1)^{p(a)p(b)}[b_{-\Lambda-\nabla} a]$,
    \item (Jacobi identity) $[a{}_{\Lambda}[b{}_{\widetilde{\Lambda}} c]]=(-1)^{p(a)+1}[[a{}_{\Lambda} b]_{\Lambda+\widetilde{\Lambda}}c]+(-1)^{(p(a)+1)(p(b)+1)}[b{}_{\widetilde{\Lambda}}[a{}_\Lambda c]],$
\end{enumerate}
for $a,b,c\in R$ and another couple $\widetilde{\Lambda}=(\widetilde{\lambda}, \widetilde{\chi})$ of variables supercommuting with $\Lambda$, then we say $R$ is a SUSY Lie conformal algebra (LCA). 
A SUSY vertex algebra $V$ is a SUSY LCA by considering the $\Lambda$-bracket given by 
\begin{equation} \label{eq:lambda vs Lambda}
    [a{}_\Lambda b]:=[Da{}_\lambda b] +\chi[a{}_\lambda b],
\end{equation}
where the $\lambda$-bracket on the RHS is induced from the LCA structure of the SUSY vertex algebra. Moreover, the SUSY LCA $\Lambda$-bracket \eqref{eq:lambda vs Lambda} directly implies the second property of $D$ in \eqref{eq:D, derivation}. Hence $\Lambda$-bracket is essential to understand a SUSY vertex algebra especially when the vertex algebra is freely generated. Indeed, whenever a SUSY LCA is given, one can get a SUSY vertex algebra by considering the universal enveloping algebra of the LCA \cite{HK07}.
Conversely, the $\Lambda$-bracket on a SUSY vertex algebra shows its LCA structure by \eqref{eq:lambda vs Lambda}. We say a SUSY vertex algebra $V$ is strongly generated by a subset $S$ if  
\[V=\text{Span}_{\mathbb{C}}\{\, :\!(D^{n_s}a_{i_s})\cdots (D^{n_2}   a_{i_2})( D^{n_1} a_{i_1})\!: \, |\ a_{i_1},\cdots, a_{i_s}\in S, \, n_1\cdots n_s\in \mathbb{Z}_+ \,\},\]
and in this case $V$ is 
 strongly generated by $S\cup DS$ as a vertex algebra. 
 
For a totally ordered subset $S$ of a SUSY vertex algebra $V$, let us define the order on $\cup_{n\in \mathbb{Z}_+} D^n S$ by $D^{n_1}a_{i_1}<D^{n_2}a_{i_2}$ (or equivalently, $(a_{i_1}, n_1)<(a_{i_2}, n_2)$) if and only if $a_{i_1}<a_{i_2}$ or $a_{i_1}=a_{i_2}$ and $n_1<n_2$ for $a_{i_1},a_{i_2}\in S.$ We say $V$ is freely generated by $S$ if $\bigsqcup_{n\in \ZZ_+}D^n S$ is a PBW basis of $V.$ When $R=\mathbb{C}[D]\otimes \g$ for a vector superspace $\g$ is a SUSY LCA, the universal enveloping SUSY vertex algebra $V$ is a SUSY vertex algebra freely generated by a basis $S$ of $\g.$ Then obviously, $V$ is freely generated by $S\sqcup DS$ as a vertex algebra.

In some interesting SUSY vertex algebras, there is an element called a superconformal vector. An element $\tau$ of a SUSY vertex algebra $V$ is called a {\it superconformal} if (i) $\tau$ generates a Neveu-Schwarz vertex algebra, that is 
    \begin{equation} \label{eq: N=1 superconformal}
        [ \tau{}_\Lambda \tau] = (2\partial + 3\lambda +\chi D) \tau + \frac{\lambda^2 \chi}{3}c
    \end{equation}
for $c\in \mathbb{C}$ called the central charge,  (ii) $\tau_{(0)}=D$, and (iii) $\frac{1}{2} D\tau$ is a conformal vector of $V$. Hence, one can consider the conformal weight $\Delta$ on $V$. In particular $\Delta_{\tau}=\frac{3}{2}$ and $\Delta_{Da}=\frac{1}{2}+\Delta_a$ for any homogeneous element $a\in V.$ 
When $V$ is a SUSY VOA freely generated by $r^{\bar{0}}_i$ (resp. $r^{\bar{1}}_i$) of conformal weight $\Delta_i$  even (resp. odd) elements, we denote 
\begin{equation}
    V=W_{N=1}(\Delta_1^{r^{\bar{0}}_1}, \Delta_2^{r^{\bar{0}}_2}, \Delta_3^{r^{\bar{0}}_3}, \cdots;\Delta_1^{r^{\bar{1}}_1}, \Delta_2^{r^{\bar{1}}_2}, \Delta_3^{r^{\bar{1}}_3}, \cdots ).
\end{equation}

\begin{example} \label{eq:affine}
    Let $\g$ be a finite dimensional Lie superalgebra with a symmetric invariant bilinear form $(\ |\ ).$ A SUSY current LCA is $R^{cur}_{N=1}(\g):= \mathbb{C}[D]\otimes \bar{\g}\oplus \mathbb{C}K$ where $D$ is the odd endomorphism with $D(K)=0$ and $\bar{\g}$ is the parity reversed vector superspace of $\g$. The $\Lambda$-bracket on $R^{cur}_{N=1}(\g)$ is defined by 
    \begin{equation}
        [\bar{a}{}_\Lambda \bar{b}]=(-1)^{p(a)+p(a)p(b)}\overline{[a,b]}+K\chi(a|b), \quad [\bar{a},K]=0.
    \end{equation}
    Let $V_{N=1}(\g)$ be the universal enveloping vertex algebra of $R_{cur}(\g).$ The level $k$ SUSY affine vertex algebra is \[V^k_{N=1}(\g):= V_{N=1}(\g)/ V_{N=1}(\g)(K-(k+ h^{\vee})).\]  
    By \eqref{eq:lambda vs Lambda}, we can derive the $\lambda$-bracket of $V^k_{N=1}(\g)$
    \begin{equation} \label{eq:affine_lambda}
    \begin{aligned}
         & [\bar{a}{}_\lambda \bar{b}]=(k+ h^{\vee})(a|b), \quad  [D\bar{a}{}_\lambda \bar{b}]=(-1)^{p(a)p(b)+p(a)}\overline{[a,b]},\\
         & [\bar{a}{}_\lambda D\bar{b}]=(-1)^{p(a)p(b)}\overline{[a,b]}, \quad [D\bar{a}{}_\lambda D\bar{b}]=(-1)^{p(a)p(b)}D{\overline{[a,b]}}+\lambda (k+ h^{\vee})(a|b), \\
    \end{aligned}
    \end{equation}
     for $a,b\in \g.$ As a SUSY vertex algebra, $V^k_{N=1}(\g)$ is strongly generated by $\bar{\g}$. As a vertex algebra, this algebra is strongly generated by $\bar{\g}\oplus D\bar{\g}$, and freely generated by $\bar{\mathcal{B}}\sqcup D\bar{\mathcal{B}}$ for a basis $\mathcal{B}$ of $\g.$ Let $h^\vee$ be the dual Coxeter number of $\g$. Then  $V^k_{N=1}(\g)$ for $k\neq -h^\vee$ has the {\it Kac-Todorov superconformal vector} given by  
    \begin{equation}\label{eq: Kac-Todorov}
        \tau_{\g}=\frac{1}{k+h^{\vee}}\sum_{i,j} (v_i|v_j):\bar{v}^i (D\bar{v}^j):+\frac{1}{3(k+h^{\vee})^2}\sum_{i,j,r} (-1)^{p(v_j)}(v_i|[v_j,v_r]):\bar{v}^i \bar{v}^j \bar{v}^r,
  \end{equation}
  where $\{v^i\}$ and $\{v_i\}$ are dual bases of $\g.$ The central charge for the vector is 
  \begin{equation}
      c_{KT}= \frac{k\, \textup{sdim}\ {\g}}{k+h^{\vee}}+\frac{1}{2}\,\textup{sdim}\ \g,
  \end{equation}
  and each $\bar{a}$ and $D\bar{a}$ for $a\in \g$ have conformal weights $\frac{1}{2}$ and $1$, respectively. Therefore,
  \begin{equation}
      V^k_{N=1}(\g)=W_{N=1}\Big( \big(\frac{1}{2}\big)^{\text{dim}\ \g_{\bar{1}}}; \big(\frac{1}{2}\big)^{\text{dim}\ \g_{\bar{0}}}\Big), \quad   V^k_{N=1}(\g)=W\Big( \big(\frac{1}{2}\big)^{\text{dim}\ \g_{\bar{1}}}, 1^{\text{dim}\ \g_{\bar{0}}}; \big(\frac{1}{2}\big)^{\text{dim}\ \g_{\bar{0}}}, 1^{\text{dim}\ \g_{\bar{1}}}\Big).
  \end{equation}
  
    \end{example}

    Recall the level $k$ affine vertex algebra $V^k(\g)$ is endowed with the $\lambda$-bracket $[a{}_\lambda b]=[a,b]+k(a|b)$. Hence $V^{k+ h^{\vee}}(\g)$ can be embedded into $V^k_{N=1}(\g)$ via the map 
    \begin{equation} \label{eq:affine, ord vs SUSY}
        a \mapsto \sqrt{-1}^{p(a)}D\bar{a}.
    \end{equation}
     % If $V$ is a vertex subalgebra of a SUSY vertex algebra $V_{SS}$ then we say $V_{SS}$ is a {\it SUSY extension} of $V$. By the observation above, we can conclude $V^k_{N=1}(\g)$ is a SUSY extension of $V^{k+ h^{\vee}}(\g).$  {\color{blue} SUSY extension?}

\section{W-superalgebras and SUSY W-algebras} \label{sec:W-alg}
In this section, we recall the basic properties of W-algebras introduced in \cite{KRW03} and of SUSY W-algebras in \cite{MRS21}. For the screening operators of ordinary and SUSY W-algebras, we refer to \cite{Genra17} and \cite{Song24free}, respectively.

Throughout this paper, let $\g$ be a finite basic simple Lie superalgebra with an $\mathfrak{osp}(1|2)$-subalgebra $\mathfrak{s}$. Note that the even subspace $\mathfrak{s}_{\bar{0}} \simeq \sll(2)$ of $\mathfrak{s}$ is spanned by the $\sll(2)$-triple $(E,H=2x,F)$. The nondegenerate supersymmetric invariant bilinear form $( \ | \ )$ is assumed to be  normalized by $(E|F)=2(x|x)=1.$ We consider the $\mathbb{Z}/2$-grading of $\g$ given as follows:
\begin{equation} \label{eq: g grading}
    \g=\bigoplus_{i\in \mathbb{Z}/2} \g_i, \quad [x,a]=i a \text{ for } a\in \g_i.
\end{equation}
Consider the subalgebras $\n = \bigoplus_{i>0} \g_i$, $\n_-:=\bigoplus_{i<0}\g_i$, and $\mathfrak{m}=\bigoplus_{i\geq 1} \g_i$ of $\g$ and denote the complimentary subspace of $\n$ by $\mathfrak{p}:=\bigoplus_{i\leq 0}\g_i.$  Set $I$ to be the set of roots of $\g$ and
\begin{equation}
    I_+=\{\alpha\in I\,|\,\textup{a root vector of }\alpha \textup{ is contained in }\n\}.
\end{equation}

\subsection{W-superalgebras} 
Recall the affine vertex algebra $V^k(\g)$ of level $k$. Consider the charged free fermion vertex algebra 
$\mathcal{F}^{\text{ch}}(\n)$ and the neutral free fermion vertex algebra $\mathcal{F}^{\text{ne}}$, which are strongly generated by  
$\phi_{\bar{\n}}\oplus \phi^{\bar{\n}_-}=\{\phi_{\bar{n}}|n\in \n\} \oplus \{\phi^{\bar{\n}_-}|n_-\in \n_-\} \simeq \bar{\n} \oplus \bar{\n}_-$ and $\Phi_{\g_{1/2}}=\{\Phi_{[m]}|m\in \g_{1/2}\}\simeq \g_{1/2},$ respectively, where $\bar{\n}$ and $\bar{\n}_-$ are parity reversed spaces of $\n$ and $\n_-$.
The OPE relations in $\mathcal{F}^{\text{ch}}(\n)$ and $\mathcal{F}^{\text{ne}}$ are determined by the following:
\begin{equation}
    [\phi_{\bar{n}}{}_\lambda \phi^{\bar{n}'}]=(n|n'), \quad [\Phi_{[m]}{}_\lambda \Phi_{[m']}]= (F|[m,m']).
\end{equation}
Take the vertex algebra 
\begin{equation}
    C^k(\g,F)= V^k(\g)\otimes \mathcal{F}^{\text{ch}}(\mathfrak{n}) \otimes \mathcal{F}^{\text{ne}}
\end{equation}
and an odd element 
\begin{equation}
    \mathrm{d}= \sum_{\alpha\in I_+}:u_\alpha \phi^\alpha: +  \sum_{\alpha\in I_{1/2}}(-1)^{p(\alpha)}:\Phi_{\alpha} \phi^\alpha:+ \phi^F + \frac{1}{2}\sum_{\alpha,\beta\in I_+}(-1)^{p(\alpha)}:\phi_{[u_\alpha,u_\beta]}\phi^\beta\phi^\alpha :
\end{equation}
in $C^k(\g,F)$, where $\{u_\alpha\,|\,\alpha\in I_+\}$ and $\{u^\alpha\,|\,I_+\}$ are bases of $\n$ and $\n_-$ such that $(u^\alpha|u_\beta)=\delta_{\alpha,\beta}$, $\phi^{\alpha}:= \phi^{\bar{u}^{\alpha}}$ and $\phi_{\alpha}:=\phi_{u_\alpha}.$
The subset $I_{1/2}\subset I_+$ consists of $\alpha$ such that $u_\alpha\in\g_{1/2}$ and $\Phi_{\alpha}:= \Phi_{[u_\alpha]}$. We further assume that each $u_\alpha$ is a root vector for $\alpha\in I_+$. For the odd differential $\mathcal{Q}:= \mathrm{d}_{(0)},$ the  W-superalgebra $W^k(\g,F)$ of level $k$ is 
\begin{equation}
     W^k(\g,F)= H(C^k(\g,F), \mathcal{Q}).
\end{equation}
Let us denote 
\begin{equation}
    \mathrm{J}_{a}:= a+\sum_{\alpha\in I_+}(-1)^{p(\alpha)}: \phi^\alpha \phi_{\pi_{+}([u_\alpha, a])}:\ \in C^k(\g,F) 
\end{equation}
for $a\in \g$ and the canonical projection map $\pi_+:\g \to  \n$. In $C^k(\g,F)$ for $k\neq -h^\vee$, there is a conformal vector $L$ with the central charge 
\begin{equation} \label{eq:conformal vector}
    c^k=\frac{k\, \textup{sdim}\ \g}{k+h^{\vee}}-6k-\sum_{\alpha\in I_+}(-1)^{p(\alpha)}(12 m_{\alpha}^2-12 m_{\alpha}+2)-\frac{1}{2}\, \textup{sdim}\ \g_{\frac{1}{2}}.
\end{equation} 
and the conformal weight induced from $L$ is given by 
\begin{equation} \label{eq:conformal weight_nonSUSY}
    \Delta_{a}=1-j_a, \quad \Delta_{\Phi_{[m]}}=\frac{1}{2}, \quad \Delta_{\phi^{\alpha}}=j_\alpha, \quad \Delta_{\phi_{\alpha}}=1-j_\alpha
\end{equation}
for $a\in \g_{j_a}$ and $u_\alpha \in \g_{j_\alpha}.$ In addition, $L$ is in the kernel of $\mathcal{Q}$ and hence it is a conformal vector of $W^k(\g,F).$ (See \cite[Theorem 2.2]{KRW03} for details.)

In \cite{DK06}, the authors showed that $ H(C^k(\g,F), \mathcal{Q})$ is isomorphic to the cohomology $ H(\widetilde{C}^k(\g,F), \widetilde{\mathcal{Q}})$, where $\widetilde{C}^k(\g,F)$ is the vertex subalgebra of $C^k(\g,F)$ generated by $\mathrm{J}_{\mathfrak{p}}\oplus \phi^{\bar{\n}_-}\oplus\Phi_{\g_{1/2}}$ and $\widetilde{\mathcal{Q}}=\mathcal{Q}|_{\widetilde{C}^k(\g,F)}.$ In particular, $W^k(\g,F)$ is a vertex subalgebra of $V^{\tau_k}(\mathrm{J}_{\mathfrak{p}})\otimes \mathcal{F}^{\text{ne}}\subset \widetilde{C}^k(\g,F)$, where $V^{\tau_k}(\mathrm{J}_{\mathfrak{p}})$ is the vertex subalgebra generated by  $\mathrm{J}_{\mathfrak{p}}$ and the $\lambda$-bracket is given by 
\begin{equation}
    [\mathrm{J}_a{}_\lambda \mathrm{J}_b] = \mathrm{J}_{[a,b]}+ \lambda \tau_k (a|b)
\end{equation}
for $a,b\in \mathfrak{p}$ and $\tau_k(a|b)= k (a|b) + \frac{1}{2}\kappa_{\g}(a|b)-\frac{1}{2}\kappa_{\g_0}(a|b)$ for the Killing form $\kappa$. Recall the conformal weight \eqref{eq:conformal weight_nonSUSY} which induces  the $\mathbb{Z}_+/2$ conformal grading on 
$V^{\tau_k}(\mathrm{J}_{\mathfrak{p}})\otimes \mathcal{F}^{\text{ne}}$:
\begin{equation}
    \Delta_{\mathrm{J}_a} = 1-j_a, \quad \Delta_{\Phi_{[m]}}=\frac{1}{2}.
\end{equation}
Additionally,  we consider the increasing filtration $\mathcal{F}^{\text{ne}}=F(\widetilde{C})^0\subset \cdots \subset F(\widetilde{C})^{p+1/2}\subset F(\widetilde{C})^{p}\subset F(\widetilde{C})^{p-1/2}\subset \cdots $ induced by the $\frac{\mathbb{Z}}{2}$ grading 
\begin{equation}
    \text{gr}(\mathrm{J}_a)= j_a-1/2, \quad   \text{gr}(\Phi_{[m]})=\text{gr}(\partial)=0
\end{equation}
for $a\in \g_{j_a}.$ 
\begin{proposition} \cite[Theorem 5.9]{DK06} \label{prop:generator_1}
    Let $\{a^F_{i}|i=1,\cdots, \mathrm{r}\}$ be a basis of $\g^F=\ker(\textup{ad} F)$ and let $j_i \in -\frac{\mathbb{Z}_+}{2}$ be given by $a^F_i \in \g_{j_i}$ so that $\mathrm{J}_{a^F_i}\in F(\widetilde{C})^{j_\alpha-1/2}$ and has the conformal weight $1-j_i.$ Then there exists a free generating set $ \{ v^F_i| i \in I^F \}\subset V^{\tau_k}(\mathrm{J}_{\mathfrak{p}})\otimes \mathcal{F}^{\text{ne}}$ of $W^k(\g,F)$ satisfying the following properties: 
\[ \text{ $\mathrm{(i)} \ \Delta_{v^F_i}= \Delta_{J_{a^F_i}}= 1-j_{i}\ ,$
        \quad $\mathrm{(ii)} \ v^F_i \in  F(\widetilde{C})^{j_i-1/2}$ and $v^F_i-\mathrm{J}_{a^F_i} \in F(\widetilde{C})^{j_i}$.}\]
\end{proposition}

By Proposition \ref{prop:generator_1}, we conclude that 
\begin{equation} \label{eq:W-type}
    W^k(\g,F)= W\Big( 1^{m_2^{\bar{0}}}, \big( \frac{3}{2} \big)^{m_3^{\bar{0}}}, 2^{m_4^{\bar{0}}} , \cdots ;1^{m_2^{\bar{1}}}, \big( \frac{3}{2} \big)^{m_3^{\bar{1}}},2^{m_4^{\bar{1}}}, \cdots \Big),
\end{equation}
where $\text{dim}(\g^F_r)_{\bar{0}}= m^{\bar{0}}_{2(1-r)}$ and $\text{dim}(\g^F_r)_{\bar{1}}= m^{\bar{1}}_{2(1-r)}.$

\subsection{Miura maps for W-superalgebras} \label{subsec: nonsusy miura}
For each $\alpha\in I_+$, we say $\alpha$ is \textit{indecomposable} if $\alpha$ cannot be written as a linear summation of elements in $I_+$. Otherwise, we say that $\alpha$ is decomposable. Denote the set of indecomposables by
\begin{equation} \label{eq: Delta indecomp roots}
    \Delta:=\{\alpha\in I_+\,|\,\alpha \textup{ is indecomposable}\}.
\end{equation}
Now, for $I_0:=\{\alpha\in I\,|\,\textup{a root vector of }\alpha \textup{ is contained in }\g_0\}$ and $S_0:= \sum_{\alpha\in I_0}\ZZ \alpha$, define the equivalence relation $\sim$ on $\Delta$ by $\alpha\sim \beta \Leftrightarrow \alpha-\beta\in S_0$. Let $[\Delta]$ be the set of equivalence classes and denote its elements by $[\alpha]$ for $\alpha\in \Delta$. Note that if we choose $F$ to be principal, $\Delta$ is equal to the set of simple roots of $\g$ and the equivalence relation $\sim$ is trivial.

Recall that the W-superalgebra $W^k(\g,F)$ is a vertex subalgebra of $V^{\tau_k}(\mathrm{J}_{\mathfrak{p}})\otimes \mathcal{F}^{\text{ne}}$. By composing with the surjective projection map $V^{\tau_k}(\mathrm{J}_{\mathfrak{p}})\simeq V^{\tau_k}(\mathfrak{p})\twoheadrightarrow V^{\tau_k}(\g_0)$, one obtains the \textit{Miura map} for W-superalgebras:
\begin{equation}
    \mu^k : W^k(\g,F)\rightarrow V^{\tau_k}(\g_0)\otimes \mathcal{F}^{\text{ne}},
\end{equation}
which is known to be injective for arbitrary $k\in \CC$. In particular, it was shown in \cite{Genra17} that, for generic $k$, the image of the Miura map is given by the intersection of the kernels of screening operators. To be explicit, the image of the Miura map $\mu^k$ is equal to
\begin{equation} \label{eq: nonsusy screening}
    W^k(\g,F)\simeq \bigcap_{[\alpha]\in [\Delta]}\text{Ker}\Big(\frac{\sqrt{-1}}{\sqrt{k+h^{\vee}}}\sum_{\alpha\in [\alpha]}\int :\!\phi^\alpha \Phi_{\alpha}\!:\!(z)dz: V^{\tau_k}(\g_0)\rightarrow \mathcal{F}^{\textup{ne}}\otimes V^{\tau_k}(\g_0)\otimes \bigoplus_{\beta\in [\alpha]}\CC \phi^{\beta}\Big),
\end{equation}
where each action of $:\!\phi^{\alpha}\Phi_{\alpha}\!:$ is given by the OPE relations in the subcomplex $\widetilde{C}^k(\g,F)$ provided that $\phi^{\gamma}$ for decomposable $\gamma$ is $0$. For each $[\alpha]\in [\Delta]$,
\begin{equation} \label{eq: nonsusy Salpha}
    S_{[\alpha]}=\frac{\sqrt{-1}}{\sqrt{k+h^{\vee}}}\sum_{\alpha\in [\alpha]}\int :\!\phi^\alpha \Phi_{\alpha}\!:\!(z)dz
\end{equation}
is called the screening operator for $W^k(\g,F)$. This is the result of applying the cohomological argument. By defining an appropriate filtration on $\widetilde{C}^k(\g,F)$, one can show that
\begin{equation*}
    W^k(\g,F)\simeq H^0(\mathcal{E}_1^k,\mathcal{Q}_1),
\end{equation*}
where $\mathcal{E}_1^k$ is the first total complex obtained by the induced spectral sequence, and $\mathcal{Q}_1$ is the part of $\mathcal{Q}$ which strictly increases the filtration. To be explicit, we have
\begin{equation} \label{eq: nonsusy first total}
    \mathcal{E}^k_1=H(\widetilde{C}^k(\g,F), (\mathrm{d}_{\textup{st}})_{(0)}), \qquad \mathcal{Q}_1=(\mathrm{d}_{F})_{(0)},
\end{equation}
where $\mathrm{d}=\mathrm{d}_{\textup{st}}+\mathrm{d}_{F}$ for
\begin{equation*}
        \mathrm{d}_{\textup{st}}:= \sum_{\alpha\in I_+}:\!u_\alpha \phi^\alpha\!: + \frac{1}{2}\sum_{\alpha,\beta\in I_+}(-1)^{p(\alpha)}:\!\phi_{[u_\alpha,u_\beta]}\phi^\beta\phi^\alpha\!:, \quad  \mathrm{d}_F:=\sum_{\alpha\in I_{1/2}}(-1)^{p(\alpha)}:\!\Phi_{\alpha} \phi^\alpha\!:+ \phi^F.
\end{equation*}
For generic $k$, the first total complex $\mathcal{E}^k_1$ can be described explicitly, in which $\phi^{\gamma}=0$ whenever $\gamma$ is decomposable. The differential $\mathcal{Q}_1$ then yields the screening operators in \eqref{eq: nonsusy Salpha}. See \cite{Genra17} for details.

In the codomain of the screening operators, we denote the subspace
\begin{equation}
    M_{[\alpha]}=V^{\tau_k}(\g_0)\otimes \bigoplus_{\beta\in [\alpha]}\CC x_{\beta}\simeq V^{\tau_k}(\g_0)\otimes \bigoplus_{\beta\in [\alpha]}\CC\phi^{\beta}
\end{equation}
to forget the vertex algebra structure and regard it as a $V^{\tau_k}(\g_0)$-module. Note that in the subcomplex $\widetilde{C}^k(\g,F)$, we have \begin{equation}
    [\mathrm{J}_u{}_{\lambda}\phi^{\beta}]=\sum_{\gamma\in I_+}(u^{\beta}|[u_{\gamma},u])\phi^{\gamma}, \quad u\in \g_0,\ \beta\in I_+.
\end{equation}
Therefore, the structure of $M_{[\alpha]}$ as a $V^{\tau_k}(\g_0)$-module is given by
\begin{equation} \label{eq: M_alpha module structure}
    [u{}_{\lambda}x_{\beta}]=\sum_{\gamma\in [\beta]}(u^\beta|[u_{\gamma},u])x_{\gamma}, \quad u\in \g_0
\end{equation}
for $\beta\in [\alpha]$, while $u_{(-n)}$ for $n>0$ acts as a left multiplication.

\begin{remark}
    In \cite{Genra17}, the screening operator $S_{[\alpha]}$ for $[\alpha]\in [\Delta]$ differs depending on whether $u_{\alpha}\in \g_{\frac{1}{2}}$ or $\g_1$. However, in our setting, one can deduce that $u_{\alpha}\in \g_{\frac{1}{2}}$ for any indecomposable $\alpha$. From the assumption that $\g$ has a subalgebra $\mathfrak{s}$ isomorphic to $\mathfrak{osp}(1|2)$, any $u_{\alpha}$ with degree $m\geq 1$ is in the image of $\text{ad}\, e$, which implies that $\alpha$ is decomposable.
\end{remark}

\begin{remark}
    In \cite{Genra17}, the formula \eqref{eq: nonsusy Salpha} does not contain the constant multiple $\frac{\sqrt{-1}}{\sqrt{k+h^{\vee}}}$. This is modification is for the comparison with the SUSY screening operators in Section \ref{sec: susy vs. nonsusy}.
\end{remark}

\subsection{SUSY W-algebras}  Let  $e\in\g_{1/2}$ and $f\in \g_{-1/2}$ be odd elements in $\mathfrak{s}_{\bar{1}}$ such that
\begin{equation} \label{eq:f and e}
    [f,f]=-2F, \quad [e,e]=2E.
\end{equation}

A {\it SUSY W-algebra} $W^k_{N=1}(\g,f)$ associated with $\g$ and $f$ is a quantum Hamiltonian reduction of the SUSY affine vertex algebra introduced in Example \ref{eq:affine}. In other words, $W^k_{N=1}(\g,f)$ is the cohomology of the SUSY BRST complex 
\begin{equation} \label{eq:SUSY BRST}
    C_{N=1}^k(\g,f) = V_{N=1}^k(\g) \otimes \mathcal{F}^{\text{ch}}_{N=1}(\n),
\end{equation}
where $\mathcal{F}^{\text{ch}}_{N=1}(\n)$ is the SUSY fermion vertex algebra freely generated by $\phi^{\bar{\n}_-}\simeq \bar{\n}_-$ and $\phi_{\n}\simeq \n$ endowed with the $\Lambda$-bracket $[\phi^{\bar{a}}{}_\Lambda \phi_b]= [\phi_b{}_\Lambda \phi^{\bar{a}}]=(a|b).$ In other words, $\mathcal{F}^{\text{ch}}_{N=1}(\n)$ is freely generated vertex algebra by $\phi^{\bar{\n}_-}\sqcup D \phi^{\bar{\n}_-} \sqcup \phi_{\n} \sqcup D \phi_{\n}$ and $[D\phi^{\bar{a}}{}_\lambda \phi_b]= [D\phi_b{}_\lambda \phi^{\bar{a}}]=(a|b).$
The differential $Q$ on $C_{N=1}^k(\g,f)$ is given by $Q=(Dd)_{(0)}$ for 
\begin{equation}
    d= \sum_{\alpha\in I_+} :(\bar{u}_\alpha-(f|u_\alpha))\phi^\alpha:+ \frac{1}{2}\sum_{\alpha,\beta\in I_+} (-1)^{p(\alpha)p(\bar{\beta})}:\phi_{[u_\alpha,u_\beta]}\phi^{\beta}\phi^{\alpha}:,
\end{equation}
where  $p(\alpha)$ and $p(\bar{\beta})$ are the parities of $u_\alpha$ and $\bar{u}_\beta$, respectively, and $\phi^{\alpha}:=\phi^{\bar{u}^{\alpha}}$.
Now the SUSY W-algebra   
\begin{equation}
    W^k_{N=1}(\g,f)= H(C_{N=1}^k(\g,f), Q)
\end{equation}
 has the SUSY vertex algebra structure induced from the $\Lambda$-bracket on $C_{N=1}^k(\g,f).$ For $k\neq -h^{\vee}$, the complex $C^k_{N=1}(\g,f)$ has a superconformal vector
\begin{equation} \label{eq:tau_W}
    \tau_C:=\tau_{\g}+\tau_{\mathcal{F}}+\partial \bar{H}\in C_{N=1}^k(\g,f)
\end{equation}
of central charge
\begin{equation} \label{eq: superconformal susy W-alg central charge}
    c_{N=1}^k=\frac{k\, \textup{sdim}{\g}}{k+h^{\vee}}+\frac{1}{2}\,\textup{sdim}\g+12\sum_{\alpha\in I_+}(-1)^{p(\alpha)}j_{\alpha}-3\, \textup{sdim}(\n)-6(k+h^{\vee}),
  \end{equation}
where $[x,u_{\alpha}]=j_{\alpha}u_{\alpha}$ for each $\alpha \in I_+$ \cite{Song24free}. The summand $\tau_{\g}$ in \eqref{eq:tau_W} is the Kac-Todorov vector in \eqref{eq: Kac-Todorov}, and $\tau_{\mathcal{F}}$ is a superconformal vector in $\mathcal{F}^{\text{ch}}_{N=1}(\mathfrak{n})$ given by
\begin{equation}
    \begin{aligned}
    \tau_{\mathcal{F}} =&\sum_{\alpha\in I_+} (-1)^{p(\alpha)}2j_{\alpha}:(\partial\phi_{\alpha})\phi^{\alpha}:-\sum_{\alpha\in I_+} (-1)^{p(\alpha)}(1-2j_{\alpha}):\phi_{\alpha}\partial \phi^{\alpha}:+\sum_{\alpha\in I_+}:(D\phi_{\alpha})(D\phi^{\alpha}):.
    \end{aligned}
  \end{equation}
The conformal weights of the generators are  
\begin{equation}
    \Delta_{\bar{a}}=\frac{1}{2}-j_a, \quad \Delta_{\phi_{\alpha}}=\frac{1}{2}-j_\alpha, \quad 
    \Delta_{\phi^{\alpha}}=j_\alpha, \quad 
    \Delta_{D}=\frac{1}{2}
\end{equation}
for $a\in \g_{j_a}$ and $u_\alpha\in \g_{j_\alpha}.$ Now one can check that $\tau_C$ is in the kernel of $Q$ and induces a superconformal vector $\tau_W$ of $W_{N=1}^k(\g,f)$.

Consider the vertex subalgebra  $\widetilde{C}_{N=1}^k(\g,f)$ of $C_{N=1}^k(\g,f)$ generated by $J_{\bar{\p}}, \ DJ_{\bar{\p}},\ \phi^{\n_-}$, and $D\phi^{\n_-}$, where $J_{\bar{\p}}=\{J_{\bar{a}}|a\in \p\}$ and  
\begin{equation} \label{eq:building block}
    J_{\bar{a}}= \bar{a}+\sum_{\beta\in I_+}(-1)^{p(\bar{a})p(\bar{\beta})} \phi^{\beta}\phi_{\pi_{>0}[u_\beta,a]}.
\end{equation}
Then $\Delta_{\bar{a}}=\Delta_{J_{\bar{a}}}=\frac{1}{2}- j_a$. Indeed, by direct computations, we get 
 \begin{equation} \label{eq:building block conf weight}
    [\tau_W {}_{\Lambda} J_{\bar{a}}]=\Big(2\partial+2\Big(\frac{1}{2}-j_a\Big)\lambda+\chi D\Big)J_{\bar{a}}-(k+h^{\vee})(e|[f,a])\lambda\chi.
 \end{equation}
The  $\Lambda$ brackets between the generators are 
\begin{equation} \label{eq:J and J}
\begin{aligned}
       &  [J_{\bar{a}}{}_\Lambda J_{\bar{b}}]=(-1)^{p(a)p(b)+p(a)}J_{\overline{[a,b]}}+(k+ h^{\vee})\chi(a|b), \\
       & [\phi^{\overline{m}}{}_\Lambda J_{\bar{a}}] = \sum_{\beta\in I_+} (-1)^{p(m)p(\bar{a})} \phi^{\overline{[m,a]}}.
\end{aligned}
\end{equation}
Equivalently, in terms of $\lambda$-bracket, we have  
\begin{equation} \label{eq:phi and J}
\begin{aligned}
        & [\phi^{m}{}_\lambda J_{\bar{a}}]=0, \quad  [D\phi^{\overline{m}}{}_\lambda J_{\bar{a}}]= (-1)^{p(m)p(\bar{a})} \phi^{\overline{[m,a]}}, \\
        & [\phi^{\overline{m}}{}_\lambda DJ_{\bar{a}}]= (-1)^{p(m)p(a)} \phi^{\overline{[m,a]}}, \quad [D\phi^{\overline{m}}{}_\lambda DJ_{\bar{a}}]=(-1)^{p(m)p(a)} D\phi^{\overline{[m,a]}}.
\end{aligned}
\end{equation}
The endomorphism
$\widetilde{Q}:= Q|_{\widetilde{C}^k_{N=1}({\g},f)}$ becomes a differential on  $\widetilde{C}^k({\g},f)$ and
\begin{equation} \label{eq:differential}
    \widetilde{Q}(J_{\bar{a}})= \sum_{\beta\in I_+}(-1)^{p(\bar{a})p(\beta)}: \phi^\beta \Big( J_{\pi_{\leq 0}[u_\beta,a]}+(f|[u_\beta,a])\Big): +\sum_{\beta\in I_+} (-1)^{p(\bar{\beta})}(k+h^\vee)D\phi^\beta(u_\beta|a),
\end{equation}
where $\pi_{\leq 0}: \g \to \p$ is the canonical projection map. Moreover, it is known by \cite{MRS21} that 
\begin{equation}
    W^k_{N=1}(\g,f)\simeq H(\widetilde{C}_{N=1}^k(\g,f), \widetilde{Q})\subset V^{\psi_k}_{N=1}(J_{\mathfrak{p}}),
\end{equation}
where $V^{\psi_k}_{N=1}(J_{\p})$ is the SUSY vertex subalgebra of $\widetilde{C}^k_{N=1}(\g,f)$ generated by $J_{\bar{\mathfrak{p}}}$. Here, $\psi_k$ is the bilinear form $\psi_k(a|b)=(k+h^{\vee})(a|b)$. By the first equation of \eqref{eq:J and J}, it is isomorphic to SUSY vertex subalgebra of $V^k_{N=1}(\g)$ generated by $\bar{\mathfrak{p}}.$ Analogous to the non-SUSY case, we consider another grading on $V^{\psi_k}_{N=1}(J_{\bar{\mathfrak{p}}})$ defined by 
\begin{equation}
    \text{gr}(J_{\bar{a}})=j_a, \quad \text{gr}(D)=0
\end{equation}
and the corresponding increasing filtration $F_{N=1}(\widetilde{C})^0 \subset F_{N=1}(\widetilde{C})^{-1/2} \subset F_{N=1}(\widetilde{C})^{-1}\cdots$. Then there is a free generating set of $W^k_{N=1}(\g,f)$ described in the following proposition.

\begin{proposition} \cite[Theorem 4.11]{MRS21} \label{prop:generator_SUSY}
     Let $\{a^f_{i}|i=1, \cdots, r\}$ be a basis of $\g^f=\textup{ker}(\textup{ad} f)$ and let $j_i \in -\frac{\mathbb{Z}_+}{2}$ be given by $a^f_i \in \g_{j_i}$ so that $J_{\bar{a}^f_i}\in F(\widetilde{C})^{j_i}$ and has the conformal weight $\frac{1}{2}-j_i.$ Then there exists a free generating set $ \{\,  v^f_{\imath},\, D v^f_{\imath}\, |\,  \i \in I^f \}\subset V^{\psi_k}_{N=1}(J_{\mathfrak{p}})$ of $W^k_{N=1}(\g,f)$ satisfying the following properties: 
\[ \text{ $\mathrm{(i)} \ \Delta_{v^f_{\imath}}= \Delta_{J_{\bar{a}^f_i}}= \frac{1}{2}-j_i\ ,$
        \quad $\mathrm{(ii)} \ v^f_{\imath} \in  F_{N=1}(\widetilde{C})^{j_\alpha}$ and $ v^f_{\imath}-J_{\bar{a}^f_i} \in F_{N=1}(\widetilde{C})^{j_i+1/2}$.}\]
\end{proposition}

By Proposition \ref{prop:generator_1}, we conclude that 
\begin{equation}
W_{N=1}^k(\g,f)= W_{N=1}\Big( \big( \frac{1}{2} \big)^{l_1^{\bar{0}}}, 1^{l_2^{\bar{0}}}, \big( \frac{3}{2} \big)^{l_3^{\bar{0}}}, 2^{l_4^{\bar{0}}} , \cdots ;\big( \frac{1}{2} \big)^{l_1^{\bar{1}}}, 1^{l_2^{\bar{1}}}, \big( \frac{3}{2} \big)^{l_3^{\bar{1}}},2^{l_4^{\bar{1}}}, \cdots \Big).
\end{equation}
where $\text{dim}(\g^f_r)_{\bar{0}}= l^{\bar{0}}_{2(\frac{1}{2}-r)}$ and $\text{dim}(\g^f_r)_{\bar{1}}= l^{\bar{1}}_{2(\frac{1}{2}-r)}.$
As a vertex algebra, 
\begin{equation}
W_{N=1}^k(\g,f)= W\Big( \big( \frac{1}{2} \big)^{l_1^{\bar{0}}}, 1^{l_2^{\bar{0}}+l_1^{\bar{1}}}, \big( \frac{3}{2} \big)^{l_3^{\bar{0}}+l_2^{\bar{1}}}, 2^{l_4^{\bar{0}}+l_3^{\bar{1}}} , \cdots ;\big( \frac{1}{2} \big)^{l_1^{\bar{1}}}, 1^{l_2^{\bar{1}}+l_1^{\bar{0}}}, \big( \frac{3}{2} \big)^{l_3^{\bar{1}}+l_2^{\bar{0}}},2^{l_4^{\bar{1}}+l_3^{\bar{0}}}, \cdots \Big).
\end{equation}
Here we remark that $m^{\bar{0}}_{r} = l^{\bar{0}}_{r}+  l^{\bar{1}}_{r-1}$ and $m^{\bar{1}}_{r} = l^{\bar{1}}_{r}+  l^{\bar{0}}_{r-1}$
for $m^{\bar{0}}_{r}$ and $m^{\bar{0}}_{r}$ in \eqref{eq:W-type} when  $r\geq 2$.

\subsection{SUSY Miura maps for SUSY W-algebras} \label{subsec: susy miura}
Recall that the SUSY W-algebra $W^k_{N=1}(\g,f)$ is a vertex subalgebra of $V^{\psi_k}_{N=1}(\mathfrak{p})$ considering the building blocks $J_{\bar{a}}$'s. Denote the SUSY vertex subalgebra of $V^{\psi_k}_{N=1}(\mathfrak{p})$ generated by $J_{\bar{\g}_0}$ by $V^{\psi_k}_{N=1}(\g_0)$. Using the projection map $V^{\psi_k}_{N=1}(\mathfrak{p})\twoheadrightarrow V^{\psi_k}_{N=1}(\g_0)$, one obtains the \textit{SUSY Miura map} for SUSY W-algebras:
\begin{equation}
    \widetilde{\mu}^k: W^k_{N=1}(\g,f)\rightarrow V^{\psi_k}_{N=1}(\g_0).
\end{equation}
For any $k\neq -h^{\vee}$, the SUSY Miura map $\widetilde{\mu}^k$ is known to be injective \cite{GSS25}. Moreover, as in nonSUSY theory, the image of the SUSY Miura map for generic $k$ is given by the intersection of the kernels of the screening operators \cite{Song24free}.

Let us inherit the notions $\Delta$ and $[\Delta]$ from Section \ref{subsec: nonsusy miura}. Then, the image of the SUSY Miura map $\widetilde{\mu}^k$ for generic $k$ is equal to
\begin{equation} \label{eq: susy screening}
     W^k_{N=1}(\g,f)\simeq \bigcap_{[\alpha]\in [\Delta]}\text{Ker}\Big(\sum_{\alpha\in [\alpha]}(f|u_{\alpha})\int D\phi^{{\alpha}}(z)dz: V^{\psi_k}_{N=1}(\g_0)\rightarrow V^{\psi_k}_{N=1}(\g_0)\otimes \bigoplus_{\beta\in [\alpha]}\CC \phi^{{\beta}}\Big).
\end{equation}
As in \eqref{eq: nonsusy screening}, the action $D\phi^{\alpha}$ in \eqref{eq: susy screening} is given by the OPE relations in the subcomplex $\widetilde{C}^k_{N=1}(\g,f)$ provided that $\phi^{\gamma}$ for decomposable $\gamma$ is $0$. For each $[\alpha]\in [\Delta]$,
\begin{equation} \label{eq: define susy screening}
    S_{[\alpha]}^{N=1}=\sum_{\alpha\in [\alpha]}(f|u_{\alpha})\int D\phi^{{\alpha}}(z)dz
\end{equation}
is called the screening operator for SUSY W-algebra $W^k_{N=1}(\g,f)$. As for W-superalgebras, it can be shown by applying the spectral sequence argument. One can show that $W^k_{N=1}(\g,f)\simeq H^0(E_1^k,Q_1)$, where
\begin{equation} \label{eq: susy first total}
    E^k_1=H(\widetilde{C}^k_{N=1}(\g,f), (Dd_{\textup{st}})_{(0)}), \qquad Q_1=(Dd_{f})_{(0)}
\end{equation}
are the first total complex and the induced differential on it. Here,  $d=d_{\textup{st}}+d_{f}$ for
\begin{equation*}
        d_{\textup{st}}:= \sum_{\alpha\in I_+} :\!\bar{u}_\alpha \phi^\alpha\!:+ \frac{1}{2}\sum_{\alpha,\beta\in I_+} (-1)^{p(\alpha)p(\bar{\beta})}:\!\phi_{[u_\alpha,u_\beta]}\phi^{\beta}\phi^{\alpha}\!:, \quad  d_f:=-\sum_{\alpha\in I_+} (f|u_\alpha)\phi^\alpha.
\end{equation*}
For generic $k$, the element $\phi^{\gamma}$ vanishes in $E^k_1$ for decomposable $\gamma$, and the differential $Q_1$ gives rise to the screening operators \eqref{eq: define susy screening}.

In the range of the screening operators, we ignore their vertex algebra structure and consider them only as $V^{\psi_k}_{N=1}(\g_0)$-modules, regarding $V^{\psi_k}_{N=1}(\g_0)$ as a vertex algebra. To emphasize this spirit, we denote the range of the screening operators by
\begin{equation}
    \widetilde{M}_{[\alpha]}=V^{\psi_k}_{N=1}(\g_0)\otimes \bigoplus_{\beta\in [\alpha]}\CC \widetilde{x}_{\beta},
\end{equation}
after identifying $\phi^{\beta}$ with $\widetilde{x}_{\beta}$. The $V^{\psi_k}_{N=1}(\g)$-module structure of $\widetilde{M}_{[\alpha]}$ is given by
\begin{equation} \label{eq: Mtild_alpha module structure}
    [D\bar{u}{}_{\lambda}\widetilde{x}_{\beta}]=\sum_{\gamma\in [\beta]}(-1)^{(p(\alpha)+1)p(u)}([u,u^{\beta}]|u_{\gamma})\widetilde{x}_{\gamma}, \quad [\bar{u}{}_{\lambda}\widetilde{x}_{\beta}]=0, \quad u\in \g_0,
\end{equation}
while $D\bar{u}_{(-n)}$ and $\bar{u}_{(-n)}$ for $n>0$ acts as a left multiplication. Note that the action \eqref{eq: Mtild_alpha module structure} follows from the $\Lambda$-bracket relation in the subcomplex $\widetilde{C}^k_{N=1}(\g,f)$:
\begin{equation*}
    [J_{\bar{u}}{}_{\Lambda}\phi^{{\beta}}]=\sum_{\gamma\in I_+}(-1)^{(p(\alpha)+1)p(u)}([u,u^{\beta}]|u_{\gamma})\phi^{{\gamma}}, u\in \g_0.
\end{equation*}

\section{Relations between W-superalgebras and SUSY W-algebras} \label{sec: susy vs. nonsusy}
Throughout this section, let $\g$ be a simple basic Lie superalgebra equipped with an odd and even nilpotent elements $f$ and $F=-\frac{1}{2}[f,f]$. For the $\mathfrak{sl}(2)$-triple formed by $F$, consider the grading \eqref{eq: g grading} on $\g$ and assume $f\in \g_{-\frac{1}{2}}$. Note that $f$ and $F$ can be completed to subalgebra $\mathfrak{s}$ of $\g$, which is isomorphic to $\mathfrak{osp}(1|2)$.
In the paper \cite{MadRag94} by Madsen and Ragoucy, the following vertex algebra isomorphism has been proposed for any $k\neq -h^{\vee}$: 
\begin{equation} \label{eq:conjecture}
    W_{N=1}^k(\g,f)\simeq W^k(\g,F) \otimes \mathcal{F}(\g^f_0),
\end{equation}
where $\g^f_0$ is the centralizer of the subalgebra $\mathfrak{s}\simeq \mathfrak{osp}(1|2)$ inside $\g$, and $\mathcal{F}(\g_0^f)$ is the free superfermion vertex algebra associated with $\g_0^f$. Namely, $\mathcal{F}(\g_0^f)$ is freely generated by a basis of $\bar{\g}_0^f$ with the $\lambda$-bracket 
\begin{equation}
    [\,  \bar{a}\, {}_\lambda \,  \bar{b} \, ] = (\, a\, | \, b\, )
\end{equation}
for $a,b\in\g_0^f$ and the bilinear form $(\ | \ )$ on $\g.$ Note that each field $\bar{a}\in \mathcal{F}(\g_0^f)$ has a reversed parity of $a\in \g_0^f$. In this section, we assume that the level is given to satisfy $k\neq -h^{\vee}$ and present a proof of \eqref{eq:conjecture}.

\subsection{Trivial nilpotent case} \label{subsec:trivial}

We observed from Example \ref{eq:affine} that $V^{k+h^\vee}(\g)$ is a vertex subalgebra of $V_{N=1}^k(\g)$ via the embedding \eqref{eq:affine, ord vs SUSY}. However, the image of this embedding does not commute with the free superfermion part. Instead, by using the following lemma, one can find another affine vertex algebra inside $V_{N=1}^k(\g)$ commuting with the superfermion part. This is the super-analogue of \cite[Lemma 3]{KT85}.

\begin{lemma} \label{lem:decomp_affine}
Let $\{u^\alpha\,|\,\alpha\in I\}$ and $\{u_\alpha\,|\,\alpha\in I\}$ be dual bases of $\g,$ that is $(u^\alpha|u_\beta)=\delta_{\alpha,\beta}$. For any $k\neq -h^\vee$ and $a\in \g$, consider the following field in $V^k_{N=1}(\g)$:
\begin{equation} \label{eq:J in affine}
    \mathcal{J}_{a}:= D\bar{a}+\frac{1}{2(k+h^\vee)}\sum_{\alpha\in I}(-1)^{\alpha} \bar{u}^{\alpha}\overline{[a, u_\alpha]}.
\end{equation}
Then for the $\lambda$-bracket in $V^k_{N=1}(\g)$, we have $[\, \bar{a}\, {}_\lambda \,  \mathcal{J}_{b}\, ]=0$ and $[\, \mathcal{J}_{a}\, {}_\lambda \,  \mathcal{J}_{b}\, ]= (-1)^{p(a)p(b)}\mathcal{J}_{
[a,b]} + (-1)^{p(a)} k\lambda(a|b).$
\end{lemma}
\begin{proof}
    This lemma can be checked by direct computations. By the Wick formula, we have 
    \begin{equation} 
        \begin{aligned}
            & [\, \bar{a}\, {}_\lambda \, \sum_{\alpha\in I} (-1)^{p(\alpha)} :\bar{u}^\alpha \overline{[b,u_\alpha]}:\, ] \\
            & = \sum_{\alpha \in I} (k+h^\vee) (u^\alpha|a) \overline{[b,u_\alpha]} - \sum_{\alpha \in I} (k+h^\vee)(-1)^{p(a)p(b)} ([a,b]|u_\alpha) \bar{u}^\alpha\\
            & = 2 (-1)^{p(a)p(b)+1}(k+h^\vee) \overline{[a,b]}.
        \end{aligned}
    \end{equation}
    Since $[\bar{a}{}_\lambda D\bar{b}]=(-1)^{p(a)p(b)}\overline{[a,b]},$ we get $[\bar{a}{}_\lambda \mathcal{J}_{b}]=0.$ Now, let us show the second equality. Since $[\bar{h}{}_\lambda \mathcal{J}_b]=0$ for any $h\in \g,$ we have  
    \begin{equation}
            [\, {\mathcal{J}}_a\,  {}_{\lambda} \, {\mathcal{J}}_b\,]= [\, D\bar{a}{}_{\lambda} {\mathcal{J}}_b\, ].
    \end{equation}
    By the Wick formula,  
    \begin{equation}
        \begin{aligned}
           &  \sum_{\alpha\in I} \ [\, D\bar{a}{}_\lambda (-1)^{p(\alpha)}:\bar{u}^\alpha \overline{[b,u_\alpha]}:\, ]\\
           & =\sum_{\alpha\in I} \ (-1)^{p(\alpha)}: [D\bar{a}{}_\lambda \bar{u}^\alpha]\overline{[b,u_\alpha]}: + \sum_{\alpha\in I} (-1)^{p(\alpha)+(\, p(a)(p(\alpha)+1)\, )}:\bar{u}^\alpha [ D \bar{a}{}_\lambda \overline{[b,u_\alpha]}]:\\
           &  \hskip 3cm + \sum_{\alpha\in I} \sum_{j\in \mathbb{N}}(-1)^{p(\alpha)}\frac{\lambda^j}{j!} \, \big( \, [D\bar{a}{}_\lambda \bar{u}^\alpha]_{(j-1)}\overline{[b,u_\alpha]}\, \big)\\
           & =  \sum_{\alpha\in I} \Big(\, (-1)^{p(a)p(b)+p(\alpha)}\bar{u}^\alpha \overline{[[a,b],u_\alpha]} + \lambda \, (-1)^{p(a)+p(\alpha)+1}(k+h^\vee)\, (a|[u^\alpha,[u_\alpha,b]]) \Big)\\
           & = \sum_{\alpha\in I} \Big(\, (-1)^{p(a)p(b)+p(\alpha)}\bar{u}^\alpha \overline{[[a,b],u_\alpha]} +  2\, \lambda \, (-1)^{p(a)+1}(k+h^\vee) h^\vee\,  (a|b) \Big)
        \end{aligned}
    \end{equation}
    and hence we have
    \begin{equation}
    \begin{aligned}\relax
            [\, D\bar{a} \,{}_{\lambda} \,\mathcal{J}_b \,] &=  [\, D\bar{a}\, {}_\lambda \, D\bar{b} \,] + \frac{1}{2(k+h^\vee)} \, \sum_{\alpha\in I} \ \big[ D\bar{a}{}_\lambda (-1)^{p(\alpha)}:\!\bar{u}^\alpha \overline{[b,u_\alpha]}\!: \big]\\
            &= (-1)^{p(a)p(b)}\mathcal{J}_{[a,b]}+(-1)^{p(a)}k\lambda(a|b)
    \end{aligned}
    \end{equation}
    which proves the lemma.
\end{proof}

The vertex subalgebra generated by $\mathcal{J}_a$ is isomorphic to $V^k(\g)$ under the isomorphism \eqref{eq:affine, ord vs SUSY}. Since SUSY and nonSUSY affine vertex algebras can be regarded as SUSY and nonSUSY W-algebras associated with the trivial nilpotent, respectively, we obtain the following proposition.

\begin{proposition} \label{prop:susy vs nonsusy affine}
    For $k\neq-h^{\vee}$, SUSY W-algebra $W^k_{N=1}(\g,0)$ has $W^k(\g,0)$ as a vertex subalgebra. Moreover, as a vertex algebra 
    \[W^k_{N=1}(\g,0)\simeq W^k(\g,0)\otimes \mathcal{F}(\g),\]
    where $\mathcal{F}(\g)$ is a free superfermion vertex algebra associated with $\g$.
\end{proposition}

\subsection{Principal nilpotent case} \label{subsec:principal}
Assume that $\g$ has an odd principal nilpotent $f_{\textup{prin}}\in \g$. In this case, it can be easily checked that $F_{\textup{prin}}=-\frac{1}{2}[f_{\textup{prin}},f_{\textup{prin}}]$ is also principal. The relationship between the corresponding SUSY W-algebra and W-algebra was recently discovered in \cite{GSS25}. Namely, they show that
\begin{equation} \label{eq:susy vs nonsusy principal}
    W^k_{N=1}(\g,f_{\textup{prin}})\simeq W^k(\g,F_{\textup{prin}})
\end{equation}
as vertex algebras for any $k\neq -h^{\vee}$. Note that \eqref{eq:susy vs nonsusy principal} is a special case of \eqref{eq:conjecture} since $\g_0^{f_{\textup{prin}}}=0$. For the reader's convenience, we briefly explain why \eqref{eq:susy vs nonsusy principal} holds. Recall from Section \ref{subsec: nonsusy miura} and \ref{subsec: susy miura} that the two algebras in \eqref{eq:susy vs nonsusy principal} have injective Miura maps for any $k\neq -h^{\vee}$:
\begin{equation} \label{eq:ff realization}
    W^k_{N=1}(\g,f_{\textup{prin}})\hookrightarrow V^{k+h^{\vee}}_{N=1}(\h), \qquad W^k(\g,F_{\textup{prin}})\hookrightarrow V^{k+h^{\vee}}(\h)\otimes \mathcal{F}^{\textup{ne}}.
\end{equation}
Note that $\g_0$ is equal to the Cartan subalgebra $\h$ of $\g$, since $F_{\textup{prin}}$ is principal. Moreover, the generic images of \eqref{eq:ff realization} can be described as the kernels of the screening operators \eqref{eq: nonsusy Salpha} and \eqref{eq: define susy screening}, and we have explicit formulas for them studied in \cite{Genra17} and \cite{Song24free}. For $\g$ allowing the existence of principal $f_{\textup{prin}}$, its simple root system can be chosen to be purely odd, whose root vectors are contained in $\g_{\frac{1}{2}}$. It implies that $[\Delta]=\Delta$, and we can construct the isomorphism
\begin{equation*}
    V^{k+h^{\vee}}(\h)\otimes \Phi(\g_{\frac{1}{2}})\rightarrow V^{k+h^{\vee}}_{N=1}(\h), \quad h\otimes 1\mapsto D\bar{h}, \quad 1\otimes \sqrt{k+h^{\vee}}\Phi_{\alpha}\mapsto \bar{h}_{\alpha},
\end{equation*}
where $h\in \h$ and $h_{\alpha}$ is the coroot of each simple root $\alpha$. Moreover, the isomorphism establishes the identification of the screening operator formulas for $W^k_{N=1}(\g,f_{\textup{prin}})$ and $W^k(\g,F_{\textup{prin}})$, which immediately implies \eqref{eq:susy vs nonsusy principal}.

% Recall that the SUSY W-algebra $W^k_{N=1}(\g,f_{\textup{prin}})$ can be viewed as a subalgebra of the SUSY affine vertex algebra $V^k_{N=1}(\g)$. By composing with the projection map $V^k_{N=1}(\g)\rightarrow V^k_{N=1}(\h)$, where $\h$ is the Cartan subalgebra of $\g$, we obtain the SUSY Miura map for the SUSY W-algebra in \eqref{eq:ff realization}. For W-algebras, the realization in \eqref{eq:ff realization} is also given by the Miura map, where $\Phi(\g_{\frac{1}{2}})$ denotes the neutral free fermion vertex algebra associated with $\g_{\frac{1}{2}}$ and the bilinear form $\langle A|B\rangle=(F_{\textup{prin}}|[A,B])$.

% Based on the results in \cite{Song24free, Genra17}, the images of \eqref{eq:ff realization} are generically described by the kernel of the screening operators associated with the simple roots of $\g$. Since the simple root system can be chosen to be purely odd, with all simple root vectors contained in $\g_{\frac{1}{2}}$, this yields an isomorphism
% \begin{equation*}
%     V^{k+h^{\vee}}(\h)\otimes \Phi(\g_{\frac{1}{2}})\rightarrow V^{k+h^{\vee}}_{N=1}(\h), \quad h\otimes 1\mapsto D\bar{h}, \quad 1\otimes \sqrt{k+h^{\vee}}\Phi_{\alpha}\mapsto \bar{h}_{\alpha},
% \end{equation*}
% where $h\in \h$ and $h_{\alpha}$ is the coroot of each simple root $\alpha$. Moreover, the isomorphism establishes the identification of the screening operator formulas for $W^k_{N=1}(\g,f_{\textup{prin}})$ and $W^k(\g,F_{\textup{prin}})$, which immediately implies \eqref{eq:susy vs nonsusy principal}.

\subsection{General cases}

In this section, let $f$ be an arbitrary odd nilpotent in an $\mathfrak{osp}(1|2)$ subalgebra  $\mathfrak{s}$ in $\g.$ We will combine the observations in Section \ref{subsec:trivial} and \ref{subsec:principal} and derive the isomorphism \eqref{eq:conjecture} for general cases.

Recall the set of indecomposable roots $\Delta$ and the set of equivalence classes $[\Delta]$ from \eqref{eq: Delta indecomp roots}. Note that
\[\Delta=\{\alpha\in I_+\,|\, \textup{a root vector of }\alpha \textup{ is contained in }\g_{\frac{1}{2}}\},\]
since any root vector of degree $\geq 1$ is in the image of $\text{ad}\, e$ and so decomposable.
% Since we take the $\mathbb{Z}/2$-grading induced from $\mathfrak{s}$ and by the $\mathfrak{osp}(1|2)$ representation, any root vector in $\Delta$ is in $\g_{1/2}.$ If $\alpha\in \Delta$ has degree $m\geq 1$ then $\alpha$ is in the image of $\text{ad}\, e$ and it is decomposable.

In Section \ref{sec:W-alg}, we introduced Miura maps for W-superalgebras and SUSY W-algebras which are generically realized as kernels of the screening operators as follows:
\begin{align}
        & \label{eq:nonSUSY-screening} W^k(\g,F)\simeq \bigcap_{[\alpha]\in [\Delta]}\text{Ker}\Big(S_{[\alpha]} : \, V^{\tau_k}(\g_0)\otimes \mathcal{F}^{\text{ne}}\to \mathcal{F}^{\text{ne}}\otimes M_{[\alpha]}\Big),\\
        & \label{eq:SUSY-screening} W_{N=1}^k(\g,f)\simeq \bigcap_{[\alpha]\in [\Delta]}\text{Ker}\Big(S^{N=1}_{[\alpha]} : \, V^{\psi_k}_{N=1}(\g_0) \to \widetilde{M}_{[\alpha]}\Big),
\end{align} 
where 
\begin{equation} \label{eq:screening}
         S_{[\alpha]}=\frac{\sqrt{-1}}{\sqrt{k+h^{\vee}}}\sum_{\alpha\in [\alpha]}\int :\!\phi^{\alpha}\Phi_\alpha\!:\!(z) dz, \qquad S^{N=1}_{[\alpha]}=\sum_{\alpha\in [\alpha]} (f|u_\alpha)\int D\phi^{\bar{\alpha}}(z)dz.
\end{equation} 
Recall that the levels of the affine vertex algebras in \eqref{eq:nonSUSY-screening} and \eqref{eq:SUSY-screening} are $\tau_k(a|b):=k(a|b)+\frac{1}{2}\kappa_{g}(a|b)-\frac{1}{2}\kappa_{\g_0}(a|b)$ and 
$\psi_k(a|b):= (k+h^\vee)(a|b)$ for $a,b\in \g_0$. Also, $M_{[\alpha]}$ and $\widetilde{M}_{[\alpha]}$ are $V^{\tau_k}(\g_0)$-module and $V^{\psi_k}_{N=1}(\g_0)$-module
\begin{equation} \label{eq: M and Mtild}
    M_{[\alpha]}=V^{\tau_k}(\g_0)\otimes \bigoplus_{\beta\in [\alpha]}\CC x_{\beta}, \qquad  \widetilde{M}_{[\alpha]}=V^{\psi_k}_{N=1}(\g_0)\otimes \bigoplus_{\beta\in [\alpha]}\CC \widetilde{x}_{\beta},
\end{equation}
whose actions are given by \eqref{eq: M_alpha module structure} and \eqref{eq: Mtild_alpha module structure}. Note that $x_{\beta}$ and $\widetilde{x}_{\beta}$ are simply alternative notations for $\phi^{\beta}$ in the BRST complexes, and the actions are derived from the $\lambda$-bracket relations on the first total complex.

% {\color{blue}Define the module structure of $M_{[\alpha]}$ and $\widetilde{M}_{[\alpha]}$ in Section 3 and recall here. Explain the action and $\lambda$-bracket, $(0)$-product}

Our strategy is to show that the actions of the screening operators in \eqref{eq:nonSUSY-screening} and \eqref{eq:SUSY-screening} coincide. More precisely, the proof consists of the following steps: 
\begin{enumerate}[]
    \item (Step 1) Compare the domains of the screening operators and show  $(V^{\tau_k}(\g_0)
\otimes \mathcal{F}^{\text{ne}}) \otimes \mathcal{F}(\g^f_0)   \simeq V^{\psi_k}_{N=1}(\g_0)$ as vertex algebras. To be precise, we find an explicit isomorphism 
    \begin{equation} \label{eq:step1-goal}
       \iota :  V^{\tau_k}(\g_0)
\otimes \mathcal{F}^{\text{ne}} \otimes \mathcal{F}(\g_0^f)  \to V^{\psi_k}_{N=1}(\g_0),
    \end{equation}
    and regard $V^{\tau_k}(\g_0)$, $\mathcal{F}^{\textup{ne}}$, and $\mathcal{F}(\g_0^f)$ as subalgebras of $V^{\psi_k}_{N=1}(\g_0)$ via the map $\iota$.
    \item (Step 2) Compare the codomains of the screening operators and show $\mathcal{F}(\g_0^f)\otimes\mathcal{F}^{\textup{ne}}\otimes M_{[\alpha]}\simeq \widetilde{M}_{[\alpha]}$ as $V^{\psi_k}_{N=1}(\g_0)$-modules. Show that $\mathcal{F}(\g^f_0)\subset V^{\psi_k}_{N=1}(\g_0)$ is contained in the kernel of the SUSY screening operators.
    \item (Step 3) Under the identifications, compare the action of $S_{[\alpha]}$ and $S^{N=1}_{[\alpha]}$ on $\mathcal{F}^{\text{ne}}$.
    \item (Step 4) Under the identifications, compare the action of $S_{[\alpha]}$ and $S^{N=1}_{[\alpha]}$ on $V^{\tau_k}(\g_0)$.
    \item (Step 5) From (Step 3) and (Step 4), we conclude that the restriction of $S^{N=1}_{[\alpha]}$ to $V^{\tau_k}(\g_0)\otimes \mathcal{F}^{\text{ne}}$ coincide with $S_{[\alpha]}$.
    \item (Step 6) By (Step 2) and (Step 5), we get the desired statement for generic $k$. For any $k\neq -h^{\vee}$, the Miura map images of $W^k_{N=1}(\g,f)$ and $W^k(\g,F)\otimes \mathcal{F}(\g_0^f)$ form continuous families of vertex subalgebras inside $V^{\psi_k}_{N=1}(\g_0)$. Therefore, the isomorphism for generic $k$ implies the isomorphism for all $k\neq -h^{\vee}$. As a result, we get the following theorem.
\end{enumerate}

\begin{theorem} \label{thm:nonSUSY vs SUSY}
As vertex algebras, the following isomorphism holds for all $k\neq -h^{\vee}$ : 
\begin{equation}
        W_{N=1}^k(\g,f) \simeq W^k(\g,F)\otimes \mathcal{F}(\g^f_0).
\end{equation}
Equivalently, $W^k(\g,F)$ is isomorphic to the coset vertex algebra
\begin{equation} \label{eq:conj coset}
    \textup{Com}(\mathcal{F}(\g_0^f),W^k_{N=1}(\g,f))\simeq W^k(\g,F),
\end{equation}
where  $\textup{Com}(\mathcal{F}(\g_0^f),W^k_{N=1}(\g,f))= \{v\in W^k_{N=1}(\g,f)\, | \, [\, v\, {}_\lambda \, \mathcal{F}(\g_0^f)\, ]=0\, \}$.
\end{theorem}

In the rest of this section, we provide a detailed proof of (Step 1)  - (Step 6).

\subsubsection{\textup{(Step 1)}} \label{Step 1}
In this section, we define the vertex algebra isomorphism
    \begin{equation} \label{iota}
        \iota :  V^{\tau_k}(\g_0)
        \otimes \mathcal{F}^{\text{ne}} \otimes \mathcal{F}(\g_0^f)  \to V^{\psi_k}_{N=1}(\g_0)
    \end{equation}
on each component. First, to define it on $V^{\tau_k}(\g_0)$, recall from Section \ref{subsec:trivial} that one can find $V^k(\g)$ inside $V^k_{N=1}(\g)$ using the $\mathcal{J}_{a}$'s in Lemma \ref{lem:decomp_affine}. Similarly, in $V^{\psi_k}_{N=1}(\g_0)$, we define
\begin{equation} \label{eq: J in susy affine step 1}
        \mathcal{J}_a:= D \bar{a}+ \frac{1}{2(k+h^\vee)}\sum_{i\in I_0}(-1)^{p(i)} \bar{u}^i\overline{[a,u_i]}
\end{equation}
for $a\in \g_0$, where $\{u^i\}$ and $\{u_i\}$ are dual bases of $\g_0$ such that $(u^i|u_j)=\delta_{i,j}$ and $p(i)$ is the parity of $u_i$. With a computation similar to the proof of Lemma \ref{lem:decomp_affine}, one can check that
\begin{equation}
    [\bar{a}{}_\lambda \mathcal{J}_b] =0, \quad [\mathcal{J}_a{}_\lambda \mathcal{J}_b] =(-1)^{p(a)p(b)}\mathcal{J}_{[a,b]} +(-1)^{p(a)}\lambda \big( (k+h^\vee)(a|b)-\frac{1}{2}\kappa_{\g_0}(a|b) \big)
\end{equation}
for $a,b\in \g_0.$ Considering the map \eqref{eq:affine, ord vs SUSY}, the following $\iota$ gives a vertex algebra embedding:
\begin{equation} \label{eq: iota affine}
    \iota|_{V^{\tau_k}(\g_0)} : V^{\tau_k}(\g_0) \rightarrow V^{\psi_k}_{N=1}(\g_0), \qquad a \mapsto \sqrt{-1}^{p(a)}\mathcal{J}_a.
\end{equation}

Next, define
\begin{equation} \label{eq: iota neutral}
    \iota|_{\mathcal{F}^{\textup{ne}}} : \mathcal{F}^{\textup{ne}} \rightarrow V^{\psi_k}_{N=1}(\g_0), \qquad \Phi_{\beta}\mapsto \frac{\sqrt{-1}^{1+p(\beta)}}{\sqrt{k+h^\vee}}\overline{[f,u_\beta]}.
\end{equation}
Then, \eqref{eq: iota neutral} is also a vertex algebra embedding due to the following lemma.

\begin{lemma}\label{lem:step1-2}
    Let $g_1,g_2\in \g_{1/2}.$ In $V^{\psi_k}_{N=1}(\g_0)$, we have 
\begin{equation} \label{eq: neutral in susy affine}
    [ \, \overline{[f,g_1]} \, {}_\lambda \, \overline{[f,g_2]}\, ]=(-1)^{p(g_1)+1}(k+h^\vee)(F|[g_1,g_2]).
\end{equation} 
In other words, the vertex subalgebra of $V^{\psi_k}_{N=1}(\g_0)$ generated by $\overline{[f,\g_{1/2}]}$ is isomorphic to the neutral fermion vertex algebra $\mathcal{F}^{\textup{ne}}.$
\end{lemma}
\begin{proof}
    We have
    \begin{equation}
    \begin{aligned}
          -2(F|[g_1,g_2]) = ( \, f\, |\, [f,[g_1,g_2]]\, )& =(\, f\, |\, [[f,g_1],g_2]+(-1)^{p(g_1)}[g_1,[f,g_2]]\, ) \\
          & = 2(-1)^{p(g_1)}( \, [f,g_1]\, | \, [f,g_2]\, ),
    \end{aligned}
    \end{equation}
    which implies \eqref{eq: neutral in susy affine}. Now, the lemma follows with the map \eqref{eq: iota neutral}.
\end{proof}

Finally, since we have $\g_0 = [f,\g_{1/2}]\oplus \g^f_0$, the map
\begin{equation} \label{eq: iota fermion}
    \iota|_{\mathcal{F}(\g_0^f)}: \mathcal{F}(\g_0^f)\rightarrow V^{\psi_k}_{N=1}(\g_0), \qquad \bar{a} \mapsto \frac{1}{\sqrt{k+h^{\vee}}}\bar{a}
\end{equation}
combined with \eqref{eq: iota affine} and \eqref{eq: iota neutral} defines a vertex algebra isomorphism $\iota$.

\begin{remark} \label{rem:parity}
    Since $\sqrt{-1}^{2}=-1 \neq 1,$ the parity in the power of $\sqrt{-1}$ should not be considered a value in $\mathbb{Z}_2$ but in $\{0,1\}\subset \mathbb{N}.$ For example,  if $a$ and $b$ are both odd then $\sqrt{-1}^{p([a,b])}=\sqrt{-1}^{0}=1 \neq \sqrt{-1}^{2}=\sqrt{-1}^{p(a)+p(b)}.$ Also, when $a$ is odd, we have $\sqrt{-1}^{1+p(a)}=\sqrt{-1}^2=-1 \neq 1 =\sqrt{-1}^0$. Here, we note the useful equality:
    \begin{equation}
        \sqrt{-1}^{p(a)+p(b)+p([a,b])}=(-1)^{p(a)p(b)+p(a)+p(b)}.
    \end{equation}
\end{remark}

\subsubsection{\textup{(Step 2)}} \label{Step 2}
Recall the modules $M_{[\alpha]}$ and $\widetilde{M}_{[\alpha]}$ in \eqref{eq: M and Mtild}. By extending the $\iota$ in \eqref{iota}, we get the $V^{\psi_k}_{N=1}(\g_0)$-modules isomorphism
\begin{equation} \label{eq:iota_module}
    \iota : \mathcal{F}(\g_0^f)\otimes\mathcal{F}^{\textup{ne}} \otimes M_{[\alpha]}  \to \widetilde{M}_{[\alpha]}, \qquad x_{\alpha}\mapsto \sqrt{-1}^{p(\alpha)}\widetilde{x}_{\alpha}.
\end{equation}
    In other words, $\iota(\Phi\otimes (A x_\alpha))= \iota(\Phi)\iota(A) \sqrt{-1}^{p(\alpha)} \widetilde{x}_\alpha$ where $\Phi\in \mathcal{F}(\g_0^f)\otimes \mathcal{F}^{\text{ne}}$  and  $A\in V^{\tau_k}(\g_0)$.   

Now, we claim that $\mathcal{F}(\g_0^f)$ is contained in the kernel of the SUSY screening operators.
In terms of the $\lambda$-bracket in the first total complex of SUSY BRST, we have 
\begin{equation}
    [D\phi^\alpha{}_\lambda \bar{a}] = \sum_{\beta\in [\alpha]} (-1)^{p(\alpha)+1}([a,u^\alpha]|u_\beta) \phi^\beta
\end{equation}
for $a\in \g_0.$ It is clear that $\bar{a}$  is in the kernel of all SUSY screening operators $S^{N=1}_{[\alpha]}=\sum_{\alpha\in [\alpha]}(f|u_\alpha) \int D \phi^\alpha dz $ for all $[\alpha]\in [\Delta]$ if and only if $[f,a]=0.$ Hence, $\mathcal{F}(\g^f_0)$ is a vertex subalgebra of $W^k_{N=1}(\g,f).$

\subsubsection{\textup{(Step 3)}} \label{Step 3}
In (Step 3) and (Step 4), we compare the screening operators $S_{[\alpha]}$ and $S^{N=1}_{[\alpha]}$. Recall from \eqref{eq: nonsusy first total} and \eqref{eq: susy first total} that the screening operator descriptions are derived from the computations in the first total complex for generic $k$. Hence, throughout Steps 3 and 4, we perform all $\lambda$-bracket computations within the first total complex. In (Step 3), we aim to show the following proposition.
\begin{proposition}
    Let $u_\beta\in \g_{1/2}$ and $\Phi_{\beta}=\Phi_{u_\beta}\in \mathcal{F}^{\textup{ne}}$. For $\alpha\in \Delta,$ we have
    \begin{equation} \label{eq:step3-goal}
   \iota \big(\, S_{[\alpha]} (\Phi_{\beta})\, \big) \ =\  S^{N=1}_{[\alpha]} (\, \iota(\Phi_{\beta}) \, ).
\end{equation}
\end{proposition}
\begin{proof}
    This proposition can be proved through Lemma \ref{lem:step3-1} and \ref{lem:step3-2} below.
\end{proof}

\begin{lemma} \label{lem:step3-1}
    Recall the screening operator in \eqref{eq:screening}. For each $\Phi_{\beta}\in \mathcal{F}^{\textup{ne}}$, we have
    \begin{equation}
        S_{[\alpha]}(\Phi_\beta)=\frac{\sqrt{-1}}{\sqrt{k+h^{\vee}}}\sum_{\alpha \in [\alpha]} (\,:\! \phi^{\alpha} \Phi_{\alpha}\!:\,)  {}_{(0)} \, (\Phi_{\beta}) \,  = \frac{\sqrt{-1}}{\sqrt{k+h^{\vee}}}\sum_{\alpha, \gamma \in [\alpha]} (f|u_\alpha) ( \,[[f,u_\beta],u^\alpha]\, |\, u_\gamma ) \,x_\gamma.
    \end{equation}
\end{lemma}
\begin{proof}
    By the Wick formula, 
    \begin{equation} \label{eq:step3-1-1}
    \begin{aligned}
        [\,:\!\phi^\alpha \Phi_\alpha\!:{}_\lambda \, \Phi_\beta] & =(F|[u_\alpha,u_\beta])\phi^\alpha = (-1)^{p(\alpha)+1} ([f,u_\alpha]|[f,u_\beta]) \phi^\alpha \\
        &= \sum_{\gamma\in [\alpha]} (-1)^{p(\alpha)+1}(f|u_\gamma)([u^\gamma,u_\alpha]|[f,u_\beta])\phi^\alpha.
    \end{aligned}
    \end{equation}
    Consider the summation of \eqref{eq:step3-1-1} over all the root $\alpha\in [\alpha]$ and exchange $\alpha$ and $\gamma$ on the RHS. Then we get
    \begin{equation}
        \begin{aligned}
             \sum_{\alpha\in [\alpha]}[\,:\!\phi^\alpha \Phi_\alpha\!:{}_\lambda \, \Phi_\beta] & = \sum_{\alpha,\gamma\in [\alpha]} (-1)^{p(\gamma)+1}(f|u_\alpha)([u^\alpha,u_\gamma]|[f,u_\beta])\phi^\gamma,
        \end{aligned}
    \end{equation}
    which proves the Lemma.
\end{proof}

\begin{lemma} \label{lem:step3-2}
    Recall the SUSY screening operator in \eqref{eq:screening} and the map \eqref{iota}. For each $\Phi_{\beta}\in \mathcal{F}^{\textup{ne}}$, we have
    \begin{equation}\label{eq:step3-2-1}
    \begin{aligned}
         \sqrt{k+h^\vee} \ S^{N=1}_{[\alpha]}(\iota(\Phi_\beta)) & =\sqrt{-1}^{p(\beta)+1}\,   \sum_{\alpha \in [\alpha]} (f|u_\alpha) (D \phi^\alpha) \, {}_{(0)} \, (\overline{[f,u_\beta]})\\
         & =\sum_{\alpha, \gamma \in [\alpha]} (f|u_\alpha) (\, [[f,u_\beta],u^\alpha]\, |\,  u_\gamma \, ) \sqrt{-1}^{p(\gamma)+1}\widetilde{x}_\gamma.
    \end{aligned}
    \end{equation}
\end{lemma}
\begin{proof}
    By the definition of the $\lambda$-bracket, we have 
    \begin{equation} \label{eq:step3-1-2}
        [D\phi^\alpha{}_\lambda\, \overline{[f,u_\beta]}] = \sum_{\gamma\in [\alpha]}(-1)^{p(\alpha)+1}([[f,u_\beta],u^\alpha]|u_\gamma)\phi^\gamma.
    \end{equation}
    Hence,
    \begin{equation}\label{eq:step3-1-3}
    \begin{aligned}
        &     [D\phi^\alpha{}_\lambda\, \sqrt{-1}^{p(\beta)+1}\overline{[f,u_\beta]}]\\
        & = \sqrt{-1}^{p(\beta)+1}\sum_{\gamma\in [\alpha]}(-1)^{p(\alpha)+1}([[f,u_\beta],u^\alpha]|u_\gamma)(-1)^{p(\gamma)}\sqrt{-1}^{p(\gamma)}(\sqrt{-1}^{p(\gamma)}\phi^\gamma)\\
        & = \sum_{\gamma\in [\alpha]}(-1)^{p(\beta)}\sqrt{-1}^{p(\beta)+p(\gamma)+1}([[f,u_\beta],u^\alpha]|u_\gamma)(\sqrt{-1}^{p(\gamma)}\phi^\gamma).
    \end{aligned}
    \end{equation}
      In the first equality in \eqref{eq:step3-1-3}, we used $\sqrt{-1}^{p(\gamma)}\sqrt{-1}^{p(\gamma)}=(-1)^{p(\gamma)}.$ Finally, we get the lemma from \eqref{eq:step3-1-3} by considering that $(f|u_\alpha)$ can be nontrivial only when $u_\alpha$ is odd and in that case $\sqrt{-1}^{p(\beta)+p(\gamma)}=(-1)^{p(\beta)}$.
\end{proof}

\subsubsection{\textup{(Step 4)}} 
Recall the map $\iota$ in \eqref{eq:iota_module}. In (Step 4), we claim the following proposition.

\begin{proposition} Let $a\in \g_0\subset V^{\tau_k}(\g_0)$ and $\alpha\in \Delta.$ Then we have 
    \begin{equation} \label{eq:step4-goal}
   \iota \big(\, S_{[\alpha]} (a)\, \big) \ =\  S^{N=1}_{[\alpha]} (\, \iota(a) \, )
\end{equation}
\end{proposition}
\begin{proof}
    This proposition can be proved by Lemma \ref{lem:step4-1} and Lemma \ref{lem:step4-4}.
\end{proof}

In the following lemma, we can compute the LHS of  \eqref{eq:step4-goal}.

\begin{lemma} \label{lem:step4-1}
    Let $a\in \g_0\subset V^{\tau_k}(\g_0)$. Then 
    \begin{equation}
        S_{[\alpha]}(a)= \frac{\sqrt{-1}}{\sqrt{k+h^{\vee}}}\sum_{\alpha\in [\alpha]} (:\!\phi^\alpha \Phi_{\alpha}\!:){}_{(0)}(a) =\frac{\sqrt{-1}}{\sqrt{k+h^{\vee}}} \sum_{\alpha,\beta\in [\alpha]}(-1)^{p(\alpha)(p(\beta)+1)}([a,u^\alpha]|u_\beta) \Phi_\alpha x_\beta.
    \end{equation}
\end{lemma}
\begin{proof}
    It is obtained by direct computations.
\end{proof}

Before see the RHS of \eqref{eq:step4-goal}, we observe the following relation in the first total complex.

\begin{lemma}\label{lem:step4-2}
    Recall the SUSY first total complex $E^k_1$ in \eqref{eq: susy first total}. For $\alpha\in \Delta$ and generic $k$, we have
    \begin{equation}
        D\phi^\alpha = \frac{1}{k+h^\vee} \sum_{\beta\in [\alpha]} (-1)^{p(\alpha)+1} : \overline{\pi_0[u_\beta,u^\alpha]}\phi^\beta:
    \end{equation}
    in $E^k_1$, where $\pi_0:\g \to \g_0$ is the canonical projection map.
\end{lemma}
\begin{proof}
    Recall that $E^k_1=H(\widetilde{C}^k_{N=1}(\g,f), (Dd_{\textup{st}})_{(0)})$. Hence, we can consider the image of $(Dd_{\text{st}}){}_{(0)}$ as a trivial element in $E^k_1$. For $\alpha \in \Delta,$ we know that 
    \begin{equation}
        (Dd_{\text{st}}){}_{(0)} (\bar{u}^\alpha) = \sum_{\beta\in I_0} (-1)^{p(\alpha)+1} :\overline{\pi_{\leq 0} [u_\beta,u^\alpha]}\phi^\beta :-(k+h^\vee) D\phi^\alpha,
    \end{equation}
    where $\pi_{\leq 0}:\g \to \g_{\leq 0}$ is the canonical projection map. Now, it is enough to show that $\pi_{\leq 0} [u_\beta,u^\alpha]=\pi_{0} [u_\beta,u^\alpha].$ If not, then $\pi_{<0}[u_\beta,u^\alpha]$ is nontrivial, which forces $\beta-\alpha = -\gamma$ for some root in $\Phi_+.$ This contradicts the indecomposability of $\alpha$.
\end{proof}

\begin{lemma}\label{lem:step4-3}
    Recall the SUSY first total complex $E^k_1$ from \eqref{eq: susy first total} and $\mathcal{J}_a$ from \eqref{eq: J in susy affine step 1}. For $a\in \g_0$, $\alpha\in \Delta$, and generic $k$, we have the following relation in $E^k_1$:
    \begin{equation}
        (D\phi^\alpha){}_{(0)}(\mathcal{J}_a) = \frac{1}{k+h^\vee} \sum_{i\in I_0, \beta\in [\alpha]}(-1)^{(p(a)+p(\beta))p(i)+1} ([[a,u_i],u^\alpha]|u_\beta) :\bar{u}^i \phi^\beta:\, . 
    \end{equation}
\end{lemma}
\begin{proof}
        We claim that 
        \begin{align}
           & \label{eq:step4-3-1} (D\phi^\alpha){}_{(0)} (D\bar{a})=\frac{1}{k+h^\vee}\sum_{i\in I_0, \, \beta\in [\alpha]}(-1)^{p(\beta)p(i)+1}([u_i,[a,u^\alpha]]|u_\beta)\bar{u}^i \phi^\beta, \\
           & \label{eq:step4-3-2} (D\phi^\alpha){}_{(0)} \Big(\sum_{i\in I_0}(-1)^{p(i)} :\bar{u}^i \overline{[a,u_i]}:\Big) \\
           & \qquad  = 2\sum_{i\in I_0, \, \beta\in [\alpha]}(-1)^{p(a)p(i)+p(\beta)p(i)+1}([[a,u_i],u^\alpha]|u_\beta):\bar{u}^i \phi^\beta:. \nonumber 
        \end{align}
    We use Lemma \ref{lem:step4-2} to see \eqref{eq:step4-3-1}. In detail, we have
    \begin{equation}\label{eq:step4-3-3}
    \begin{aligned}
                & (k+h^\vee)(D\phi^\alpha){}_{(0)} (D\bar{a})  = -(k+h^\vee)\sum_{\beta\in [\alpha]}([a,u^\alpha]|u_\beta) D\phi^\beta \\
                & = \sum_{\beta,\gamma \in [\alpha]}(-1)^{p(\beta)}([a,u^\alpha]|u_\beta) : \overline{\pi_0[u_\gamma,u^\beta]}\phi^\gamma:\\
                & = \sum_{\beta,\gamma\in [\alpha], i\in I_0}(-1)^{p(\beta)}([a,u^\alpha]|u_\beta)([u_\gamma, u^\beta]|u_i) :\bar{u}^i \phi^\gamma: \\
                & =  \sum_{\gamma\in [\alpha], i\in I_0}(-1)^{p(\gamma)p(i)+1} ([u_i,[a,u^\alpha]]|u_\gamma) :\bar{u}^i \phi^\gamma:.
    \end{aligned}
    \end{equation}
    The second equation \eqref{eq:step4-3-2} can be obtained as follows:
    \begin{align}
        & \sum_{i\in I_0} (-1)^{p(i)} (D\phi^\alpha) \, {}_{(0)} \, \big( :\bar{u}^i \overline{[a,u_i]}:\big)\nonumber\\
        & = \sum_{i\in I_0}(-1)^{p(i)} : (D\phi^\alpha {}_{(0)}\bar{u}^i)\overline{[a,u_i]}: +\sum_{i\in I_0}(-1)^{p(i)}(-1)^{p(i)+p(\alpha)+p(i)p(\alpha)}:\bar{u}^i(D\phi^\alpha{}_{(0)}\overline{[a,u_i]}): \\
        & \label{eq:step4-3-4}= \sum_{\substack{\beta\in [\alpha]\\ i\in I_0}}(-1)^{p(i)+p(\alpha)+1}([u^i,u^\alpha]|u_\beta):\phi^\beta \overline{[a,u_i]}:\\
        & \label{eq:step4-3-5} \hskip 1em +  \sum_{\substack{\beta\in [\alpha]\\ i\in I_0}}(-1)^{p(i)+p(i)p(\alpha)+1}([[a,u_i],u^\alpha]|u_\beta):\bar{u}^i \phi^\beta:
    \end{align}
    One can show that 
    \begin{equation}
        \eqref{eq:step4-3-4}=\eqref{eq:step4-3-5}= \sum_{\beta\in [\alpha], i\in I_0}(-1)^{p(a)p(i)+p(\beta)p(i)+1}([[a,u_i],u^\alpha]|u_\beta) :\bar{u}^i \phi^\beta: \, ,
    \end{equation}
    which shows \eqref{eq:step4-3-2}. Now by the Jacobi identity, we get the lemma.
\end{proof}

\begin{lemma}\label{lem:step4-4}
     Let $a\in \g_0$ and $\alpha\in \Delta$ and recall that
     \begin{equation*}
         \iota(\Phi_\alpha)=\frac{\sqrt{-1}^{p(\alpha)+1}}{\sqrt{k+h^\vee}}\overline{[f,u_\alpha]}, \qquad \iota(x_\beta)={\sqrt{-1}^{p(\beta)}}\widetilde{x}_\beta.
     \end{equation*}
     We have
     \begin{equation*}
         S^{N=1}_{[\alpha]}\big( \sqrt{-1}^{p(a)} \mathcal{J}_a\big) = \frac{\sqrt{-1}}{\sqrt{k+h^{\vee}}}\sum_{\alpha,\beta\in [\alpha]}(-1)^{p(\alpha)(p(\beta)+1)}([a,u^\alpha]|u_\beta)\,\iota(\Phi_{\alpha})\,\iota(x_{\beta}).
     \end{equation*}
\end{lemma}

\begin{proof}
    By Lemma \ref{lem:step4-3}, we have 
    \begin{align}
           &  (k+h^\vee)\sum_{\alpha \in [\alpha]}(f|u_\alpha) (D\phi^\alpha)_{(0)}\mathcal{J}_a \nonumber \\
            & \label{eq:step4-4-1} = \sum_{i\in I_0, \alpha,\beta\in [\alpha]}(-1)^{p(a)p(i)+p(\beta)p(i)+1}(f|u_\alpha)([a,[u_i,u^\alpha]]|u_\beta): \bar{u}^i \phi^\beta:\\
            & \label{eq:step4-4-2} = \sum_{i\in I_0, \beta\in [\alpha]}(-1)^{p(a)p(i)+p(\beta)p(i)+1}([a,[u_i,f]]|u_\beta): \bar{u}^i \phi^\beta:\, .
    \end{align}
    Since $\g_0= \g^f_0 \oplus [f,\g_{1/2}]$, the basis $\{u_i\}$ of $\g_0$ in \eqref{eq:step4-4-2} can be replaced by a basis of $[f,\g_{1/2}].$ We take a basis $\{[f,u_\alpha]|\alpha \in I_{1/2}\}$ of $[f,\g_{1/2}]$ and consider its dual $\{w_\alpha\}$, i.e. $([f,u_\alpha]|w_\beta)=\delta_{\alpha\beta}.$ Then we have 
    \begin{equation}\label{eq:step4-4-3}
        [f,w_\gamma]=u^\gamma.
    \end{equation}
    We replace $u^i$ by $[f,u_\gamma]$ and $u_i$ by $w_\gamma$ in \eqref{eq:step4-4-2} for $i\in I_0$ and $\gamma\in I_{1/2}$. Then 
    \begin{equation*}
        \begin{aligned}
             \eqref{eq:step4-4-2} & = \sum_{\beta\in [\alpha], \gamma \in I_{1/2}}(-1)^{(p(a)+p(\beta))(p(\gamma)+1)+1}([a,[w_\gamma,f]]|u_\beta) :\overline{[f,u_\gamma]} \phi^\beta : \\
            & = \sum_{\beta,\gamma\in [\alpha]}(-1)^{p(\gamma)+1}([a,u^\gamma]|u_\beta) :\overline{[f,u_\gamma]} \phi^\beta : \, .
        \end{aligned}
    \end{equation*}
    Hence,
    \begin{equation*}
        \begin{aligned}
            & \sum_{\alpha \in [\alpha]}(f|u_\alpha) (D\phi^\alpha)_{(0)}\big( \sqrt{-1}^{p(a)} \mathcal{J}_a\big) \\
            & = 
             \frac{\sqrt{-1}}{\sqrt{k+h^{\vee}}}\sum_{\alpha, \beta\in [\alpha]}(-1)^{p(\beta)}\sqrt{-1}^{p(a)+p(\alpha)+p(\beta)}([a,u^\alpha]|u_\beta) :\Big(\frac{\sqrt{-1}^{p(\alpha)+1}}{\sqrt{k+h^{\vee}}}\overline{[f,u_{\alpha}]}\Big) \Big(\sqrt{-1}^{p(\beta)}\phi^{\beta}\Big):
        \end{aligned}
    \end{equation*}
    Finally, we get the statement using Remark \ref{rem:parity} so that $\sqrt{-1}^{p(\alpha)+p(\beta)+p([u_\alpha,u_\beta])}=(-1)^{p(\alpha)p(\beta)+p(\alpha)+p(\beta)}.$
\end{proof}

\subsubsection{\textup{(Step 5) and (Step 6)}} 
In (Step 3) and (Step 4), we showed that the injective homomorphism $\iota$ satisfies the property 
\begin{equation}
    \iota(S_{[\alpha]}(A))= S_{[\alpha]}^{N=1}(\iota(A))
\end{equation} for any $A\in  V^{\tau_k}(\g_0)\otimes \mathcal{F}^{\text{ne}}$ and $[\alpha]\in [\Delta].$ Hence, $\iota(A)$ is in the kernel of all the SUSY screening operators $S^{N=1}_{[\alpha]}$ if and only if $A$ is in the in the kernel of all the screening operators $S_{[\alpha]}$. In other words, if we identify W-algebras with their Miura map images, the map $\iota$ induces the inclusion map 
\begin{equation} \label{eq:iota-between W's}
    \iota : W^k(\g,F) \xhookrightarrow{} W^k_{N=1}(\g,f)
\end{equation}
for generic $k$. Moreover, by (Step 2), we have $\mathcal{F}(\g^f_0)\subset  W^k_{N=1}(\g,f)$ and hence $W^k(\g,F) \otimes \mathcal{F}(\g^f_0)$ is embedded in $W^k_{N=1}(\g,f)$. Finally, since the dimension of each conformal weight space of $W^k(\g,F) \otimes \mathcal{F}(\g^f_0)$ is the same as the dimension of the corresponding conformal weight space of $W^k_{N=1}(\g,f)$, we get the statement \eqref{eq:conjecture} for generic $k$. This result can be extended to arbitrary $k\neq -h^{\vee}$ by applying \cite[Lemma 5.14]{CGN21}. Note that for any $k\neq -h^{\vee}$, the Miura map images of $W^k_{N=1}(\g,f)$ and $W^k(\g,F)\otimes \mathcal{F}(\g_0^f)$ form continuous families of vertex subalgebras inside $V^{\psi_k}_{N=1}(\g_0)$. Forget their vertex algebra structure and regard them as vector spaces so that the ambient space $V^{\psi_k}_{N=1}(\g_0)$ is now independent of the parameter $k$. Now, by applying the proof of \cite[Lemma 5.14]{CGN21}, we get Theorem \ref{thm:nonSUSY vs SUSY}.

\subsection{Corollaries}
Here we give several corollaries of Theorem \ref{thm:nonSUSY vs SUSY}. Throughout this subsection, $\mathfrak{g}$, $f$, and $F$ will be as in Theorem \ref{thm:nonSUSY vs SUSY} and $k$ noncritical. Recall the large level limit $W^{\text{free}}(\g,F)$ of a W-algebra introduced in \cite[Section 6]{DK06} and \cite[Section 3]{CL22a}. By Theorem \ref{thm:nonSUSY vs SUSY}, we can also consider a large level limit of $W^k_{N=1}(\g,f)$ and get some properties of SUSY W-algebras and their cosets.

\begin{corollary} \label{cor:largelevel} The SUSY W-algebra $W^k_{N=1}(\mathfrak{g}, f)$ admits a large level limit $W^{\text{free}}_{N=1}(\mathfrak{g}, f)$ which is a tensor product of the simple free field algebras $\mathcal{O}_{\text{ev}}(n,r)$, $\mathcal{O}_{\text{odd}}(n,r)$, $\mathcal{S}_{\text{ev}}(n,r)$, and $\mathcal{S}_{\text{odd}}(n,r)$ introduced in \cite{CL22a}.
\end{corollary}

\begin{proof} By \cite[Corollary 3.4]{CL22a}, the large level limit $W^{\text{free}}(\mathfrak{g}, F)$ of $W^k(\mathfrak{g}, F)$ decomposes as a tensor product of these free field algebras. The claim is immediate from Theorem \ref{thm:nonSUSY vs SUSY} together with the fact that $\mathcal{F}(\mathfrak{g}^f_0)$ is a tensor product of a free fermion algebra $\mathcal{F}(m) = \mathcal{O}_{\text{odd}}(m,1)$ and a $\beta\gamma$-system $\mathcal{S}(n) = \mathcal{S}_{\text{ev}}(n,1)$. 
\end{proof}

\begin{corollary} \label{cor:simplicity} $W^k_{N=1}(\mathfrak{g}, f)$ is simple for generic values of $k$.
\end{corollary}

\begin{proof} This immediate from Theorem \ref{thm:nonSUSY vs SUSY} and with Theorem 3.6 of \cite{CL22a}, since $\mathcal{F}(\mathfrak{g}^f_0)$ is simple.
\end{proof} 

\begin{corollary} \label{cor:ordinary module}
$W^k_{N=1}(\mathfrak{g}, f)$ and $W^k(\mathfrak{g}, F)$ have equivalent categories of ordinary modules, i.e., positive energy modules with finite-dimensional weight spaces.
\end{corollary}

\begin{proof} In view of Theorem \ref{thm:nonSUSY vs SUSY}, it suffices to prove that the category of ordinary modules of $\mathcal{F}(\mathfrak{g}_0^f)$ is semisimple and has only one irreducible module. Since $\mathcal{F}(\mathfrak{g}_0^f)$ is a tensor product of a free fermion algebra and a $\beta\gamma$-system, and the free fermion algebra is well known to have this property, it suffices to prove it for the $\beta\gamma$-system $\cS$ of rank $1$.

First, $\cS$ has an action of $\mathbb{Z}_2$ by automorphisms, and the orbifold $\cS^{\mathbb{Z}_2}$ which is generated by all quadratics in $\beta, \gamma$, and their derivatives, is isomorphic to the simple affine vertex algebra $L_{-1/2}(\mathfrak{sl}(2))$. As a module over $L_{-1/2}(\mathfrak{sl}(2))$, $\cS$ is the sum of $L_{-1/2}(\mathfrak{sl}(2))$ and the module $L_{\omega_1}(\mathfrak{sl}(2))$, which is just the simple quotient of the Weyl module corresponding to the first fundamental weight $\omega_1$. The category of ordinary modules for $L_{-1/2}(\mathfrak{sl}(2))$ is semisimple, and these two modules (call them $M_0$ and $M_1$) are the only two irreducible ordinary modules for $L_{-1/2}(\mathfrak{sl}(2))$ \cite[Thm. 3.5.3]{AM95}.
 
Given any ordinary $\cS$-module $U$, if we view it as a module over $L_{-1/2}(\mathfrak{sl}(2))$, it is completely reducible over $L_{-1/2}(\mathfrak{sl}(2))$, and it decomposes as a sum of copies of $M_0$ and $M_1$. For each copy of $M_0$ that appears in $U$, the $\cS$-module generated by $M_0$ must contain a copy of $M_1$, since otherwise the action by all modes of both $\beta$ and $\gamma$ would be zero. Since $L_{-1/2}(\mathfrak{sl}(2))$ is generated by all quadratics in $\beta, \gamma$, and their derivatives, the $\cS$-module generated by $M_0$ is exactly $M_0 \oplus M_1$, so it is isomorphic to $\cS$ as an $\cS$-module. Similarly, for each copy of $M_1$ that appears in $U$, the $\cS$-module generated by $M_1$ must contain a copy of $M_0$, and hence must also be a copy of $\cS$. It follows that $U$ is a direct sum of copies of $\cS$.
 \end{proof}

We also get some corollories on coset vertex algebras of SUSY W-algebras. First, recall that the coset construction is a basic way to construct new vertex algebras from old ones. It was introduced by Frenkel and Zhu in \cite{FZ92}, generalizing earlier constructions in representation theory \cite{KP81} and physics \cite{GKO85}, where it was used to construct the unitary discrete series representations of the Virasoro algebra. Given a vertex algebra $V$ and a subalgebra $A\subseteq V$, the coset $\mathcal{C} = \text{Com}(A,V)$ is the subalgebra of $V$ which commutes with $A$. The most important feature of $\mathcal{C}$ is the following.

\begin{lemma} \label{cosetvirasoro} \cite[Theorem 5.1]{FZ92} Suppose that $V$ and $A$ have conformal vectors $L^{V}$ and $L^{A}$ respectively, and ${L^{V}}_{(2)} L^{A} = 0$. Then $\mathcal{C}$ has conformal vector $L^{V} - L^{A}$.
\end{lemma} In particular, this implies that $V$ is a conformal extension of $A\otimes \mathcal{C}$. The theory of vertex (super)algebra extensions then allows the representation theory of $V$ to be studied using the representation theory of $\mathcal{A}$ and $\mathcal{C}$, and vice versa \cite{CKM24}. 

We have the following SUSY analogue of Lemma \ref{cosetvirasoro}. To emphasize the forms of the conformal and superconformal vectors, we say that a SUSY vertex algebra $V$ with $N=1$ structure $(L,G)$, when $G$ is a superconformal vector of $V$ and $L=\frac{1}{2}DG$.

\begin{lemma} \label{cosetSUSY} Suppose that $V$ is a SUSY vertex algebra with $N=1$ structure $(L^{V}, G^{V})$, and $\cA \subseteq \mathcal{V}$ is a SUSY subalgebra with $N=1$ structure $(L^{A}, G^{A})$. Suppose that 
\begin{enumerate}[(i)]
    \item ${L^{V}}_{(2)} L^{A} = 0$,
    \item ${G^{V}}_{(0)} G^{A} = 2L^{A}$,
    \item ${G^{V}}_{(1)} G^{A} = 0$.
\end{enumerate}
Then $\mathcal{C} = \textup{Com}(A, V)$ is a SUSY vertex algebra with $N=1$ structure $(L^{V} - L^{A}, G^{V} - G^{A})$.
\end{lemma}
\begin{proof} This is straightforward computation using our hypotheses.
\end{proof}

 In general, the structure of cosets is quite mysterious, but in the case where $\mathcal{V} = W^k_{N=1}(\mathfrak{g}, f)$ and $\mathcal{A}$ is a SUSY affine vertex algebra, we have methods to describe this structure for generic levels. Let $\mathfrak{a} \subseteq \mathfrak{g}^f_0$ be a Lie sub(super)algebra, and suppose that
\begin{enumerate}[(i)]
\item $\mathfrak{a} = \oplus_{i=1}^r \mathfrak{a}_{i}$, where each summand $\mathfrak{a}_i$ is either an abelian or simple Lie algebra, or the Lie superalgebra $\mathfrak{osp}(1|2n)$ for some $n$.
\item The restriction of the bilinear form on $\mathfrak{g}^f_0$ to $\mathfrak{a}$ is nondegenerate.
\end{enumerate}
Under the assumption, we have an embedding $V^{k_1}_{N=1}(\mathfrak{a}_1) \otimes \cdots \otimes V^{k_r}_{N=1}(\mathfrak{a}_r) \hookrightarrow W^k_{N=1}(\mathfrak{g}, f)$ of SUSY affine vertex algebras, and we use the shorthand $V^{\bar{k}}_{N=1}(\mathfrak{a}) = V^{k_1}_{N=1}(\mathfrak{a}_1) \otimes \cdots \otimes V^{k_r}_{N=1}(\mathfrak{a}_r)$. 

\begin{corollary} \label{cor:simplicityofcosets} For any Lie sub(super)algebra $\mathfrak{a}$ satisfying (i) and (ii), the coset $\textup{Com}(V^{\bar{k}}_{N=1}(\mathfrak{a}), W^k_{N=1}(\mathfrak{g}, f))$ is a simple SUSY vertex algebra for generic values of $k$.
\end{corollary}

\begin{proof} First, we give $V^{k_i}_{N=1}(\mathfrak{a}_i)$ its Kac-Todorov $N=1$ structure when $\mathfrak{a}_i$ is simple, and in the case where $\mathfrak{a}_i$ is abelian, we give the corresponding Heisenberg algebra the $N=1$ structure where the generators are primary of weight one. Then the hypotheses of Lemma \ref{cosetSUSY} are clearly satisfied, so $\text{Com}(V^{\bar{k}}_{N=1}(\mathfrak{a}), W^k_{N=1}(\mathfrak{g}, f))$ is a SUSY vertex algebra with the above $N=1$ structure.

Since the bilinear for on $\mathfrak{a}$ is nondegenerate, we have decompositions 
$$V^{\bar{k}}_{N=1}(\mathfrak{a}) = V^{\bar{k}}(\mathfrak{a}) \otimes \mathcal{F}(\mathfrak{a}),\qquad V^{\bar{k}}(\mathfrak{a}) =  V^{k_1}(\mathfrak{a}_1) \otimes \cdots \otimes V^{k_r}(\mathfrak{a}_r),$$ and $\mathcal{F}(\mathfrak{g}^f_0) \cong \mathcal{F}(\mathfrak{a}) \otimes  \mathcal{F}(\mathfrak{a}^{\perp})$, where $\mathfrak{a}^{\perp}$ is the orthogonal complement of $\mathfrak{a}$ in $\mathfrak{g}^f_0$, and both $\mathcal{F}(\mathfrak{a})$ and $\mathcal{F}(\mathfrak{a}^{\perp})$ are simple. Then
\begin{equation} \text{Com}(V^{\bar{k}}_{N=1}(\mathfrak{a}), W^k_{N=1}(\mathfrak{g}, f)) \cong \text{Com}(V^{\bar{k}}(\mathfrak{a}), W^k(\mathfrak{g}, F) \otimes \mathcal{F}(\mathfrak{a}^{\perp})),
\end{equation} which is simple by Corollary \ref{cor:simplicity} together with \cite[Lemma 2.1]{ACKL17} in the case when $\mathfrak{a}$ is a Lie algebra, and \cite[Theorem 5.3]{CL22b} in the case when $\mathfrak{a}$ has some factors of the form $\mathfrak{osp}(1|2n)$.
\end{proof}

\begin{corollary} \label{cor:SFGproperty} Suppose that $\mathfrak{g}$ and $f$ are as in Theorem \ref{thm:nonSUSY vs SUSY}. 

\begin{enumerate}
\item Let $G$ be any reductive group of automorphisms of $W^k_{N=1}(\mathfrak{g}, f)$ which preserves the conformal vector. Then the orbifold $W^k_{N=1}(\mathfrak{g}, f)^G$ is strongly finitely generated.
\item Let $\mathfrak{a}\subseteq \mathfrak{g}^f_0$ and $V^{\bar{k}}_{N=1}(\mathfrak{a})$ be as in Corollary \ref{cor:simplicityofcosets}. Then the coset $\textup{Com}(V^{\bar{k}}_{N=1}(\mathfrak{a}), W^k_{N=1}(\mathfrak{g}, f))$ is strongly finitely generated for generic values of $k$.
\end{enumerate}
\end{corollary}

\begin{proof} The proof of the first statement is the same as the proof of \cite[Theorem 4.1 (1)]{CL22a}, using the fact that the large level limit of $W^k_{N=1}(\mathfrak{g}, f)^G$ is $(W^{\text{free}}(\mathfrak{g}, F) \otimes \mathcal{F}(\mathfrak{g}^f_0))^G$. In the case that $\mathfrak{a}$ is a Lie algebra, the proof of the second statement is the same as the proof of \cite[Theorem 4.1 (2)]{CL22a}. Recall that the large level limit of $\text{Com}(V^{\bar{k}}(\mathfrak{a}), W^k(\mathfrak{g}, F))$ is an orbifold of the form $\mathcal{V}^A$, where $\mathcal{V}$ is a tensor product of the free field algebras $\mathcal{O}_{\text{ev}}(n,r)$, $\mathcal{O}_{\text{odd}}(n,r)$, $\mathcal{S}_{\text{ev}}(n,r)$, and $\mathcal{S}_{\text{odd}}(n,r)$, and $A$ is a Lie group with Lie algebra $\mathfrak{a}$. Then $\text{Com}(V^{\bar{k}}_{N=1}(\mathfrak{a}), W^k_{N=1}(\mathfrak{g}, f))$ has large level limit $(\mathcal{V} \otimes \mathcal{F}(\mathfrak{a}^{\perp}))^A$. This is strongly finitely generated because $A$ is reductive (\cite[Corollary 4.2]{CL22a}), which implies the claim by \cite[Theorem 6.10]{CL19}. Finally, in the case when $\mathfrak{a}$ has some factors of the form $\mathfrak{osp}(1|2n)$, the same argument applies using in addition \cite[Theorem 5.3]{CL22b}.
\end{proof}

\section{Minimal W-superalgebras and SUSY W-algebras} \label{sec:minimal}

\begin{comment}
    Applying the above lemma to $W^k_{N=1}(\g,f)$ and its subalgebra $\mathcal{F}(\g_0^f)$, we can deduce that \eqref{eq:conjecture} is equivalent to
\begin{equation} \label{eq:conj coset}
    \textup{Com}(\mathcal{F}(\g_0^f),W^k_{N=1}(\g,f))\simeq W^k(\g,F).
\end{equation}
\end{comment}

In this section, we investigate SUSY W-algebras associated with odd minimal nilpotent elements $f$, where the corresponding even nilpotent $F=-\frac{1}{2}[f,f]$ is an even minimal nilpotent. We classify minimal odd nilpotents $f$ in a basic Lie superalgebra $\g$, and determine the $\g^f_0$-module structure on $\g^F_0$. Furthermore, we provide explicit descriptions of the SUSY vertex algebra structures for the minimal SUSY W-algebras.

\subsection{List of minimal $f$ and $F$} \label{subsec:list of minimal}
In \cite[Section 5]{KW04}, there is a list of Lie superalgebras that admit an even minimal nilpotent $F$. In this section, we show that every Lie superalgebra in the list also possesses an odd minimal nilpotent $f$ satisfying $F=-\frac{1}{2}[f,f]$ when $\g$ is basic. Furthermore, we describe the structure of $\g_0^F$ as a $\g_0^f$-module, which decomposes as
\begin{equation} \label{eq:gF decomp}
    \mathfrak{g}^F_0= \mathfrak{g}^f_0 \oplus M^{\mathfrak{g}}_{(F,f)}.
\end{equation}
for some $\g_0^f$-module $M^{\mathfrak{g}}_{(F,f)}$ under the adjoint action of $\g_0^f$. To begin with the conclusion, the two summands $\mathfrak{g}_0^f$  and $M^{\mathfrak{g}}_{(F,f)}$ in \eqref{eq:gF decomp} are described as follows:

\begin{table}[h!]
\centering
\begin{tabular}{||c | c | c | c||} 
 \hline 
 $\mathfrak{g}$ & $\mathfrak{g}^F_0$ & $\mathfrak{g}^f_0$ & $M^{\mathfrak{g}}_{(F,f)}$ \\ [0.5ex] 
 \hline\hline
 $\mathfrak{sl}(2|m)$ $(m\neq 2)$  & $\mathfrak{gl}_m$ & $\mathfrak{gl}_{m-1}$ & $\mathbb{C}\oplus \mathbb{C}^{m-1}\oplus (\mathbb{C}^{m-1})^*$\\ 
  \hline
 $\mathfrak{psl}(2|2)$ & $\mathfrak{sl}(2)$ & $0$  & $\mathbb{C}\oplus \mathbb{C}\oplus \mathbb{C}$ \\
  \hline
 $\mathfrak{spo}(2|m)$ & $\mathfrak{so}(m)$ & $\mathfrak{so}(m-1)$ & $\mathbb{C}^{m-1}$ \\
  \hline
 $\mathfrak{osp}(4|m)$ & $\mathfrak{sl}(2)\oplus \mathfrak{sp}(m)$ & $\mathfrak{sl}(2)\oplus \mathfrak{sp}(m-2)$ & $\mathbb{C}^3 \oplus \mathbb{C}^2 \otimes \mathbb{C}^{m-2}$ \\
  \hline
 $D(2,1;\alpha)$ $(\alpha\neq 0,-1)$ & $\mathfrak{sl}(2)\oplus \mathfrak{sl}(2)$ & $\mathfrak{sl}(2)$ & $\mathbb{C}^3$ \\ [0.5ex] 
 \hline
\end{tabular}
\vskip 2mm
\caption{$\mathfrak{g}^F_0$ is a Lie algebra $(m\geq 1)$}
\label{table1}
\end{table}
\vskip -2em
\begin{table}[h!]
\centering
\begin{tabular}{||c | c | c | c||} 
 \hline 
 $\mathfrak{g}$ & $\mathfrak{g}^F_0$ & $\mathfrak{g}^f_0$ & $M^{\mathfrak{g}}_{(F,f)}$ \\ [0.5ex] 
 \hline\hline
 $\mathfrak{sl}(m|n)$ $(m\neq n, m> 2)$ & $\mathfrak{gl}(m-2|n)$ &  $\mathfrak{gl}(m-2|n-1)$ & $\mathbb{C}^{1|0}\oplus \mathbb{C}^{m-2|n-1}\oplus (\mathbb{C}^{m-2|n-1})^*$ \\
 \hline
 $\mathfrak{psl}(m|m)$ $(m>2)$ & $\mathfrak{sl}(m-2|m)$ & $\mathfrak{sl}(m-2|m-1)$ & $\mathbb{C}^{1|0}\oplus \mathbb{C}^{m-2|m-1}\oplus (\mathbb{C}^{m-2|m-1})^*$ \\
 \hline
 $\mathfrak{spo}(n|m)$ $(n\geq 4)$ & $\mathfrak{spo}(n-2|m)$ & $\mathfrak{spo}(n-2|m-1)$ & $\mathbb{C}^{n-2|m-1}$ \\
 \hline
 $\mathfrak{osp}(m|n)$ $(m\geq 5)$ & $ \mathfrak{sl}(2)\oplus \mathfrak{osp}(m-4|n)$ & $\mathfrak{sl}(2)\oplus\mathfrak{osp}(m-4|n-2)$ & $\mathbb{C}^{3|0}\oplus \mathbb{C}^{2|0}\otimes \mathbb{C}^{m-4|n-2}$ \\[0.5ex] 
 \hline
\end{tabular}
\vskip 2mm
\caption{$\mathfrak{g}^F_0$ is not a Lie algebra $(m,n\geq 1)$}
\label{table2}
\end{table}

\vskip 10mm

In the tables above, the distinction between $\mathfrak{spo}(m|n)$ and $\mathfrak{osp}(n|m)$ lies in the placement of an even nilpotent $F$. In the former case, $F$ belongs to $\mathfrak{sp}(m)$ subalgebra, whereas in the latter, it is contained in $\mathfrak{so}(n)$. Since the proof is similar for other cases, we only explain $\mathfrak{spo}(2|m)$, $D(2,1;\alpha)$ and $\mathfrak{osp}(m|n)$.

For the proof, we briefly introduce the embeddings of Lie superalgebras inside free field algebras. In this section, let $\mathcal{S}(n)$ be the rank $n$ $\beta\gamma$-system generated by $\{\beta^i, \gamma^i\,|\,i=1, \cdots, n \}$ and $\mathcal{F}(n)$ be the free fermion algebra generated by the fermions $\{\phi^i\,|\,i=1, \cdots, n\}$ satisfying $[\phi^i\, {}_\lambda \, \phi^j]=\delta_{i,j}$. Also, let $\mathcal{E}(n)$ be the rank $n$ $bc$-system generated by $\{b^i,c^i\,|\,i=1, \cdots,n\}$. We assume that $\beta^i$, $\gamma^i$ and $\phi^i$ have conformal weight $\frac{1}{2}$. Note that $\mathcal{F}(2n)\simeq \mathcal{E}(n)$ and $\mathcal{F}(2n+1)\simeq \mathcal{F}(1)\otimes \mathcal{E}(n)$ under the identification
\begin{equation} \label{eq:bc and fermion}
    b^i\equiv \frac{1}{\sqrt{2}}(\phi^i+\sqrt{-1}\phi^{n+i}), \quad c^i\equiv \frac{1}{\sqrt{2}}(\phi^i-\sqrt{-1}\phi^{n+i}), \quad \phi \equiv \phi^{2n+1}
\end{equation}
for $i=1, \cdots, n$.

Recall from \cite{goodmanwallach,CL19} that the following Lie algebras have embeddings into free field algebras:
\begin{equation}
    \begin{aligned}
        \mathfrak{gl}(n)&=\textup{Span}_{\CC}\{e_{i,j}\,|\,i,j=1, \cdots, n\},\\
          \mathfrak{sp}(2n)&=\textup{Span}_{\CC}\{e_{j,k+n}+e_{k,j+n}, e_{j+n,k}+e_{k+n,j}, e_{j,k}-e_{n+k,n+j}\,|\,j,k=1, \cdots,n\},\\
        \mathfrak{so}(2n)&=\textup{Span}_{\CC}\{e_{j,k+n}-e_{k,j+n},e_{j+n,k}-e_{k+n,j},e_{j,k}-e_{k+n,j+n}\,|\,j,k=1,\cdots,n\},\\
        \mathfrak{so}(2n+1)&=\textup{Span}_{\CC}\{e_{0,j}-e_{n+j,0},e_{0,n+j}-e_{j,0}\,|\,j=1,\cdots,n\}\cup \mathfrak{so}(2n).
    \end{aligned}
\end{equation}
First, $\mathfrak{gl}(n)$ embeds into either $\mathcal{S}(n)$ or $\mathcal{E}(n)$ as follows:
\begin{equation} \label{eq:gl(n) free}
    \begin{aligned}
    \mathfrak{gl}(n)&\rightarrow \mathcal{S}(n), \quad e_{j,k}\mapsto \,:\!\gamma^j\beta^k\!:,\\
    \mathfrak{gl}(n)&\rightarrow \mathcal{E}(n), \quad e_{j,k}\mapsto \,:\!c^jb^k\!:.
    \end{aligned}
\end{equation}
Note that the images of the embeddings in \eqref{eq:gl(n) free} are the weight $1$ subspaces with charge $0$. The $\mathfrak{sp}$ and $\mathfrak{so}$ type Lie algebras are embedded as
\begin{equation} \label{eq:sp(n) free}
    \begin{aligned}
        \mathfrak{sp}(2n)\rightarrow \mathcal{S}(n), \quad -e_{j,k+n}-e_{k,j+n}&\mapsto\, :\!\gamma^j\gamma^k\!:,\\
        e_{j+n,k}+e_{k+n,j}&\mapsto\, :\!\beta^j \beta^k\!:,\\
        e_{j,k}-e_{n+k,n+j}&\mapsto\, :\!\gamma^j \beta^k\!:,
    \end{aligned}
\end{equation}

\begin{equation} \label{eq:so(2n) free}
    \begin{aligned}
        \mathfrak{so}(2n)\rightarrow \mathcal{F}(2n)\simeq \mathcal{E}(n), \quad
        e_{j,k+n}-e_{k,j+n}&\mapsto \,:\!b^jb^k\!:,\\
        e_{j+n,k}-e_{k+n,j}&\mapsto \,:\!c^j c^k\!:,\\
        e_{j,k}-e_{k+n,j+n}&\mapsto \,:\!b^j c^k\!:,
    \end{aligned}
\end{equation}

\begin{equation} \label{eq:so(2n+1) free}
    \begin{aligned}
        \mathfrak{so}(2n+1)\rightarrow \mathcal{F}(2n+1)\simeq \mathcal{F}(1)\otimes \mathcal{E}(n), \quad
        e_{0,j}-e_{n+j,0}&\mapsto\, :\!\phi^0 c^j\!:,\\
        e_{0,n+j}-e_{j,0}&\mapsto\, :\!\phi^0 b^j\!:.
    \end{aligned}
\end{equation}
In \eqref{eq:so(2n+1) free}, the $\mathfrak{so}(2n)$ subalgebra is mapped as in \eqref{eq:so(2n) free}. Note that in the above cases, the images of the embeddings are the weight $1$ subspaces. For later use, we remark here that $\mathfrak{sl}(2)\simeq \mathfrak{sp}(2)$ and the $\mathfrak{sl}(2)$-triple $(X,Y,H)$ are embedded to $\mathcal{S}(1)$ as 
\begin{equation} \label{eq:sl(2) free}
    X\mapsto \frac{1}{2}:\!\gamma\gamma\!:,\quad Y\mapsto -\frac{1}{2}:\!\beta\beta\!:,\quad  H \mapsto \ :\!\gamma\beta\!:.
\end{equation}

Furthermore, one can generalize these embeddings to $\mathfrak{gl}$ and $\mathfrak{osp}$ type Lie superalgebras. Consider $\mathfrak{gl}(m|n)$ with the index set $I=\{1, \cdots,m, \bar{1}, \cdots, \bar{n}\}$, where the barred indices are odd. Then, one can easily check that
\begin{equation*}
    \begin{gathered}
    \mathfrak{gl}(m|n)\hookrightarrow \mathcal{S}(m)\otimes \mathcal{E}(n),\\
    e_{j,k}\mapsto\, :\!\gamma^j \beta^k\!:\, , \quad e_{j,\bar{k}}\mapsto\, :\!\gamma^j b^k\!:\,, \quad e_{\bar{j},k}\mapsto \, :\!c^j\beta^k\!:\, , \quad e_{\bar{j},\bar{k}}\mapsto \,:\!c^j b^k\!:
    \end{gathered}
\end{equation*}
is an embedding as a Lie superalgebra. Note that the image consists of weight $1$ elements of charge $0$. Now, consider $\g=\mathfrak{osp}(2m+1|2n)$ and the index set $I=\{0,1, \cdots, 2m,\bar{1}, \cdots, \overline{2n}\}$. Note that $\g$ is spanned by the matrices of the form
\begin{equation} \label{eq:osp Lie super}
    \left(
    \begin{array}{ccc|cc}
       0&-u^t&-v^t& x&x_1\\
       v& a  & b  & y&y_1\\
       u& c  &-a^t& z&z_1\\
        \hline
       -x_1^t& -z_1^t&-y_1^t&d &e\\
       x^t   & z^t   & y^t  &f &-d
    \end{array}\right),
\end{equation}
where $b$ and $c$ are skew-symmetric $m\times m$ matrices, and $e$ and $f$ are symmetric $n\times n$ matrices. Under the maps \eqref{eq:so(2n+1) free} and \eqref{eq:sp(n) free}, the even subalgebra $\mathfrak{so}(2m+1)\oplus \mathfrak{sp}(2n)$ can be embedded in $\mathcal{F}(2m+1)\otimes \mathcal{S}(n)$. By mapping the odd part of $\g$ as
\begin{equation} \label{eq:osp free field}
    \begin{gathered}
        \mathfrak{osp}(2m+1|2n)\hookrightarrow \mathcal{F}(2m+1)\otimes \mathcal{S}(n), \\
        % e_{1,1+j}-e_{1+m+j,1}\mapsto\, :\!\phi^0 c^j\!:\, , \quad
        % e_{1,1+m+j}-e_{1+j,1}\mapsto \,:\!\phi^0 b^j\!:\,,\\
        % e_{1+j,1+m+k}-e_{1+k,1+m+j}\mapsto\, :\!b^j b^k\!:\, , \quad
        % e_{1+m+j,1+k}-e_{1+m+k,1+j}\mapsto \, :\!c^jc^k\!:\, , \\
        % e_{1+j,1+k}-e_{1+m+k,1+m+j}\mapsto\, :\!b^j c^k\!:\,, \\
        % -e_{\bar{p},\overline{q+n}}-e_{\bar{q},\overline{p+n}}\mapsto \, :\!\gamma^p \gamma^q\!:\, , \quad
        % e_{\overline{p+n},\bar{q}}+e_{\overline{q+n},\bar{p}}\, \mapsto \,:\!\beta^p \beta^q\!:\,, \\
        % e_{\bar{p},\bar{q}}-e_{\overline{q+n},\overline{p+n}}\mapsto \, :\!\gamma^p \beta^q\!:\,,\\
        e_{0,\bar{p}}+e_{\overline{p+n},0}\mapsto \, :\!\phi^0 \beta^p\!:\, , \quad e_{\bar{p},0}-e_{0, \overline{p+n}}\mapsto \,:\!\phi^0 \gamma^p\!:\,,\\
        e_{j,\bar{p}}+e_{\overline{n+p},m+j}\mapsto \, :\!b^j \beta^p\!:\,, \quad
        -e_{j,\overline{p+n}}+e_{\bar{p},m+j}\mapsto \, :\!b^j \gamma^p\!:\,,\\
        e_{m+j,\bar{p}}+e_{\overline{p+n},j}\mapsto \, :\!c^j\beta^p\!:\,, \quad e_{\bar{p},j}-e_{m+j,\overline{p+n}}\mapsto \, :\!c^j \gamma^p\!:\,,
    \end{gathered}
\end{equation}
we get the realization of $\mathfrak{osp}(2m+1|2n)$ inside free fields. In \eqref{eq:osp free field}, we denoted the generators of $\mathcal{F}(2m+1)\simeq \mathcal{F}(1)\otimes \mathcal{E}(m)$ by $\{\phi^0, b^i,c^i\,|\,i=1, \cdots,m\}$. By ignoring the first index in \eqref{eq:osp Lie super} and $\eqref{eq:osp free field}$, we get the realization $\mathfrak{osp}(2m|2n)\hookrightarrow \mathcal{F}(2m)\otimes \mathcal{S}(n)$. As in Lie algebra cases, the images of the embeddings are the weight $1$ subspaces.

\subsubsection{$\mathfrak{spo}(2|m)$}
Consider the case when $m$ is even. Let $\g=\mathfrak{spo}(2|2n)$ for $n\geq 1$ and $I=\{1, \cdots, 2n, \bar{1},\bar{2}\}$ be the index set. Then, $\g$ is spanned by the matrices of the form \eqref{eq:osp Lie super} with the first row and column deleted.

Take a even minimal nilpotent $F=-e_{\bar{2},\bar{1}}$. Then
\begin{equation} \label{eq:minimal odd f}
    f=e_{\bar{2},i}+e_{n+i,\bar{1}}+e_{i,\bar{1}}+e_{\bar{2},n+i}
\end{equation}
for any $1\leq i \leq n$ satisfies $F=-\frac{1}{2}[f,f]$. Fix $i=1$. Note that $\g_0^F$ and $\g_0^f$ are
\begin{equation}\label{eq:so(n-1) inside so(n)}
    \begin{aligned}
        \g_0^F=&\left\{\left(\begin{array}{cc} a& b\\
                                 c& -a^t
        \end{array}\right)\right\}\simeq \mathfrak{so}(2n),\\
        \g_0^f=&\,\textup{Span}_{\CC}\{e_{1,j}-e_{n+j,n+1}+e_{n+j,1}-e_{n+1,j}\}_{2\leq j \leq n} \\
        &\cup\textup{Span}_{\CC} \{e_{p,q}-e_{n+p,n+q}, e_{p,n+q}-e_{q,n+p}, e_{n+p,q}-e_{n+q,p}\}_{2\leq p,q \leq n}\\
        =&\left\{ \left(\begin{array}{cccc}
                0  & u & 0  &v \\ 
                v^t& a &-v^t&b \\ 
                0  &-u & 0  &-v\\ 
                u^t& c &-u^t&-a^t
        \end{array}\right)\middle\vert \begin{array}{l}a:(n-1)\times (n-1) \text{ matrix}\\b,c:\text{skew-symmetric } (n-1)\times (n-1) \text{ matrix} \\ u,v:1\times (n-1)\text{ matrix} \end{array} \right\}\\
        \simeq &\,  \mathfrak{so}(2n-1).
    \end{aligned} 
\end{equation}

If we use the realization of $\g_0^F\simeq \mathfrak{so}(2n)$ inside $\mathcal{F}(2n)$ in \eqref{eq:so(2n) free}, then $\g_0^f$ in \eqref{eq:so(n-1) inside so(n)} is exactly the weight $1$ subspace of $\mathcal{F}(2n-1)$ generated by the fermions $\phi^2, \cdots,\phi^{2n}$. In other words, $\g_0^f$ is the weight $1$ subspace generated by $\phi^{n+1}, b^i, c^i$ for $2\leq i \leq n$ using the identification \eqref{eq:bc and fermion}. Observe that
\[\{:\!\phi^{n+1}\phi^1\!:\,, :\!b^2\phi^1\!:\,, \cdots, :\!b^n\phi^1\!:\,, :\!c^2\phi^1\!:\,,\cdots, :\!c^n \phi^1\!:\}\subset \g_0^F\setminus \g_0^f\]
forms a basis for the standard representation of $\mathfrak{so}(2n-1)$. Therefore, as a $\g_0^f$-module,
\begin{equation}
    \g_0^F\simeq \g_0^f\oplus \CC^{2n-1}
\end{equation}
for the standard representation $\CC^{2n-1}$ of $\mathfrak{so}(2n-1)$. Similarly, one can show the analogous result for odd $m$.

\subsubsection{$\mathfrak{osp}(m|n)$ $(m\geq 5)$}
Other cases can be checked similarly, so we only consider $\g=\mathfrak{osp}(2m+1|2n)$ for $m\geq 2$. We denote its element as in \eqref{eq:osp Lie super}.
Take a minimal even nilpotent $F=-e_{2m,1}+e_{m+1,m}$. Then,
\begin{equation*}
    f=e_{2m,\bar{r}}+e_{\overline{n+r},m}+e_{\bar{r},1}-e_{m+1,\overline{n+r}}
\end{equation*}
for any $1\leq r \leq n$ satisfies $F=-\frac{1}{2}[f,f]$. Fix $r=1$. Note that the matrices of the form \eqref{eq:osp Lie super} is contained in $\g_0^F$, when its $(i,j)$ entry is $0$ whenever $i$ or $j$ is contained in $\{1,m,m+1,2m\}$. It immediately follows that $\mathfrak{osp}(2m-3|2n)\subset \g_0^F$. In addition, $\g_0^F$ contains 
\begin{equation*}
   X_F:=e_{1,m}-e_{2m,m+1},\ Y_F:=e_{m,1}-e_{m+1,2m},\ H_F:=e_{1,1}-e_{m+1,m+1}-e_{m,m}+e_{2m,2m},
\end{equation*}
which form an $\mathfrak{sl}(2)$-triple. Thus, $\g_0^F\simeq \mathfrak{osp}(2m-3|2n)\oplus \mathfrak{sl}(2)$. Similarly, the matrices of the from \eqref{eq:osp Lie super} is contained in $\g_0^f$, when its $(i,j)$ entry is $0$ whenever $i$ or $j$ is contained in $\{1,m,m+1,2m, \bar{1}, \overline{n+1}\}$. In addition,
\begin{equation*}
    \begin{aligned}
    X_f:=&e_{1,m}-e_{2m,m+1}+e_{\bar{1}.\overline{n+1}},\ Y_f:=e_{m,1}-e_{m+1,2m}+e_{\overline{n+1},\bar{1}}\\
    H_f:=&e_{1,1}-e_{m+1,m+1}-e_{m,m}+e_{2m,2m}+e_{\bar{1},\bar{1}}-e_{\overline{n+1},\overline{n+1}},
 \end{aligned}
 \end{equation*}
 are contained in $\g_0^f$ forming an $\mathfrak{sl}(2)$-triple.Thus, $\g_0^f \simeq \mathfrak{osp}(2m-3|2n-2)\oplus \mathfrak{sl}(2)$. Now, we embed $\g_0^F$ inside $\mathcal{F}(2m-3)\otimes \mathcal{S}(n)\oplus \mathcal{S}(1)$ using \eqref{eq:osp free field} and \eqref{eq:sl(2) free}. Under this embedding, $\mathfrak{osp}(2m-3|2n-2)$ is realized as the weight $1$ subspace of $\mathcal{F}(2m-3)\otimes \mathcal{S}(n)$ excluding $\beta^1$ and $\gamma^1$ in $\mathcal{S}(n)$. Moreover, the $\mathfrak{sl}(2)$-triple is realized as
 \begin{equation} \label{eq:proof sl(2) free}
        X_f\mapsto \frac{1}{2}:\!\gamma \gamma\!:-\frac{1}{2}:\!\gamma^1 \gamma^1\!:, \quad Y_f\mapsto -\frac{1}{2}:\!\beta \beta\!:+\frac{1}{2}:\!\beta^1\beta^1\!:, \quad H_f\mapsto\ :\!\gamma \beta\!:+:\!\gamma^1 \beta^1\!:.
 \end{equation}
 Consider the following three sets
 \begin{gather}
    \{:\!\phi \beta^1\!:\, , :\!b^j\beta^1\!:\,, :\!c^j\beta^1\!:\,,:\!\gamma^l\beta^1\!:\,,:\!\beta^t \beta^1\!:\,|\,1\leq j \leq m-2,\, 2\leq l\leq n\}, \label{eq:proof standard1}\\
    \{:\!\phi \gamma^1\!:\, , :\!b^j\gamma^1\!:\,, :\!c^j\gamma^1\!:\,,:\!\gamma^l\gamma^1\!:\,,:\!\beta^t \gamma^1\!:\,|\,1\leq j \leq m-2,\, 2\leq l\leq n\}, \label{eq:proof standard2} \\
    \{:\!\beta^1 \beta^1\!:\,, :\!\gamma^1 \beta^1\!:\,, :\!\gamma^1\gamma^1\!:\}, \label{eq:proof adjoint}
 \end{gather}
which are linearly independent subsets of $\g_0^F \setminus \g_0^f$. Observe that both of \eqref{eq:proof standard1} and \eqref{eq:proof standard2} forms a basis of standard representation of $\mathfrak{osp}(2m-3|2n-2)$, whereas \eqref{eq:proof adjoint} generates the adjoint representation of $\mathfrak{sl}(2)$ in \eqref{eq:proof sl(2) free}. Since $\{\beta^1, \gamma^1\}$ generates the standard representation of $\mathfrak{sl}(2)$ in \eqref{eq:proof sl(2) free}, one can write
\begin{equation}
\g_0^F \simeq \g_0^f \oplus \mathbb{C}^{3|0}\oplus \mathbb{C}^{2|0}\otimes \mathbb{C}^{2m-3|2n-2}
\end{equation}
as an $\mathfrak{\g_0^f}$-module, where $\mathbb{C}^{2m-3|2n-2}$ is the standard representation of $\mathfrak{osp}(2m-3|2n-2)$. Note that the superspace $\mathbb{C}^{2m-3|2n-2}$ has dimension $(2n-2|2m-3)$.
\subsubsection{$D(2,1;\alpha)$ $(\alpha\neq 0,-1)$}
For the definition of $\g=D(2,1;\alpha)$, refer to \cite{Musson12}. We denote $\g_{\bar{0}}=\mathfrak{sl}(2)_1\oplus \mathfrak{sl}(2)_2\oplus \mathfrak{sl}(2)_3$, where $\mathfrak{sl}(2)_i \simeq \mathfrak{sl}(2)$ is spanned by $E_i,H_i$, and $ F_i$. The odd subspace of $\g$ is $\g_{\bar{1}}=V_1 \boxtimes V_2\boxtimes V_3$, where $V_i\simeq V=\textup{Span}_{\CC}\{u_1, u_{-1}\}$ is the standard representation of $\mathfrak{sl}(2)$. Here, $u_i$'s correspond to the matrices
\begin{equation*}
    u_1\equiv \left( \begin{array}{c} 1\\0 \end{array}\right), \quad u_{-1}\equiv \left( \begin{array}{c} 0\\1 \end{array}\right).
\end{equation*}
Take an even minimal element $F=F_1\in \mathfrak{sl}(2)_1$. Then,
\begin{equation} \label{eq: D(2,1;a) minimal}
    f= \tfrac{1}{\sqrt{2(\alpha+1)}}(u_{-1}\otimes u_{-1}\otimes u_{-1}+ u_{-1}\otimes u_1 \otimes u_1), \text{ or } f=\tfrac{1}{\sqrt{-2(\alpha+1)}}(u_{-1}\otimes u_{-1}\otimes u_1+ u_{-1}\otimes u_1 \otimes u_{-1})
\end{equation}
satisfies $F=-\frac{1}{2}[f,f]$. In \eqref{eq: D(2,1;a) minimal}, $\sqrt{\alpha+1}$ is a complex number determined up to sign by the equality $(\sqrt{1+\alpha})^2=1+\alpha$. Similarly, $\sqrt{-(\alpha+1)}$ is determined up to sign. In either case, one can easily compute that
\begin{equation*}
    \begin{aligned}
    \g_0^F&=\mathfrak{sl}(2)_2 \oplus \mathfrak{sl}(2)_3, \\
    \g_0^f&=\textup{Span}_{\CC}\{E_2-F_3, E_3-F_2,H_2-H_3\}\simeq \mathfrak{sl}(2).
    \end{aligned}
\end{equation*}
Note that $\{E_2+F_3, E_3+F_2, H_2+H_3\}\subset \g_0^F \setminus \g_0^f$ forms a basis of the adjoint representation of $\g_0^f$. Hence,
\begin{equation*}
    \g_0^F\simeq \g_0^f\oplus \CC^3
\end{equation*}
as a $\g_0^f$-module, where $\CC^3$ is the adjoint representation of $\mathfrak{sl}(2)$.

\subsection{W-algebra $W^k(\g,F)$ inside $W_{N=1}^k(\g,f)$}
In the rest of Section \ref{sec:minimal}, we assume $k\neq -h^\vee.$
In this section, we describe the coset vertex algebra $\textup{Com}(\mathcal{F}(\g^f_0), W^k_{N=1}(\g,f))$ which is isomorphic to $W^k(\g,F)$. 
Recall that every element of $W^k_{N=1}(\g,f)$ can be expressed in terms of the building blocks $J_{\bar{a}}$ for $a\in \mathfrak{p}$ introduced in \eqref{eq:building block}. To simplify the notations, let us denote $J_{\bar{a}}$ and $DJ_{\bar{a}}$ simply by $\bar{a}$ and $D\bar{a}$. Thanks to Theorem \ref{thm:nonSUSY vs SUSY}, it is enough to find all elements in $W^k_{N=1}(\g,f)$ which commute with the free field part $\mathcal{F}(\g^f_0)$. 

\subsubsection{Weight 1 fields} In the weight 1 subspace of the vertex algebra $W^k_{N=1}(\g,f)$, there are two types of elements. A weight 1 field is called a {\it Type 1 element} if it lies in the odd differential algebra generated by weight $\frac{1}{2}$
 fields. Otherwise, it is called a {\it Type 2 element}. The weight 1 subspace of $\textup{Com}(\mathcal{F}(\g^f_0), W^k_{N=1}(\g,f))\simeq W^k(\g,F)\subset W^k_{N=1}(\g,f)$ has dimension $\text{dim}(\g^F_0)$ and Type 1 subspace in the weight 1 space has dimension $\text{dim}(\g^f_0)$.
A Type 1 weight 1 generator in the centralizer of $\mathcal{F}(\g^f_0)$ is characterized by the following proposition.

\begin{proposition} \label{Prop:type1 weight 1 generator}
Let $a\in \g^f_0$. For the set of index $I^f_0$,  let $\{u^\alpha\}_{\alpha\in I^f_0}$ and $\{u_\alpha\}_{\alpha\in I^f_0}$ be dual bases of $\g^f_0.$
Then 
\begin{equation}
    \mathcal{J}_{\{a\}}=D\bar{a}+\frac{1}{k+h^\vee}\sum_{\alpha\in I^f_0}(-1)^{p(\alpha)}:\bar{u}^\alpha \overline{[a,u_\alpha]}: \, \in \, W^k_{N=1}(\g,f)
\end{equation}
is in the centralizer of $\mathcal{F}(\g^f_0).$
\end{proposition}
\begin{proof}
    Since any $\bar{a}$ and $D\bar{a}$ for $a\in \g^f_0$ is in $W^k_{N=1}(\g,f)$, it is clear that $\mathcal{J}_{\{a\}}$ is also in $W^k_{N=1}(\g,f)$.
    By an argument analogous to that of Lemma \ref{lem:decomp_affine}, one can show that $\mathcal{J}_{\{a\}}$ is in the centralizer of $\mathcal{F}(\g^f_0).$
\end{proof}

For any subspace $\mathfrak{m}$ of $\g$ and $A,B\in \g$, let us denote $\kappa_{\mathfrak{m}}(A|B):=\textup{str}_{\mathfrak{m}}\big((\textup{ad}A)(\textup{ad}B)\big)$. For $a_1$, $a_2\in \g^f_0$, 
by the similar computations to the proof of Lemma \ref{lem:decomp_affine}, one can check that 
    \begin{equation} \label{eq:level check}
         [\mathcal{J}_{\{a_1\}}{}_\lambda \mathcal{J}_{\{a_2\}}] = (-1)^{p(a_1)p(a_2)}\big(\mathcal{J}_{\{[a_1,a_2]\}} + (k+h^\vee)(a_1|a_2)-\frac{1}{2} \kappa_{\g^f_0}(a_1|a_2)\big).
    \end{equation}
Observe that $\g_0= \g^f_0 \oplus [f,\g_{1/2}]$ and thus  $\kappa_{\g_0}=\kappa_{\g^f_0}+\kappa_{[f,\g_{1/2}]}.$
Moreover, since 
$\text{ad}\, f$ is injective on $\g_{1/2}$ and   
$ [a_1, [a_2, [f,g]]]= (-1)^{p(a_1)+p(a_2)}[f,[a_1,[a_2,g]]]$
for $g\in \g_{1/2}$, we can say that 
   \begin{equation} \label{eq:level compare}
        \kappa_{[f,\g_{\frac{1}{2}}]}(a_1|a_2)=-\kappa_{\g_{\frac{1}{2}}}(a_1|a_2).
   \end{equation}
Here, the sign on the right-hand side of \eqref{eq:level compare} arises from the fact that $f$
 is an odd element. Finally, we get 
\begin{equation}
     \kappa_{\g_0}(a_1|a_2)+\kappa_{\g_{1/2}}(a_1|a_2)= \kappa_{\g^f_0}(a_1|a_2) 
\end{equation}
and hence \eqref{eq:level check} can be rewritten as 
\begin{equation}
    [\mathcal{J}_{\{a_1\}}{}_\lambda \mathcal{J}_{\{a_2\}}] = (-1)^{p(a_1)p(a_2)}\big(\mathcal{J}_{\{[a_1,a_2]\}} + \kappa(a_1|a_2)\big),
\end{equation}
where 
\begin{equation} \label{eq:kappa for minimal}
    \kappa(a_1|a_2):= (k+h^\vee) (a_1|a_2) -\frac{1}{2} \kappa_{\g_0}(a_1|a_2)-\frac{1}{2} \kappa_{\g_{1/2}}(a_1|a_2).
\end{equation}
We will see in Proposition \ref{prop:affine} that the level $\kappa$ coincides with the level of the weight 1 component of $W^k(\g,F)$ as stated in \cite[Theorem 2.1]{KRW03}, which aligns with the expectation from Theorem \ref{thm:nonSUSY vs SUSY}.
\vskip 2mm

Now, in the following proposition, we describe Type 2 generators of weight 1 in $W^k_{N=1}(\g,f)$ which commute with $\mathcal{F}(\g^f_0).$

\begin{proposition} \label{prop:weight 1 elt}
    Let $b\in \g^f_{-1/2}$ and choose $v_b\in [e,\g^f_{-1/2}]$ such that $[f,v_b]=b$. Then,
    \begin{equation} \label{eq: G_b}
     \mathcal{G}_{\bar{b}}=\bar{b}+(-1)^{p(b)}(k+h^\vee) D(\bar{v}_b) -   \sum_{\beta\in I_{1/2}} \frac{1}{2} :\!\overline{[e,u^\beta]}(\, \overline{[b,u_\beta]}+\overline{[b,u_\beta]}^\sharp\, )\!:+(-1)^{p(b)}:\!\bar{x}\overline{v}_b\!:
    \end{equation}
    is contained in $W^k_{N=1}(\g,f)$, where 
    \begin{itemize}
        \item  $e\in \g_{1/2}$ is an element in \eqref{eq:f and e} such that $[e,f]=[f,e]=-2x$,
        \item  $\{u_\beta|\beta\in I_{1/2}\}$ and $\{u^\beta|\beta\in I_{1/2}\}$ are bases of $\g_{1/2}$ and $\g_{-1/2}$ such that $(u^\alpha|u_\beta)=\delta_{\alpha,\beta}$,
        \item $ a^\sharp$ for $a\in \g_0$ is the orthogonal projection of $a$ onto $\g^f_0$.
    \end{itemize}
    In addition, $\mathcal{G}_{\bar{b}}$ is in the centralizer of $\mathcal{F}(\g^f_0).$
\end{proposition}

\begin{proof}
    For each summand $A$ of \eqref{eq: G_b}, we have
    \begin{equation} \label{eq: G_b proof}
        \widetilde{Q}(A)= \sum_{\beta \in I_{1/2}} :\phi^{\beta}K^\beta(A): +\sum_{\beta \in I_{1/2}} (-1)^{p(\beta)+1}:(D\phi^{\beta})K^{D\beta}(A): 
    \end{equation}
    for some $K^\beta(A)$ and $K^{D\beta}(A)\in V_{N=1}^k(\mathfrak{p})$, where $\widetilde{Q}$ is the differential introduced in \eqref{eq:differential}. Analogously, we denote the coefficients of $\widetilde{Q}(\mathcal{G}_{\bar{b}})$ by $K^{\alpha}(\mathcal{G}_{\bar{b}})$ and $K^{D\alpha}(\mathcal{G}_{\bar{b}})$ after summing up \eqref{eq: G_b proof}. We claim that $K^\alpha(\mathcal{G}_{\bar{b}})=K^{D\alpha}(\mathcal{G}_{\bar{b}})=0$ for every $\alpha\in I_{1/2}$. Fix $\alpha\in I_{1/2}$. Then, we have 
   \begin{equation} 
       \begin{aligned}
           & K^{D\alpha}(\bar{b})= (k+h^\vee) (u_\alpha|b),\\
           & K^{D\alpha}(D\bar{v}_b)= (f|[u_\alpha,v_b])=(-1)^{p(b)+1}(u_\alpha|b),\\
           & K^{D\alpha}(:\!\bar{a}_1 \bar{a}_2\!:) =0 \quad \text{for } a_1, a_2\in \g_0. 
       \end{aligned}
   \end{equation}
   Hence $K^{D\alpha}(\mathcal{G}_{\bar{b}})=0.$ We also have 
       \begin{align}
           &  \label{eq:Prop_weight1_3}
           K^{\alpha}(\bar{b}+ (-1)^{p(b)}(k+h^\vee) D\bar{v}_b) = (-1)^{p(\alpha)p(b)+p(\alpha)}  \overline{[u_\alpha,b]}, \\
           & K^{\alpha} \big(\, -\frac{1}{2} \sum_{\beta\in I_{1/2}} :\overline{[e,u^\beta]} \ \overline{[b,u_\beta]}: \   \big) \\
            &  \label{eq:Prop_weight1_1} =\frac{1}{2}\sum_{\beta\in I_{1/2}}(-1)^{p(\alpha)p(\beta)+1} (f|[u_{\alpha},[e,u^\beta]])\overline{[b,u_\beta]}  
            \\
            & \label{eq:Prop_weight1_2} +\frac{1}{2}\sum_{\beta\in I_{1/2}}(-1)^{p(\alpha)p(b)+p(\alpha)+1}\overline{[e,u^\beta]}(f|[u_\alpha, [b,u_\beta]]) .
       \end{align}
    Let us denote by $\g^e$ the centralizer of $e$ in $\g$ and $\g^e_{i}= \g^e \cap \g_i.$ Since $F$ is a minimal nilpotent, $\g^e=\g^e_0 \oplus \g^e_{1/2} \oplus \mathbb{C}E$ and $\n= \g^e_{1/2} \oplus \mathbb{C}e \oplus \mathbb{C}E.$
    By direct computation, one can check that 
    \begin{equation}
        \eqref{eq:Prop_weight1_1} = \left\{\begin{array}{ll} (-1)^{p(b)p(\alpha)+p(\alpha)+1}\frac{1}{2}\overline{[u_\alpha,b]} & \text{ if } u_\alpha \in \g^e_{1/2}, \\
        (-1)^{p(b)+1}\frac{1}{2}\overline{v}_b & \text{ if } u_\alpha =e. \end{array} \right.
    \end{equation}
    Here we used $[e,b]=v_b$.
    By the representation theory of $\mathfrak{osp}(1|2)$, the $\mathfrak{osp}(1|2)$ subalgebra $\mathfrak{s}$ decomposes $\g$ by $\g= \g[1] \oplus \g[3] \oplus \g[5]$ where $\g[1]=\g^f_0$ and $\g[3]$ are the sum of dimension $1$ and $3$ irreducible modules, respectively, and $\g[5]= \mathfrak{s}$ is the dimension $5$ irreducible module. For $a\in \g_0$, let us denote by $a^\sharp$,  $a_{[3]}$ and $a_x$ the orthogonal projections of $a$ onto $\g[1]$, $\g[3]$ and $\g[5].$ Note that $\g[5]\cap \g_0=\mathbb{C}x$ and hence $a_x$ is the projection of $a$ onto $\mathbb{C}x.$ Then 
    \begin{equation}
        \eqref{eq:Prop_weight1_2} = \left\{\begin{array}{ll} 
        (-1)^{p(b)p(\alpha)+p(\alpha)}\frac{1}{2}\overline{[u_\alpha,b]}_x+ (-1)^{p(b)p(\alpha)+p(\alpha)+1}\frac{1}{2}\overline{[u_\alpha,b]}_{[3]}  & \text{ if } u_\alpha \in \g^e_{1/2}, \\
        (-1)^{p(b)}\frac{1}{2}\overline{v}_b  & \text{ if } u_\alpha=e.
        \end{array}\right.
        \end{equation}
    Hence ,
    \begin{equation} \label{eq:Prop_weight1_4}
    \begin{aligned}
          & K^{\alpha}  \big(\, -\frac{1}{2} \sum_{\beta\in I_{1/2}} :\overline{[e,u^\beta]} \ \overline{[b,u_\beta]}:\   \big) \\
          & = \left\{\begin{array}{ll} (-1)^{p(b)p(\alpha)+p(\alpha)+1}\frac{1}{2} \, \big( \ \overline{[u_\alpha,b]}^\sharp+ 2 \overline{[u_\alpha,b]}_{[3]} \ \big) & \text{ if } u_\alpha \in \g^e_{1/2}, \\
           0  & \text{ if } u_\alpha=e.
          \end{array}\right.
    \end{aligned}
    \end{equation}
   By similar computations, we get 
       \begin{equation}  \label{eq:Prop_weight1_5}
    \begin{aligned}
          & K^{\alpha}  \big(\, -\frac{1}{2} \sum_{\beta\in I_{1/2}} :\overline{[e,u^\beta]} \ \overline{[b,u_\beta]}^\sharp:+ (-1)^{p(b)} :\bar{x} \overline{v}_b:\   \big) \\
          & = \left\{\begin{array}{ll} (-1)^{p(b)p(\alpha)+p(\alpha)+1}\frac{1}{2} \, \big( \ \overline{[u_\alpha,b]}^\sharp+2\overline{[u_\alpha,b]}_x \ \big) & \text{ if } u_\alpha \in \g^e_{1/2}, \\
           (-1)^{p(b)}\overline{v}_b  & \text{ if } u_\alpha=e.
          \end{array}\right.
    \end{aligned}
    \end{equation}
    By  \eqref{eq:Prop_weight1_3},  \eqref{eq:Prop_weight1_4} and   \eqref{eq:Prop_weight1_5}, we conclude the proposition. Similarly, we can check that $\mathcal{G}_{\bar{b}}$ is in the centralizer of $\mathcal{F}(\g^f_0)$ by direct computations. 
\end{proof}

 Let $a\in \g^F_0$ and $a=a^\sharp+a^\flat$ for $a^\sharp\in \g^f_0$ and $a^\flat\in [e,\g^f_{-1/2}].$ We define 
\begin{equation} \label{eq:weight 1 generator, general}
    \mathcal{J}_{\{a\}}:= \mathcal{J}_{\{a^\sharp\}}+ \mathcal{J}_{\{a^\flat\}},
\end{equation}
where $\mathcal{J}_{\{a^\sharp\}}$ is the element introduced in Proposition \ref{Prop:type1 weight 1 generator} and $\mathcal{J}_{\{a^\flat\}}=\frac{(-1)^{p(a^\flat)+1}}{k+h^\vee}\mathcal{G}_{\overline{[f,a^\flat]}}$ for $\mathcal{G}_{\overline{[f,a^\flat]}}$ is given by Proposition \ref{prop:weight 1 elt}. Then the following theorem is deduced from Theorem \ref{thm:nonSUSY vs SUSY}.

\begin{proposition} \label{prop:affine}
    The set $\mathcal{J}_{\{\g^F_0\}}$ is the weight 1 space of $\textup{Com}(\mathcal{F}(\g^f_0),W^k_{N=1}(\g,f)).$ Moreover, 
    for $a,b\in \g^F_0,$ we have  
    \begin{equation}
        [\, \mathcal{J}_{\{a\}}\, {}_\lambda \, \mathcal{J}_{\{b\}}\, ]=(-1)^{p(a)p(b)}\big(\mathcal{J}_{\{[a,b]\}}+\lambda\kappa(a|b)\big),
    \end{equation}
    where $\kappa$ is \eqref{eq:kappa for minimal}.
\end{proposition}
\begin{proof}
    It is clear that  $\mathcal{J}_{\{\g^F_0\}}$ is the set of weight 1 elements in $W^k_{N=1}(\g,f)$ which commute with $\mathcal{F}(\g^f_0).$  Hence $\mathcal{J}_{\{\g^F_0\}}$ generate the vertex algebra  which is isomorphic to the affine vertex subalgebra of $W^k(\g,F).$
    
    Indeed, according to \cite{KRW03}, the ordinary W-algebra $W^k(\g,F)$ has an affine vertex algebra $V^{\psi_k}(\g^F_{0})$ with the $\lambda$-bracket $[a{}_\lambda b] = [a,b]+\psi_k(a|b)$ for $a,b\in \g_0^F$, where $\psi_k(a|b)= k(a|b)+\kappa_{\g_+}(a|b)-\frac{1}{2}\kappa_{\g_{\frac{1}{2}}}(a|b).$ Recall the basis $\{u_{\alpha}\}_{\alpha\in I_+}$ of $\g_+$ and its dual $\{u^{\alpha}\}_{\alpha\in I_+}$ satisfying $(u^{\alpha}|u_{\beta})=\delta_{\alpha,\beta}$. Then for any $a,b\in \g_0^F$, we have
    \begin{equation} \label{eq: affine level proof}
        \begin{aligned}
        \kappa_{\g_+}(a|b)=&\sum_{\alpha\in I_+}(-1)^{p(\alpha)}(u^{\alpha}|[a,[b,u_{\alpha}]])=\sum_{\alpha\in I_+}(-1)^{p(\alpha)+p(a)p(\alpha)+1}(a|[u^{\alpha},[b,u_{\alpha}]])\\
        =&\sum_{\alpha\in I_+}(u_{\alpha}|[a,[b,u^{\alpha}]])+(a|[b,2\rho_{>0}])= \kappa_{\g_-}(a|b)+(a|[b,2\rho_{>0}]),
        \end{aligned}
    \end{equation}
    where $\rho_{>0}:=\frac{1}{2}\sum_{\alpha\in I_+}[u_{\alpha},u^{\alpha}]$. By \cite[Section 6.1]{KW22}, we have $\rho_{>0}=(h^{\vee}-1)x$ for minimal $F$, which implies that $\kappa_{\g_+}=\kappa_{\g_-}$ as bilinear forms on $\g_0^F$. Thus, on $\g_0^F$, we have $2h^{\vee}(\cdot|\cdot)=\kappa_{\g}=\kappa_{\g_0}+2\kappa_{\g_+}$. Therefore, we obtain 
\begin{equation}
    \psi_k(a|b)=(k+h^{\vee})(a|b)-\frac{1}{2}\kappa_{\g_0}(a|b)-\frac{1}{2}\kappa_{\g_{\frac{1}{2}}}(a|b)=\kappa(a|b),
\end{equation}
 which proves the proposition.
\end{proof}

Recall the isomorphism
\begin{equation} \label{eq:weight 1 correspondence}
   \iota: W^k(\g,F) \to \textup{Com}(\mathcal{F}(\g^f_0),W^k_{N=1}(\g,f))\subset W^k_{N=1}(\g,f)
\end{equation}
introduced in \eqref{eq:iota-between W's}. In [34], the weight $1$ elements of $W^k(\g,F)$ are denoted by $J^{\{a\}}$ for $a\in \g^F_0.$ Explicitly, we have the correspondence 
\begin{equation}
    \iota \, : \, J^{\{a\}} \mapsto \sqrt{-1}^{p(a)}\mathcal{J}_{\{a\}}
\end{equation}
for $\mathcal{J}_{\{a\}}$ defined in \eqref{eq:weight 1 generator, general}.

\subsubsection{Weight $\frac{3}{2}$ fields}

As in the weight 1 case, we refer to a weight $\frac{3}{2}$ field in $W^k_{N=1}(\g,f)$ as a {\it Type 1 element} if it belongs to the odd differential algebra generated by weight $\frac{1}{2}$ and $1$ fields. Otherwise, we denote the weight $\frac{3}{2}$ field by a {\it Type 2} element. 

The weight $\frac{3}{2}$ subspace of  $\textup{Com}(\mathcal{F}(\g^f_0),W^k_{N=1}(\g,f))$ has dimension $\text{dim}(\g^F_{-1/2})$.
In the minimal nilpotent case, we have the decomposition $\g^F_{-1/2}=\g^f_{-1/2}\oplus \mathbb{C}f.$ The subspace of Type 1 weight $\frac{3}{2}$ elements in $W^k_{N=1}(\g,f)$ has dimension $\text{dim} (\g^f_{-1/2})=\text{dim}(\g^F_{-1/2})-1$ and hence the entire weight $\frac{3}{2}$ subspace is spanned by Type 1 fields together with a single Type 2 field. In particular, there exists a distinguished Type 2 field of weight $\frac{3}{2}$ induced from the superconformal vector of $W^k_{N=1}(\g,f)$.

\begin{proposition}
    Let $G$ be the superconformal vector of $W^k_{N=1}(\g,f)$ and 
    \[\tau_{\g^f_0}=\frac{1}{k+h^\vee}\sum_{i,j\in I^f_0}(u_i|u_j) :\bar{u}^i D\bar{u}^j:+\frac{1}{3(k+h^\vee)^2}\sum_{i,j,r\in I^f_0} (-1)^{p(j)}(u_i|[u_j,u_r]):\bar{u}^i\bar{u}^j\bar{u}^r:
    \]
    be the Kac-Todorov vector 
    of the SUSY affine vertex algebra of $\g^f_0.$ Then $G-\tau_{\g^f_0}$ is a Type 2 element in $\textup{Com}(\mathcal{F}(\g^f_0),W^k_{N=1}(\g,f))$.
\end{proposition}
\begin{proof}
    Obviously, both $G$ and $\tau_{\g^f_0}$ are in $W^k_{N=1}(\g,f)$ and hence $G-\tau_{\g^f_0}\in W^k_{N=1}(\g,f).$
    Moreover, by the property of a superconformal vector, we know that 
    \[\, [G{}_{\lambda} \bar{a}]= [\tau_{\g^f_0}\, {}_\lambda \bar{a}]=D\bar{a}\]
    for any $a\in \g^f_0.$ Hence $G-\tau_{\g^f_0}$ is in the centralizer of $\mathcal{F}(\g^f_0).$ Now, it is easy to see that $G$ has the term $\frac{2}{(k+h^\vee)^2} \bar{F}-\frac{1}{k+h^\vee}D\bar{f}$  which means $G-\tau_{\g^f_0}$ is a Type 2 element.
\end{proof}

We remark that $G-\tau_{\g^f_0}$ is the element in $W^k_{N=1}(\g,f)$ corresponding to a nonzero scalar multiple of 
    $G^{\{f\}}$ in \cite{KW04}, i.e. $\iota(G^{(\{f\}})= c(G-\tau_{\g^f_0})$ for some $c\in \mathbb{C}\setminus \{0\}$.
    This identification is justified by the following observation. The linear part of both $G$ and $ G-\tau_{\g^f_0}$  is given by $\frac{2}{(k+h^\vee)^2} \bar{F}-\frac{1}{k+h^\vee}D\bar{f}+c_H\partial\bar{H}$ for a nontrivial constant $c_H$.  Hence the inverse image $\iota^{-1}(G-\tau_{\g^f_0})$ under the inclusion  $\iota$ defined in \eqref{eq:iota-between W's}, must be a scalar multiple of $G^{\{f\}}$. This follows from the explicit form of $\iota$ and the fact that the $\partial \mathcal{F}^{\textup{ne}}$ part of $G^{\{a\}}$ in \cite{KW04} for $a\in \g^F_{-1/2}$ is a constant multiple of $\partial \Phi_{[\tilde{a}]}$, where $\tilde{a}$ denotes the dual of $a$
 with respect to the bilinear form $(F|[\, \cdot\, , \, \cdot \, ]).$ 
 
Moreover, the precise value of the constant 
$c$
 can be determined. In \cite{KW04}, it is shown that $G^{\{f\}}{}_{(0)}G^{\{f\}}$ has the term $2(k+h^\vee)\,L_{N=0}$  where $L_{N=0}$ is the conformal vector of $W^k(\g,F).$ On the other hand, we know that $G_{(0)}G=2L$ where $G$ and $L$ are superconformal and conformal vectors of $W^k_{N=1}(\g,f)$ and $L=L_{N=0}+(\text{conformal vector of } \mathcal{F}(\g^f_0)\, ).$ Hence, we deduce that $c$ should be $\pm\sqrt{k+h^\vee}.$ Finally, by comparing the sign of $c_H$ in $\iota(G^{\{f\}})$ and in $G,$ we conclude $c=-\sqrt{k+h^\vee}$. In other words, if we denote 
\begin{equation} \label{eq:J_f}
    \mathcal{G}_{\{f\}}:= G-\tau_{\g^f_0},
\end{equation}
then the inclusion $\iota$ gives the correspondence
\begin{equation} \label{eq:weight 3/2 correspondence}
\iota:\frac{-1}{\sqrt{k+h^\vee}}G^{\{f\}}\mapsto   \mathcal{G}_{\{f\}}.
\end{equation}

Next, in the following proposition,  Type 1  weight $\frac{3}{2}$ elements in $\textup{Com}(\mathcal{F}(\g^f_0),W^k_{N=1}(\g,f))$ are described with weight 1 and $\frac{1}{2}$ fields in $W^k_{N=1}(\g,f).$ 

\begin{proposition}\label{prop:Type 1 weight 3/2}   
The set of Type 1 weight $\frac{3}{2}$ elements in $\textup{Com}(\mathcal{F}(\g^f_0),W^k_{N=1}(\g,f))$ consists of 
\begin{equation} \label{eq:Type 1 weight 3/2-1}
    \mathcal{G}_{\{[f,a]\}}:=(-1)^{p(a)}\mathcal{G}_{\{f\}}{}_{(0)} \mathcal{J}_{\{a\}}=(-1)^{p(a)}\Big(D\mathcal{J}_{\{a\}}+ \frac{1}{k+h^\vee}\sum_{j\in I^f_0}:\bar{u}_j \mathcal{J}_{\{[a,u^j]\}} :\Big)
\end{equation}
for $a\in [e, \g^f_{-1/2}]$ and $\mathcal{J}_{\{[u^j,a]\}}$ given by \eqref{eq:weight 1 generator, general}. 
\end{proposition}

\begin{proof}
    Since both $G-\tau_{\g_0^f}$ and $\mathcal{J}_{\{a\}}$ are in the commutant, it is clear that $\mathcal{G}_{\{[f,a]\}}$ is also in the centralizer of $\mathcal{F}(\g^f_0).$ Hence, we only need to show the second equality in \eqref{eq:Type 1 weight 3/2-1}. Note that $G_{(0)}\mathcal{J}_{\{a\}}=D\mathcal{J}_{\{a\}}$ and 
    \begin{equation}
        \begin{aligned}
            \tau_{\g^f_0}{}_{(0)}\mathcal{J}_{\{a\}}& = 
            \frac{1}{k+h^\vee}\sum_{i,j\in I^f_0}(u_i|u_j)\bar{u}^i \big( D\bar{u}^j{}_{(0)}\mathcal{J}_{\{a\}}\big) = \frac{1}{k+h^\vee}\sum_{j\in I^f_0} (-1)^{p(a)p(j)}:\bar{u}_j \mathcal{J}_{\{[u^j,a]\}}:.
        \end{aligned}
    \end{equation}
    Here, we used $\mathcal{J}_{\{u^j\}}{}_{(0)}\mathcal{J}_{\{a\}}=D\bar{u}^j{}_{(0)}\mathcal{J}_{\{a\}}.$ Hence, we get the proposition.
\end{proof}

Let $e=[E,f]$ and $b\in \g^f_{-1/2}$. Then $[f,e]=-2x$ and $[f,[e,b]]=b$. Hence 
\eqref{eq:Type 1 weight 3/2-1} can be rewritten as follows:
\begin{equation}\label{eq:Type 1 weight 3/2}
    \mathcal{G}_{\{b\}}:=(-1)^{p(b)+1}\Big(D\mathcal{J}_{\{[e,b]\}}+ \frac{1}{k+h^\vee}\sum_{j\in I^f_0}:\bar{u}_j \mathcal{J}_{\{[[e,b],u^j]\}}:\Big).
\end{equation}
Recall the weight $1$ and $\frac{3}{2}$ elements $J^{\{a\}}$ and $G^{\{b\}}$ of $W^k(\g,F)$ in \cite{DK06}. Since  $G^{\{f\}}{}_{(0)}J^{\{a\}}= G^{\{[f,a]\}}$
in the ordinary minimal W-algebra, we can deduce 
\begin{equation} \label{eq:weight 3/2 fields, corresp}
\begin{aligned}
      & \iota^{-1}\big((-1)^{p(a)}  \mathcal{G}_{\{f\}}{}_{(0)} \mathcal{J}_{\{a\}}\big)  
      = -\frac{(-1)^{p(a)}}{\sqrt{-(k+h^\vee)}}\sqrt{-1}^{p(f)}G^{\{f\}}{}
_{(0)}\sqrt{-1}^{p(a)}J^{\{a\}}      \\& =-\frac{(-1)^{p(a)}}{\sqrt{-(k+h^\vee)}}\sqrt{-1}^{p(f)+p(a)} G^{\{[f,a]\}}
       =-\frac{\sqrt{-1}^{p([f,a])}}{\sqrt{-(k+h^\vee)}}   G^{\{[f,a]\}}
\end{aligned}
\end{equation}
from  \eqref{eq:weight 1 correspondence} and \eqref{eq:weight 3/2 correspondence}.
Hence we get the correspondence between weight $\frac{3}{2}$ fields
\begin{equation}
    \iota: - \frac{\sqrt{-1}^{p(b)}}{\sqrt{-(k+h^\vee)}}\, G^{\{b\}}\quad \mapsto \quad \mathcal{G}_{\{b\}}
\end{equation}
for $b\in \g^F_{-1/2}$.

\begin{proposition} \label{prop:uniq min virasoro}
Let $W^k(\mathfrak{g}, F)$ be a minimal W-algebra for a simple Lie superalgebra $\mathfrak{g}$. For  $k \neq -h^{\vee}$, there is a unique conformal vector $L$ on $W^k(\mathfrak{g}, F)$ for which the generators of the affine subalgebra have conformal weight $1$.

\end{proposition}

\begin{proof} Recall from \cite{KW04}, that $\mathfrak{g}_0^F$ is either:
\begin{enumerate}[(i)]
    \item A simple Lie superalgebra,
    \item A sum of two simple ideals $\mathfrak{a}_1 \oplus \mathfrak{a}_2$,
    \item $\mathfrak{a} \oplus \mathbb{C}$, for $\mathfrak{a}$ a simple Lie algebra,
    \item $\mathbb{C}$.
\end{enumerate}
Accordingly, will write the affine subalgebra in the form 
\begin{enumerate}[(i)]
    \item $V^{k'}(\mathfrak{g}^F_0)$ when $\mathfrak{g}^F_0$ is simple,
       \item $V^{k_1}(\mathfrak{a}_1) \otimes V^{k_2}(\mathfrak{a}_2)$ when $\mathfrak{g}^F_0 = \mathfrak{a}_1 \oplus \mathfrak{a}_2$ is a sum of two simple ideals,
       \item $V^{k_1}(\mathfrak{a}_1) \otimes \mathcal{H}^{k_2}$ when $\mathfrak{g}^F_0 = \mathfrak{a} \oplus \mathbb{C}$ and $\mathfrak{a}$ is simple. Here $\mathcal{H}^{k_2}$ denotes a Heisenberg algebra of level $k_2$,
       \item $\mathcal{H}^{k'}$ when $\mathfrak{g}^F_0 = \mathbb{C}$.
\end{enumerate}
Let $\tilde{L}$ be another conformal vector on $W^k(\mathfrak{g}, F)$ such that the generators of $V^{k'}(\mathfrak{g}_0^F)$ have conformal weight $1$. Note that this condition automatically implies that $\tilde{L}$ is $\mathfrak{g}_0^F$-invariant, that is, $a_{(0)} \tilde{L} = 0$ for all $a \in \mathfrak{g}_0^F$. 

We begin with Case (i) where $\mathfrak{g}^F_0$ is simple.  Consider the affine coset 
$$C^k = \text{Com}(V^{k'}(\mathfrak{g}_0^F), W^k(\mathfrak{g}, F)),$$ which can be given conformal vector $\tilde{L}^C = \tilde{L} - L^{\mathfrak{g}_0^F}$ or $L^C = L -  L^{\mathfrak{g}_0^F}$, where $L^{\mathfrak{g}_0^F}$ is the Sugawara conformal vector on $V^{k'}(\mathfrak{g}_0^F)$. 

First, assume $k'$ is noncritical for $\mathfrak{g}^F_0$, so that the center of $V^{k'}(\mathfrak{g}_0^F)$ is trivial, and $C$ has no fields in weight $1$. Any field in $C$ of weight $2$ is necessarily $\mathfrak{g}_0^F$-invariant, hence it is a linear combination of $L$ and the Sugawara conformal vector $L^{\mathfrak{g}^F_0}$. Since under both $L$ and $L^{\mathfrak{g}^F_0}$, the generators of $V^{k'}(\mathfrak{g}^F_0)$ have conformal weight $1$, it follows that the only element of weight $2$ in $C$ is (up to scaling) $L - L^{\mathfrak{g}^F_0}$. Therefore, $\tilde{L}^C$ is a scalar multiple of $L^C$, and since both are Virasoro fields we must have $L^C = \tilde{L}^C$, so that $L = \tilde{L}$ as well.

Next, assume $k'$ is critical for $\mathfrak{g}^F_0$. Then $L^{\mathfrak{g}^F_0}$ does not exist, but the rescaled field $A = (k' + h^{\vee}_{\mathfrak{g}^F_0})L^{\mathfrak{g}^F_0}$ does exist and is the unique (up to scalar) element of the Feigin-Frenkel center of $V^{k'}(\mathfrak{g}_0^F)$ in weight $2$. Since $\tilde{L}$ is $\mathfrak{g}_0^F$-invariant, it is a linear combination of $L$ and $A$. Since $A$ is central in $V^{k'}(\mathfrak{g}_0^F)$ and the generators of $V^{k'}(\mathfrak{g}_0^F)$ are primary of weight $1$ with respect to $\tilde{L}$, this forces the coefficient of $L$ in $\tilde{L}$ to be $1$. The condition that $\tilde{L}$ satisfies the Virasoro OPE relation forces the coefficient of $A$ in $\tilde{L}$ to be zero, so again we have $\tilde{L} = L$.

Consider Case (ii) where the affine subalgebra is $V^{k_1}(\mathfrak{a}_1) \otimes V^{k_2}(\mathfrak{a}_2)$ for simple ideals  $\mathfrak{a}_1$ and $\mathfrak{a}_2$. The argument is exactly the same: if $k_1$ and $k_2$ are noncritical for $\mathfrak{a}_1$ and $\mathfrak{a}_2$, respectively, the affine coset $C = \text{Com}(V^{k_1}(\mathfrak{a}_1) \otimes V^{k_2}(\mathfrak{a}_2), W^k(\mathfrak{g}, F))$ has $1$-dimensional weight $2$ space, so $\tilde{L} = L$. Otherwise, one of the two levels, say $k_2$ is critical for $\mathfrak{a}_2$. We now consider the partial affine coset
$$C^k = \text{Com}(V^{k_1}(\mathfrak{a}_1), W^k(\mathfrak{g}, F)),$$ which has conformal vector $L^C = L - L^{\mathfrak{a}_1}$ or $\tilde{L}^C = \tilde{L} - L^{\mathfrak{a}_1}$. As above, $\tilde{L}^C$ is $\mathfrak{a}_2$-invariant, and hence it must be a linear combination of $L^C$ and the element $A = (k' + h^{\vee}_{\mathfrak{a}_2})L^{\mathfrak{a}_2}$ in the Feigin-Frenkel center of $V^{k_2}(\mathfrak{a}_2)$. By the same argument as above, the coefficient of $L^C$ is $1$ and the coefficient of $A$ is zero, so $\tilde{L} = L$. The same argument applies if $k_1$ is critical for $\mathfrak{a}_1$, and both levels cannot be critical simultaneously, so Case (ii) is complete.

Next, we consider Case (iii) where the affine subalgebra is $V^{k_1}(\mathfrak{a}_1) \otimes \mathcal{H}^{k_2}$, and $\mathfrak{a}$ is simple. If $k_1$ is noncritical for $\mathfrak{a}$ and $k_2 \neq 0$, $V^{k_1}(\mathfrak{a}_1) \otimes \mathcal{H}^{k_2}$ has trivial center. As above, consider the coset
$$C^k = \text{Com}(V^{k_1}(\mathfrak{a}) \otimes \mathcal{H}^{k_2}, W^k(\mathfrak{g}, F)),$$ which has conformal vector either $L^C = L - L^{\mathfrak{a}} - L^{\mathcal{H}}$ or $\tilde{L}^C = \tilde{L} - L^{\mathfrak{a}} - L^{\mathcal{H}}$. Here $L^{\mathcal{H}}$ is the conformal vector on $\mathcal{H}$ such that the generator $b$ of $\mathcal{H}$ is primary of weight $1$. As above, any field in $C$ of weight $2$ is $\mathfrak{g}^F_0$-invariant, and hence is a linear combination of $L$, $L^{\mathfrak{a}}$, $L^{\mathcal{H}}$, and $\partial b$. Since $(\partial b)_{(2)} b \neq 0$, whereas $L_{(2)} b= 0$, $L^{\mathfrak{a}}_{(2)} b = 0$, and $L^{\mathcal{H}}_{(2)} b = 0$, it follows that the coefficient of $\partial b$ in $\tilde{L}^C$ is zero. By the same argument as above, we get $\tilde{L}^C = L^C$, and $\tilde{L} = L$.
If either $k_1$ is critical for $\mathfrak{a}$ or $k_2=0$ (both cannot happen simultaneously), so that the center of the affine subalgebra is nontrivial, the argument is similar to Case (ii) when either $k_1$ or $k_2$ is critical, and is omitted. 

Finally, Case (iv) when $\mathfrak{g}^F_0 = \mathbb{C}$ is similar and is omitted as well. \end{proof}
\vskip 2mm

Finally, we get the following theorem directly from Proposition \ref{prop:affine},  \ref{prop:Type 1 weight 3/2} and \ref{prop:uniq min virasoro}.  

\begin{theorem}\label{thm:minimal-generator}
    The minimal SUSY W-algebra $W^k_{N=1}(\g,f)$ for $k\neq -h^\vee$ is strongly generated as a vertex algebra by the weight $1/2$ fields in $\g^f_0$ and a set of centralizer 
    \begin{equation} \label{eq:minimal-generator-VA}
        \{\mathcal{J}_{\{a\}},\mathcal{G}_{\{b\}}, L_{N=0} |\, a\in \g^F_{0},\, b\in \g^F_{-1/2} \},
    \end{equation}
    of $\mathcal{F}(\g^f_0)$, 
    where $\mathcal{J}_{\{ a \}}$ and $\mathcal{G}_{\{b\}}$ are given by  \eqref{eq:weight 1 generator, general}, \eqref{eq:J_f} and \eqref{eq:Type 1 weight 3/2}, respectively, and $L_{N=0}$ denotes the conformal vector of the centralizer. Specifically,  $L_{N=0}=\frac{1}{2}DG-L_{\mathcal{\g}^f_0}$, where $L_{\mathcal{\g}^f_0}$ is the conformal vector of the free field vertex algebra $\mathcal{F}(\g^f_0)$. On the other hand, as a SUSY vertex algebra, $W^k_{N=1}(\g,f)$ is strongly generated by 
    \begin{equation}\label{eq:minimal-generator-SUSY}
        \{\, \bar{h}, \mathcal{J}_{\{a\}}, G\, |\, h\in \g^f_0, \, a\in [e,\g^f_{-1/2}]\}.
    \end{equation} 
\end{theorem}
\begin{proof}
    The first statement is clear since \eqref{eq:minimal-generator-VA} corresponds to a strong generating set of $W^k(\g,F).$ The second statement follows from the fact that Type 1 generators of weight 1 and weight $\frac{3}{2}$ can be removed when we consider a strong generating set as a SUSY vertex algebra. By the uniqueness of conformal vector in Proposition \ref{prop:uniq min virasoro}, we also know the conformal vector of $W^k_{N=1}(\g,f)$ is $L=\frac{1}{2}DG=L_{N=0}+L_{\g^f_0}$. 
\end{proof}

\begin{remark}
    When $k$ is critical, we do not know the structure of $\textup{Com}(\mathcal{F}(\g^f_0),W^k_{N=1}(\g,f)).$ However, we can still find a strong generating set of $W^k_{N=1}(\g,f)$ by the  computations similar to noncritical cases. More precisely, $W^{-h^\vee}_{N=1}(\g,f)$ is a SUSY vertex algebra generated by $\{\, \bar{h},  \mathcal{G}_{\bar{b}}, (k+h^\vee)^2 G \, |\, h\in \g^f_0, b\in \g^f_{-1/2}\, \}$ where $\mathcal{G}_{\bar{b}}$ is the element in Proposition \ref{prop:weight 1 elt}.
\end{remark}

\subsection{Structure of minimal SUSY W-algebra; via superfield formalism }

In this section, we present $\Lambda=(\lambda, \chi)$ brackets between the generators in \eqref{eq:minimal-generator-SUSY} of $W^k_{N=1}(\g,f)$ when $k\neq -h^\vee$. Recall that the $\Lambda$ bracket is determined by the $\lambda$ and the odd derivation $D$ in the following way:
\begin{equation} \label{eq: Lambda bracket}
    [\, a \, {}_\Lambda \, b \, ]= [\, Da\, {}_\lambda \, b \, ]+ \chi [\, a\, {}_\lambda \, b\, ].
\end{equation}
Let $G$ be the superconformal vector of $W^k_{N=1}(\g,f).$ Then the conformal vector  $L:=\frac{1}{2}DG= L_{N=0}+L_{\g^f_0}$. The equality follows from the uniqueness of a conformal vector of the minimal W-algebra which assigns the conformal weight $1-j_a$ to a building block $J_{a}$ for $a\in \g_{j_a}$.

\begin{proposition} \label{prop:minimal Lambda braket 1}
    Let $h\in \g^f_0$ and $a\in [e,\g^f_{-1/2}]$. Then we have 
    \begin{equation} \label{eq:Lambda bracket with G}
    \begin{aligned}
             & [ \, G\, {}_{\Lambda}\, {}\bar{h}\, ]= (2\partial+ \lambda+\chi D) \bar{h}, \\
             & [ \, G\, {}_{\Lambda}\, {}\mathcal{J}_{\{a\}}\, ]= (2\partial + 2\lambda +\chi D) \mathcal{J}_{\{a \}} + \lambda^2 \textup{str}_{\n}(\textup{ad}\, a )\\
             & [ \, G\, {}_{\Lambda} \, G\, ]= (2\partial+3\lambda +\chi D) G+ \frac{\lambda^2 \chi}{3}c
    \end{aligned}
    \end{equation}
    for $c$ in \eqref{eq: superconformal susy W-alg central charge}.
\end{proposition}

\begin{proof}
    The first and third equalities of \eqref{eq:Lambda bracket with G} follows from \cite{Song24free}. To get the second equality of \eqref{eq:Lambda bracket with G}, recall that  $\mathcal{J}_{\{a\}}$ corresponds to an element $J^{\{a\}}$ in $W^k(\g,F)$ in \cite{KRW03}. In  Theorem 2.4 (b) of \cite{KRW03}, they showed 
    \begin{equation}
        [ \, L \, {}_\lambda \, \mathcal{J}_{\{a\}} \, ]= (\partial+\lambda) \mathcal{J}_{\{a\}} + \frac{1}{2} \lambda^2 \text{str}_{\n}(\text{ad}\, a).
    \end{equation}
    Now, it is enough to show $G_{(1)}\mathcal{J}_{\{a\}}=0$ which implies
    $[\, G\, {}_\lambda \, \mathcal{J}_{\{a\}}\, ]=D\mathcal{J}_{\{a\}}.$ Since $G=\mathcal{G}_{\{f\}}+\tau_{\g^f_0}$ and $\mathcal{G}_{\{f\}}{}_{(1)}\mathcal{J}_{\{a\}}$ should be a weight $1/2$ field in $\textup{Com}(\mathcal{F}(\g^f_0),W^k_{N=1}(\g,f))$, we have $\mathcal{G}_{\{f\}}{}_{(1)}\mathcal{J}_{\{a\}}=0$ and hence $G_{(1)}\mathcal{J}_{\{a\}}=\tau_{\g^f_0}{}_{(1)}\mathcal{J}_{\{a\}}.$ In addition, again, since $[\bar{u}_i{}_\lambda \, \mathcal{J}_{\{a\}}]=0,$ we have  
    \begin{equation}
        (\bar{u}_i D\bar{u}_j)_{(1)}\mathcal{J}_{\{a\}}=0.
    \end{equation}
    Finally, we conclude $G_{(1)}\mathcal{J}_{\{a\}}=0$ and the second equality of \eqref{eq:Lambda bracket with G} follows.
\end{proof}
    
The $\Lambda$ brackets between two superfields of weight $(\frac{1}{2},1)$ and $(\frac{1}{2},1)$ or $(1,\frac{3}{2})$ are as follows.

\begin{proposition} \label{prop:minimal Lambda braket 2}
    Let $h,h_1, h_2\in \g^f_0$ and $a\in [e,\g^f_{-1/2}]$. Then we have
    \begin{equation} \label{eq:Lambda bracket, rest}
    \begin{aligned}
        & [\, \bar{h}_1\, {}_\Lambda\, \bar{h}_2\, ]= (-1)^{p(h_1)p(h_2)+p(h_1)}\overline{[h_1,h_2]}+\chi (k+h^\vee)(h_1|h_2), \\
        & [\, \bar{h}\, {}_\Lambda \, \mathcal{J}_{\{a\}}\, ]= (-1)^{p(h)p(a)}\mathcal{J}_{[h,a]}.\\
    \end{aligned}
    \end{equation}
\end{proposition}

\begin{proof}
    The first equality of \eqref{eq:Lambda bracket, rest} is obvious since $\bar{h}_1$ and $\bar{h}_2$ are building blocks of SUSY BRST. The second equality of \eqref{eq:Lambda bracket, rest} follows from the fact that $[\, \bar{h}\, {}_\lambda\,\mathcal{J}_{\{a\}}]=0$ and $[\, D\bar{h}\, {}_\lambda\,\mathcal{J}_{\{a\}}]=(-1)^{p(h)p(a)}\mathcal{J}_{[h,a]}.$
\end{proof}

The last case is the $\Lambda$ bracket between two weight $(1,\frac{3}{2})$ superfields. For $a_1,a_2\in [e,\g^f_{-1/2}],$ we already know that 
\begin{equation} \label{eq:lambda (JJ)}
[\, \mathcal{J}_{\{a_1\}}\, {}_\lambda \, \mathcal{J}_{\{a_2\}}\, ]= (-1)^{p(a_1)p(a_2)}\mathcal{J}_{\{[a_1,a_2]\}}+(-1)^{p(a_1)}\lambda \kappa (a_1|a_2).
\end{equation}
In order to find $[\, D\mathcal{J}_{\{a_1\}}\, {}_\lambda \, \mathcal{J}_{\{a_2\}}\, ]$, recall the result in \cite{KW04} that $[G^{\{[f,a_1]\}}{}_\lambda J^{\{a_2\}}]=G^{\{[[f,a_1],a_2]\}}$ in the ordinary W-algebra. Now, by \eqref{eq:weight 3/2 fields, corresp}, the following $\lambda$-bracket holds in the SUSY W-algebra : 
\begin{equation}
    [\, \mathcal{G}_{\{[f,a_1]\}}\, {}_\lambda \, \mathcal{J}_{\{a_2\}}\, ]= (-1)^{p([f,a_1])p(a_2)}\mathcal{G}_{\{[[f,a_1],a_2]\}},
\end{equation}
where 
\begin{equation} \label{eq:weight 3/2 general}
    \mathcal{G}_{\{b\}}=\frac{1}{2}(b|e)\mathcal{G}_{\{f\}}+\mathcal{G}_{\{b-\frac{1}{2}(b|e)f\}}
\end{equation} for $b\in \g^F_{-1/2}$. Here, the two weight $\frac{3}{2}$ fields $\mathcal{G}_{\{f\}}$ and $\mathcal{G}_{\{b-\frac{1}{2}(b|e)f\}}$ are  given by \eqref{eq:J_f} and \eqref{eq:Type 1 weight 3/2} since $b-\frac{1}{2}(e|b)f\in \g^f_{-1/2}$. In conclusion, we get the following proposition.

\begin{proposition} \label{prop:minimal Lambda braket 3}
Let $a_1,a_2\in [e,\g^f_{-1/2}].$ Then 
\begin{equation}\label{eq:lambda (GJ)}
    \begin{aligned}
            [\, \mathcal{J}_{\{a_1\}}\, {}_\Lambda\, \mathcal{J}_{\{a_2\}}\, ] & =\Big(\mathcal{G}_{\{[a_2,[a_1,f]]\}}+\frac{1}{k+h^\vee} \sum_{j\in I^f_0} :\bar{u}_j\mathcal{J}_{\{[a_2,[a_1,u^j]]\}}:\Big) \\
            & \hskip 1cm +\chi \Big((-1)^{p(a_1)p(a_2)} \mathcal{J}_{\{[a_1,a_2]\}}+(-1)^{p(a_1)}\lambda \kappa(a_1|a_2) \Big),
    \end{aligned}
\end{equation}
for $\mathcal{G}_{\{[a_2,[a_1,f]]\}}$ in \eqref{eq:weight 3/2 general}.
\end{proposition}

\begin{proof}
    By Proposition \ref{prop:Type 1 weight 3/2}, $D\mathcal{J}_{\{a_1\}}$ can be written as 
    \begin{equation}
        D\mathcal{J}_{\{a_1\}}=(-1)^{p(a_1)}\mathcal{G}_{\{[f,a_1]\}}-\frac{1}{k+h^\vee}\sum_{j\in I^f_0} :\bar{u}_j \mathcal{J}_{\{[a_1,u^j]\}}:.
    \end{equation}
    Now using \eqref{eq:lambda (JJ)} and \eqref{eq:lambda (GJ)}, we get the proposition.
\end{proof}

\section{ $N=2,3$ and $4$ superconformal algebras} \label{sec:superconformal}

Recall that a SUSY vertex algebra is a vertex algebra equipped with a supersymmetry $D$. Generalizing the notion, we define an \textit{$N=n$ SUSY vertex algebra} as a vertex algebra equipped with $n$ supercommuting supersymmetries $D_1, \cdots, D_n$. Namely, they satisfy $[D_i,D_j]=2\delta_{i,j}\partial$.

In this section, we recall the $N=n$ superconformal algebras from \cite{KW04} and explain their relation to $N=n$ superconformal vectors introduced in \cite{HK07} for $n=2,3$, or $4$. These superconformal algebras were originally introduced as tensor products of W-algebras with free fermions or free bosons. However, in view of Theorem \ref{thm:nonSUSY vs SUSY}, we can construct them as SUSY W-algebras, which more naturally explains why the W-algebras should be tensored with extra fields.

The superconformal vectors were initially defined in terms of the $\mathbf{\Lambda}$-brackets for $\mathbf{\Lambda}=(\lambda, \chi_1, \cdots, \chi_n)$. For the reader's convenience, we present an equivalent formulation using the usual $\lambda$-bracket or the $\Lambda$-bracket in \eqref{eq: Lambda bracket}. These equivalent definitions are obtained by using \cite[Theorem 2.3.1]{Songthesis}. In the following sections, we describe the generating types of the superconformal algebras. Unless otherwise specified, we assume that the generators with integer conformal weights are even, while those with half-integer weights are odd.

% Recall an $N=1$ superconformal vector $G$ introduced in \eqref{}
% {\color{red} $N=1$ superconformal associated with $D$}

\subsection{$N=2$ superconformality} \label{sec: N=2 superconformal}
The $N=2$ superconformal algebra is the SUSY W-algebra $W^k_{N=1}(\mathfrak{sl}(2|1),f)$ for non-critical $k$, where $f$ is the odd minimal nilpotent of $\mathfrak{sl}(2|1)$. As a vertex algebra, $W^k_{N=1}(\mathfrak{sl}(2|1),f)$ has a generating type $W(1,\big(\frac{3}{2}\big)^2,2)$, and is freely generated by
\begin{equation} \label{eq: N=2 generators}
    J,\quad G,\quad \widetilde{G}=DJ,\quad L:=\frac{1}{2}DG,
\end{equation}
where $G$ is an $N=1$ superconformal vector of central charge $c=-3(1+2k)$, and $L$ is a conformal vector. With respect to $L$, the even field $J$ and the odd fields $G$ and $\widetilde{G}$ are primary of weight $1$ and $\frac{3}{2}$, respectively. The remaining non-zero $\lambda$-brackets between the generators are
\begin{equation} \label{eq: N=2 conditions}
    \begin{gathered}
        [\widetilde{G}{}_{\lambda}\widetilde{G}]=2L+\frac{c}{3}\lambda^2, \quad [G{}_{\lambda}\widetilde{G}]=(\partial+2\lambda) J,\\
        [G{}_{\lambda}J]=\widetilde{G},\quad [\widetilde{G}{}_{\lambda}J]=-G, \quad [J{}_{\lambda}J]=-\frac{c}{3}\lambda.
    \end{gathered}
\end{equation}
From the above relations, one can see that $D_1=D=G_{(0)}$ and $D_2=\widetilde{G}_{(0)}$ satisfy
\begin{equation} \label{eq: check N=2 susy}
    \begin{aligned}
    [D_2,D_2]=&[(DJ)_{(0)},(DJ)_{(0)}]=\big((DJ)_{(0)}(DJ)\big)_{(0)}=(2L)_{(0)}=2\partial,\\
    [D_1,D_2]=&[G_{(0)},(DJ)_{(0)}]=\big(G_{(0)}(DJ)\big)_{(0)}=(\partial J)_{(0)}=0,
    \end{aligned}
\end{equation}
that is, the $N=2$ superconformal algebra is an $N=2$ SUSY vertex algebra with two supersymmetries $D_1$ and $D_2$. To emphasize the relation with the two supersymmetries, rename the generators as $J_{12}:=J$, $G_1:=G$, and $G_2:=\widetilde{G}$. Then, the action of the two supersymmetries can be drawn as follows:
\begin{equation} \label{tikz: N=2}
\begin{tikzpicture}[baseline=(current  bounding  box.center)]
    \node (L) at (0,2) {$2L$};
    \node (G) at (-2,0) {$G_1$};
    \node (Gtild) at (2,0) {$G_2$};
    \node (J) at (0,-2) {$J_{12}$};
    \node (diff) at (-1.5,-1.3) {\small{$-D_2$}};
    \node (diff2) at (1.5,1.3) {\small{$D_2$}};
    \node (diff3) at (-1.5,1.3) {\small{$D_1$}};
    \node (diff4) at (1.5,-1.3) {\small{$D_1$}};
    \draw[->] (J)--(G);
    \draw[->] (G)--(L);
    \draw[->] (J)--(Gtild);
    \draw[->] (Gtild)--(L);
\end{tikzpicture}
\end{equation}
Note that reversing the order of the supersymmetries changes the sign of $J$. In our notation, one can write $J_{21}=-J_{12}$. Moreover, the relation \eqref{eq: check N=2 susy} only depends on the $\lambda$-brackets \eqref{eq: N=2 conditions} and $L_{(0)}=\partial$. Thus, any vertex algebra $V$ containing the $N=2$ superconformal algebra with a total Virasoro $L$ is an $N=2$ SUSY vertex algebra with $D_1=G_{(0)}$ and $D_2=\widetilde{G}_{(0)}$.

When provided that the two supersymmetries $D_1$ and $D_2$ are already given on a vertex algebra, one can define the $N=2$ superconformal vector using the $N=2$ superconformal algebra. Let $V$ be an $N=2$ SUSY vertex algebra with supersymmetries $D_1$ and $D_2$. Then, we call an even $J_{12}\in V$ an \textit{$N=2$ superconformal vector (associated with $D_1$ and $D_2$) of central charge $c$} if
\begin{equation*}
    J_{12},\quad G_1:=-D_2 J_{12}, \quad G_2:=D_1 J_{12}, \quad L=-\frac{1}{2}D_1 D_2 J_{12}
\end{equation*}
form an $N=2$ superconformal algebra, $L$ is a total Virasoro of central charge $c$ inside $V$, and satisfy $D_1=(G_1)_{(0)}$ and $D_2=(G_2)_{(0)}$.

\subsection{$N=3$ superconformality} \label{sec: N=3 superconformal}
In this section, consider $\mathfrak{spo}(2|3)$ and its odd minimal nilpotent $f$. The $N=3$ superconformal algebra is the SUSY W-algebra $W^k_{N=1}(\mathfrak{spo}(2|3),f)$ for non-critical $k$. Recall from Section \ref{appendix: N=3 structure} that the $N=3$ superconformal algebra has a generating type $W(\frac{1}{2}, 1^3, \big(\frac{3}{2}\big)^3,2)$, and is freely generated by
\begin{equation} \label{eq: N=3 generators}
    K,\quad J_{12}, \quad J_{13}, \quad J_{23}, \quad G_1, \quad G_2, \quad G_3, \quad L,
\end{equation}
where the four fields $(J_{ij},G_i,G_j,L)$ generate the $N=2$ superconformal algebra for each choice of a pair of distinct indices $(i,j)$. It implies that this algebra is an $N=3$ SUSY vertex algebra with $D_i:=(G_i)_{(0)}$ for $i=1,2,$ and $3$. Moreover, $K$ is related to other fields via
\begin{equation} \label{eq: N=3 vector relation}
    \begin{gathered}
    [K{}_{\lambda}K]=-\frac{c}{3},\quad [J_{ij}{}_{\lambda}K]=0,\quad [L{}_{\lambda}K]=\big(\partial+\frac{1}{2}\lambda\big)K\\
    \quad [G_1{}_{\lambda}K]=J_{23}, \quad [G_2 {}_{\lambda}K]=-J_{13}, \quad [G_3 {}_{\lambda}K]=J_{12},
    \end{gathered}
\end{equation}
where $c$ is the central charge of the Virasoro field $L$. Since reversing the supersymmetries changes the sign of an $N=2$ superconformal vector, we let $J_{ij}:=-J_{ji}$ when $i>j$. Then one can draw the action of the supersymmetries on the fields \eqref{eq: N=3 generators} as follows:
\begin{equation}\label{tikz: N=3}
    \begin{tikzpicture}[baseline=(current  bounding  box.center)]
        \node(L) at (0,3) {$2L$};
        \node(G1) at (-3,1.5) {$G_1$};
        \node(G2) at (-0,1.5) {$G_2$};
        \node(G3) at (3,1.5) {$G_3$};
        \node(J1) at (-3,0) {$J_{12}$};
        \node(J2) at (0,0) {$J_{31}$};
        \node(J3) at (3,0) {$J_{23}$};
        \node(K) at (0,-1.5) {$K$};
        \draw[->] (J1)--(G1);
        \draw[->] (J1)--(G2);
        \draw[->] (J2)--(G1);
        \draw[->] (J2)--(G3);
        \draw[->] (J3)--(G2);
        \draw[->] (J3)--(G3);
        \draw[->] (G1)--(L) node[midway,left=8pt] {$D_1$};
        \draw[->] (G2)--(L) node[midway,left=2pt] {$D_2$};
        \draw[->] (G3)--(L) node[midway,right=2pt] {$D_3$};
        \draw[->] (K)--(J1) node[midway,left=8pt] {$D_3$};
        \draw[->] (K)--(J2) node[midway,left=2pt] {$D_2$};
        \draw[->] (K)--(J3) node[midway,right=8pt] {$D_1$};
        \node(J) at (6,-1) {$J_{ij}$};
        \node(G) at (6,0.75) {$G_j$};
        \node(Jprim) at (6,2.5) {$\partial J_{rj}$};
        \draw[->] (J)--(G) node[midway,left] {$D_i$};
        \draw[->] (G)--(Jprim) node[midway,left] {$D_r$};
    \end{tikzpicture}
\end{equation}

In the above left, the arrows in the middle are appropriately determined as in \eqref{tikz: N=2}, since each parallelogram formed by $(J_{ij}, G_i, G_j, 2L)$ should be the $N=2$ superconformal algebra. In the above right, $i,j$ and $r$ are distinct indices with $i<j$.

As in the $N=2$ superconformality, any vertex algebra containing the $N=3$ superconformal algebra with a total Virasoro $L$ is an $N=3$ SUSY vertex algebra. Conversely, when a vertex algebra $V$ is initially equipped with supersymmetries $D_1, D_2$, and $D_3$, one can define an $N=3$ superconformal vector following \cite{HK07}. We call an odd field $K\in V$ an \textit{$N=3$ superconformal vector (associated with $D_1$, $D_2$, and $D_3$) of central charge $c$} if
\begin{enumerate}[(i)]
    \item $L:=-\frac{1}{2}D_1 D_2 D_3 K$ is a total Virasoro of central charge $c$ and $K$ is primary of conformal weight $\frac{1}{2}$,
    \item $K$ satisfies \eqref{eq: N=3 vector relation} for
    \begin{equation} \label{eq: N=3 vector D_i relation}
        \begin{gathered}
        J_{12}:=D_3 K, \quad J_{13}=-D_2 K, \quad J_{23}:=D_1 K,\\
        G_1:=-D_2 D_3 K, \quad G_2:= D_1D_3 K, \quad G_3:=-D_1 D_2 K,
        \end{gathered}
    \end{equation}
    \item $D_i=(G_i)_{(0)}$ for $G_i$'s in \eqref{eq: N=3 vector D_i relation}.
\end{enumerate}
Note that in this case, the fields $(J_{ij},G_i,G_j,L)$ again form the $N=2$ superconformal algebra, determined by the properties of $D_i$'s.

\subsection{$N=4$ superconformality} \label{subsec:N=2superconformal}
From Section \ref{sec: N=2 superconformal} and \ref{sec: N=3 superconformal}, we can detect the features of $N=n$ superconformal vector $U$ for $n=1,2$, and $3$ as follows:
\begin{enumerate}[(a)]
    \item $U$ has parity $n \Mod{2}$,
    \item $U$ is primary of conformal weight $\frac{4-n}{2}$,
    \item $U$ along with $D_1, \cdots, D_n$ generates the $N=n$ superconformal algebra, whose generating type is 
    \begin{equation}\label{eq: N=4 with wt0}
    W\Big(\tfrac{4-n}{2}, \big(\tfrac{5-n}{2}\big)^{n \choose 1}, \big(\tfrac{6-n}{2}\big)^{n \choose 2}, \cdots, \big(\tfrac{3}{2}\big)^{n \choose n-1},2\Big),
    \end{equation}
    \item for $n\geq 2$, $D_i U$ is an $N=n-1$ superconformal vector up to sign,
    \item $\frac{1}{2}D_1 \cdots D_n U$ is a conformal vector up to sign.
\end{enumerate}

One might wish the generalize the concepts to $n=4$ to obtain the notion of $N=4$ superconformal vector, which leads to the definition of it in \cite{HK07}.

\begin{definition}[\cite{HK07}] \label{def: HK N=4 superconformal}
    Let $V$ be an $N=4$ SUSY vertex algebra with supersymmetries $D_i$'s for $i=1,2,3,$ and $4$. An even field $P\in V$ is called an $N=4$ superconformal vector if
    \begin{equation} \label{eq: HK N=4 lambda-brackets}
        \begin{gathered}
        [D_1 D_2 D_3 D_4 P {}_{\lambda} P]=2\partial P, \quad [D_1 D_2 D_3 P {}_{\lambda}P]=-D_4 P, \quad [D_1 D_2 D_4 P {}_{\lambda}P]=D_3 P, \\
        [D_1 D_3 D_4 P {}_{\lambda} P]=-D_2 P, \quad [D_2 D_3 D_4 P {}_{\lambda}P]=D_1 P,\quad  [D_i D_j P {}_{\lambda}P]=[D_i P{}_{\lambda}P]=[P{}_{\lambda}P]=0
        \end{gathered}
    \end{equation}
    for any $i\neq j$, $L:=\frac{1}{2}D_1 D_2 D_3 D_4 P$ is a total Virasoro of $V$, and
    \begin{equation*}
        (D_1 D_2 D_3 P)_{(0)}=-D_4, \ (D_1 D_2 D_4 P)_{(0)}=D_3, \ (D_1 D_3 D_4 P)_{(0)}=-D_2, \ (D_2 D_3 D_4 P)_{(0)}=D_1.
    \end{equation*}
\end{definition}
Note that the even field $P$ in Definition \ref{def: HK N=4 superconformal} satisfies the conditions (a)-(e) for $n=4$. However, using this concept for $N=4$ superconformality may cause a difficulty. Specifically, $\lambda$-bracket \eqref{eq: HK N=4 lambda-brackets} allows the central extension only by adding $\lambda c$ for $c\in \CC$ to the first relation in \eqref{eq: HK N=4 lambda-brackets}. Due to this extension, $P$ is no longer an eigenvector of $L_{(1)}$, which implies that $L$ is not a conformal vector anymore. Moreover, the central extension does not affect the $\lambda$-bracket between $L$ and itself, leading to $[L{}_{\lambda}L]=(\partial+2\lambda)L,$  which does not cover Virasoro elements with nonzero central charge.

Thus, for the notion of $N=4$ superconformality, we consider appropriate superconformal algebras in place of a superconformal vector. As candidates for such replacements, we introduce the small and big $N=4$ superconformal algebras in the following sections. 
These algebras do not satisfy all the conditions proposed earlier, but they satisfy weaker forms of conditions (c) and (d). In addition, they admit a Virasoro field with arbitrary central charge.

\subsubsection{Small $N=4$ superconformal algebra} \label{sec: small N=4}
The small $N=4$ algebra is the W-algebra $W^k_{N=1}(\mathfrak{psl}(2|2),f)$, where $f$ is an odd minimal nilpotent in $\mathfrak{psl}(2|2)$. Recall from \eqref{eq: modified generators of small N=4} that this algebra has a generating type $W\big(1^3, \big(\frac{3}{2}\big)^4,2\big)$ with free generators
\begin{equation} \label{eq: small N=4 generators}
    J_{12}=-J_{34},\quad  J_{13}=J_{24},\quad  J_{14}=-J_{23},\quad G_1,\quad G_2,\quad G_3,\quad  G_4,\quad L,
\end{equation}
where $L$ is a Virasoro field. This algebra is an $N=4$ SUSY vertex algebra with $D_i:=(G_i)_{(0)}$, $i=1,2,3,4$, and each $(J_{ij}, G_i,G_j,L)$ forms the $N=2$ superconformal algebra having $J_{ij}$ as an $N=2$ superconformal vector. Now, let $J_{ij}:=-J_{ji}$ for $i>j$. Then, one can draw the action of the supersymmetries as follows:
\begin{equation}\label{tikz: small N=4}
    \begin{tikzpicture}[baseline=(current  bounding  box.center)]
        \node(L) at (0,3) {$2L$};
        \node(G1) at (-3,1) {$G_1$};
        \node(G2) at (-1,1) {$G_2$};
        \node(G3) at (1,1) {$G_3$};
        \node(G4) at (3,1) {$G_4$};
        \node(J1) at (-2,-1) {$J_{12}$};
        \node(J2) at (0,-1) {$J_{13}$};
        \node(J3) at (2,-1) {$J_{14}$};
        \draw[->] (J1)--(G1);
        \draw[->] (J1)--(G2);
        \draw[->] (J2)--(G1);
        \draw[->] (J2)--(G3);
        \draw[->] (J3)--(G1);
        \draw[->] (J3)--(G4);
        \draw[->, dashed] (J1)--(G3);
        \draw[->, dashed] (J1)--(G4);
        \draw[->, dashed] (J2)--(G2);        
        \draw[->, dashed] (J2)--(G4);
        \draw[->, dashed] (J3)--(G2);
        \draw[->, dashed] (J3)--(G3);
        \draw[->] (G1)--(L) node[midway,left=8pt] {$D_1$};;
        \draw[->] (G2)--(L) node[midway,left=2pt] {$D_2$};
        \draw[->] (G3)--(L) node[midway,right=2pt] {$D_3$};
        \draw[->] (G4)--(L) node[midway,right=8pt] {$D_4$};
        \node(J2) at (6,-1) {$J_{ij}$};
        \node(G3) at (6,1) {$G_j$};
        \node(J3prim) at (6,3) {$\partial J_{rj}$};
        \draw[->] (J2)--(G3) node[midway,left] {$D_i$};
        \draw[->] (G3)--(J3prim) node[midway,left] {$D_r$};
    \end{tikzpicture}
\end{equation}

In the above left, we denoted the action of $D_i$ and $D_j$ on $J_{ij}$ with solid arrows, and the action of the remaining supersymmetries $D_r$ and $D_l$ with dashed arrows. In the above right, $i,j,$ and $r$ are distinct indices with $i<j$.

The small $N=4$ superconformal algebra can be considered as a natural replacement of an $N=4$ superconformal vector in the following sense. First, it is equipped with four supersymmetries, and for each pair $(G_i, G_j)$, there exists an associated $N=2$ superconformal vector $J_{ij}$. Moreover, this algebra is minimal among those with such properties, since the $N=2$ superconformal vectors $J_{ij}$ and $J_{rl}$ overlap for distinct indices $i,j,r,$ and $l$. Lastly, the action of the  supersymmetries shown in the RHS of \eqref{tikz: small N=4} agrees with that in \eqref{tikz: N=3}.

% However, this algebra does not contain $N=3$ superconformal algebra, since it has no fields of weight $\frac{1}{2}$. Moreover, the generating type of this vertex algebra is significantly different from our expectation \eqref{eq: N=4 with wt0}, and even the number of strong generators differs. Therefore, we consider the following big $N=4$ superconformal algebra as a more suitable replacement.

\subsubsection{Big $N=4$ superconformal algebra}
The big $N=4$ superconformal algebra is defined to be the W-algebra $W^k_{N=1}(D(2,1;\alpha)\oplus \CC,f)$, where $f$ is an odd minimal nilpotent in $D(2,1;\alpha)$. Recall from Section \ref{appendix: N=4 structure of big} that this algebra has a generating type $W\big(\big(\frac{1}{2}\big)^4, 1^7,\big(\frac{3}{2}\big)^4,2\big)$, and is freely generated by
\begin{equation*}
    \sigma_i, \ \tilde{\xi},\ J_{ij},\ G_i, \ \tilde{L},
\end{equation*}
where $i,j=1,2,3,4$ with $i<j$, and $\tilde{L}$ is a Virasoro field. With the odd derivations $D_i:=(G_i)_{(0)}$, this algebra is an $N=4$ SUSY vertex algebra. As in the previous sections, let $J_{ij}:=-J_{ji}$ if $i>j$. Then, each subalgebra generated by $(J_{ij}, G_i,G_j,\tilde{L})$ is isomorphic to the $N=2$ superconformal algebra. Here, $J_{ij}$ is an $N=2$ superconformal vector associated with $D_i$ and $D_j$. The remaining generators $\tilde{\xi}$ and $\sigma_i$ are primary with respect to $\tilde{L}$ of conformal weight $1$ and $\frac{1}{2}$, respectively. On the weight $1$ and $\frac{1}{2}$ fields, the supersymmetries act as follows:
\begin{equation} \label{nontikz: big N=4}
    \begin{aligned}
    J_{ij} &\xmapsto{\quad D_r\quad }\tau(i,j,r,l)\Big(\frac{1-\alpha}{1+\alpha}G_l+\frac{\sqrt{\alpha}}{1+\alpha}\partial \sigma_l\Big),\\
    \sigma_i&\xmapsto{\quad D_i\quad }\tilde{\xi}\xmapsto{\quad D_j\quad }\partial \sigma_j, \\
    \sigma_i&\xmapsto{\quad D_j\quad }\frac{1}{\sqrt{\alpha}}\Big((1-\alpha)J_{ij}-\tau(i,j,r,l)(1+\alpha)J_{rl}\Big).
    \end{aligned}
\end{equation}
In \eqref{nontikz: big N=4}, the four indices $i,j,r,l$ are distinct, and $\tau(i,j,r,l)$ is either $1$ or $-1$ determined by the equality
\begin{equation*}
    \theta_i\theta_j\theta_r\theta_l=\tau(i,j,r,l)\theta_1\theta_2\theta_3 \theta_4,
\end{equation*}
where $\theta_i$'s are the odd Grassmannian variables. The action of $D_i$'s on the weight $\frac{3}{2}$ fields are determined by the equalities \eqref{nontikz: big N=4}, and $G_j=D_i J_{ij}$. To be explicit, for $r \neq i,j$
\begin{equation}
    \begin{aligned}
    D_r(G_j)&=-D_i(D_r J_{ij})=-D_i\Big(\tau(i,j,r,l)\Big(\frac{1-\alpha}{1+\alpha}G_l+\frac{\sqrt{\alpha}}{1+\alpha}\partial \sigma_l\Big)\Big)\\
    &=\tau(i,j,r,l)\Big(-D_i \Big(\frac{1-\alpha}{1+\alpha}D_i J_{il}\Big)-\frac{\sqrt{\alpha}}{1+\alpha}\partial D_i \sigma_l\Big)=\partial J_{rj},
    \end{aligned}
\end{equation}
which coincides with the $N=3$ and small $N=4$ cases.

Note from \eqref{eq: N=4 with wt0} that the $N=4$ superconformal vector in Definition \ref{def: HK N=4 superconformal} should generate an algebra of generating type $W\big(0,\big(\frac{1}{2}\big)^4, 1^6, \big(\frac{3}{2}\big)^4,2\big)$. One can observe that the number of free generators coincides with that of the big $N=4$ superconformal algebra. The main difference between the two is the coupling of weight $\frac{1}{2}$ fields. In \eqref{eq: N=4 with wt0}, these are coupled with the weight $0$ element with supersymmetries, while in the big $N=4$ superconformal algebra, they are all coupled with a weight $1$ element $\tilde{\xi}$ through the second equation in \eqref{nontikz: big N=4}. Moreover, the big $N=4$ superconformal algebra satisfies weakened form of condition (d). To be explicit, this algebra admits conformal embeddings of $N\leq 3$ superconformal algebras, as well as the small $N=4$ superconformal algebra, after suitably modifying the conformal vector. These embeddings are realized in Section \ref{appendix: embedding of N=3} and \ref{appendix: embedding of small N=4}.

%by not requiring $L$ to be the total Virasoro.

\section{$N\geq 2$ SUSY Vertex Operator Algebras}
% We say a vertex algebra $V$ is a conformal extension of $W$ if $W$ is embedded in $V$ as a vertex subalgebra, and the Virasoro field in $W$ is embedded as a conformal vector of $V$. Note that we do allow non-canonical embedding of a Virasoro field. For example, in the big $N=4$ superconformal algebra, the small $N=4$ superconformal algebra is embedded as a subalgebra with choice of different Virasoro field. Even in this case, we say that the big $N=4$ superconformal algebra is a conformal extension of the small $N=4$ superconformal algebra. {\color{red} Do we allow the case where the conformal grading is different?}

Using the superconformal algebras introduced in the previous sections, we define $N=2,3,$ or $4$ SUSY vertex operator algebras (VOAs) as follows. For $n=2$ or $3$, we say that a vertex algebra $V$ is an $N=n$ SUSY vertex operator algebra(VOA) if it is a conformal extension of the $N=n$ superconformal algebra. For $n=4$, we define $V$ a small (resp. big) $N=4$ SUSY vertex operator algebra if $V$ is a conformal extension of the small (resp. big)  $N=4$ superconformal algebra. Here, we say a vertex algebra $V$ is a conformal extension of $W$ if $W$ is a vertex subalgebra of $V$, and $V$ and $W$ have the same conformal vector. Note from Section \ref{appendix: embedding of N=3} that any big $N=4$ SUSY VOA automatically becomes a small $N=4$, $N=3$, and $N=2$ SUSY VOA. 

\subsection{Examples of $N=2$ SUSY VOAs}
There is a large class of $N=2$ SUSY VOAs that were studied in \cite{CKLSS25}, namely
\begin{enumerate}
\item For $n,m \geq 0$, $$\cE^{\psi}_{N=2}(n,m) = \text{Com}(V^{k+1}_{N=1}(\mathfrak{gl}(m)), W^k_{N=1}(\mathfrak{sl}(m+n+1|n), f_{n+1|n})),\qquad \psi = k+m+1,$$
\item For $n\geq 0$ and $m \geq 1$, 
$$\cD^{\psi}_{N=2}(n,m) = \text{Com}(V^{-k-1}_{N=1}(\mathfrak{gl}(m+1)), W^k_{N=1}(\mathfrak{sl}(n+1|n+m+1), f_{n+1|n})), \qquad \psi = k-m,$$
\item For $n\geq 0$ and $m =0$, 
$$\cD^{\psi}_{N=2}(n,0) = W^{k}_{N=1}(\mathfrak{psl}(n+1|n+1), f_{n+1|n})^{U(1)},\qquad \psi = k.$$
\end{enumerate}
Here $f_{n+1|n}$ is the odd principal nilpotent of the subalgebra isomorphic to $\mathfrak{sl}(n+1|n)$ in each Lie superalgebra.
Note that $\cE^{\psi}_{N=2}(1,0) = W^k_{N=1}(\mathfrak{sl}(2|1))$, so this family generalizes the $N=2$ algebra. Also, for $n=0$, 
\begin{equation*}
    \begin{aligned}
        & \cE^{\psi}_{N=2}(0,m) = \text{Com}(V^{k+1}_{N=1}(\mathfrak{gl}(m)), V^k_{N=1}(\mathfrak{sl}(m+1))),\\
        & \cD^{\psi}_{N=2}(0,m) = \text{Com}(V^{-k-1}_{N=1}(\mathfrak{gl}(m+1)), V^k_{N=1}(\mathfrak{sl}(1|m+1))).
    \end{aligned}
\end{equation*}
All $\cE^{\psi}_{N=2}(n,m)$ and $\cD^{\psi}_{N=2}(n,m)$  arise as $1$-parameter quotients of a universal $2$-parameter $N=2$ superconformal VOA $\cW^{N=2}_{\infty}$ which is freely generated of type
$$W\bigg(1,2^2,3^2,\dots; \bigg(\frac{3}{2}\bigg)^2, \bigg(\frac{5}{2}\bigg)^2,\dots\bigg).$$ In fact, $\cW^{N=2}_{\infty}$ is a conformal extension of a Heisenberg algebra and two commuting copies of the universal $2$-parameter VOA  $\cW_{\infty}$ algebra of type $W(2,3,\dots)$. It was conjectured to exist by Candu and Gaberdiel \cite{CG2013}, and realizing the above $N=2$ SUSY VOAs as $1$-parameter quotients of $W^{N=2}_{\infty}$ allowed many dualities of Feigin-Frenkel type among them to be proven, namely $\cD^{\psi}_{N=2}(n,m) \cong \cD^{\psi^{-1}}_{N=2}(m,n)$ and $\cE^{\psi}_{N=2}(n,m) \cong \cE^{\psi^{-1}}_{N=2}(m,n)$, for all $n,m \geq 0$.

\subsection{Examples of big N=4 SUSY VOAs}

There is a similar family of big $N=4$ SUSY vertex algebras which we now describe. For $n\geq 1$, consider $\mathfrak{g} = \mathfrak{osp}(4n|2(2n-1)) \oplus \mathbb{C}$, let $f = f_{2n,2n|2n-1,2n-1}$ be the odd nilpotent whose square $F_{2n,2n|2n-1,2n-1}$ is the nilpotent which is the sum of the rectangular nilpotents $F_{2n,2n} \in \mathfrak{so}(4n)$ and $F_{2n-1, 2n-1} \in \mathfrak{sp}(2(2n-1))$ which were studied in \cite{CKL24}.
Then $W^k_{N=1}(\go\gs\gp(4n|2(2n-1)) \oplus \mathbb{C},f_{2n,2n|2n-1,2n-1})$
has strong generating type
 \begin{equation} \label{N=4:secondattempt} W\bigg(1^7, 2^2, 3^6, 4^2,\dots, (2n-1)^6, 2n; \bigg(\frac{1}{2}\bigg)^4, \bigg(\frac{3}{2}\bigg)^4, \bigg(\frac{5}{2}\bigg)^4, \dots, \bigg(\frac{4n-1}{2}\bigg)^4\bigg)\end{equation}
and is generated by superfields of type
$$W_{N=1}\bigg(1^3, 2, 3^3, 4, \cdots, (2n-1)^3; \bigg(\frac{1}{2}\bigg)^4, \frac{3}{2}, \bigg( \frac{5}{2}\bigg)^3, \frac{7}{2}, \bigg( \frac{9}{2}\bigg)^3, \cdots, \bigg( \frac{4n-1}{2}\bigg)\bigg).$$
%$$\bigg(\frac{1}{2}, 1\bigg)^4,  \bigg(1,\frac{3}{2}\bigg)^3, \bigg(\frac{3}{2}, 2\bigg), \bigg(2,\frac{5}{2}\bigg), \bigg(\frac{5}{2}, 3\bigg)^3, \bigg(3, \frac{7}{2}\bigg)^3, \dots, \bigg(\frac{4n-3}{2}, 2n-1\bigg)^3, \bigg(2n-1, \frac{4n-2}{2}\bigg)^3, \bigg(\frac{4n-1}{2}, 2n\bigg)^3.$$
It can be checked that the Virasoro field together with the fields in weights at most $\frac{3}{2}$ close under OPE and generate a vertex algebra $\mathcal{F}(4) \otimes \mathcal{H} \otimes W$ where $W$ has the same generating type as $W^k_{N=1}(D(2,1;\alpha), f_{\text{min}})$. By uniqueness of minimal $W$-algebras, it is a homomorphic image of the big $N=4$ superconformal algebra.

Similarly, for $n\geq 1$, we consider $\mathfrak{g} = \mathfrak{osp}(4n|2(2n+1) )\oplus \mathbb{C}$, let $f = f_{2n,2n|2n+1,2n+1}$ be the odd nilpotent whose square $F_{2n,2n|2n+1,2n+1}$ is the nilpotent which is the sum of $F_{2n,2n} \in \mathfrak{so}(4n)$ and $F_{2n+1, 2n+1} \in \mathfrak{sp}(2(2n+1))$. 
Then $W^k_{N=1}(\go\gs\gp(4n|2(2n+1)) \oplus \mathbb{C},f_{2n,2n|2n+1,2n+1})$
has strong generating type
 \begin{equation} \label{N=4:secondattempt} W\bigg(1^7, 2^2, 3^6, 4^2,\dots, (2n-1)^6, (2n)^2, (2n+1)^3; \bigg(\frac{1}{2}\bigg)^4, \bigg(\frac{3}{2}\bigg)^4, \bigg(\frac{5}{2}\bigg)^4, \dots, \bigg(\frac{4n+1}{2}\bigg)^4\bigg).\end{equation}
and is generated by superfields of type
$$W_{N=1}\bigg(1^3, 2, 3^3, 4, \cdots, 2n; \bigg(\frac{1}{2}\bigg)^4, \frac{3}{2}, \bigg( \frac{5}{2}\bigg)^3, \frac{7}{2}, \bigg( \frac{9}{2}\bigg)^3, \cdots, \bigg( \frac{4n+1}{2}\bigg)^3\bigg).$$
%$$\bigg(\frac{1}{2}, 1\bigg)^4,  \bigg(1,\frac{3}{2}\bigg)^3, \bigg(\frac{3}{2}, 2\bigg), \bigg(2,\frac{5}{2}\bigg), \bigg(\frac{5}{2}, 3\bigg)^3, \bigg(3, \frac{7}{2}\bigg)^3, \dots,  \bigg(\frac{4n-1}{2}, 2n\bigg), \bigg(2n, \frac{4n+1}{2}\bigg), \bigg(\frac{4n+1}{2}, 2n+1\bigg)^3.$$
As above, this is an extension of the big $N=4$ algebra.

These two families of examples strongly suggest the existence of a $2$-parameter VOA $\cW^{N=4}_{\infty}$ which is a conformal extension of the big $N=4$ algebra, and is freely generated of type 
\begin{equation} \label{N=4:firstattemptuniversal} W\bigg(1^7, 2^2, 3^6, 4^2,\dots; \bigg(\frac{1}{2}\bigg)^4, \bigg(\frac{3}{2}\bigg)^4, \bigg(\frac{5}{2}\bigg)^4, \dots\bigg), \end{equation}
We expect $\cW^{N=4}_{\infty}$ to admit all the above examples as $1$-parameter quotients. Just as $\cW^{N=2}_{\infty}$ is a conformal extension of $\cH$ tensored with two commuting copies of $\cW_{\infty}$, we expect $W^{N=4}_{\infty}$ to be an extension of $\cH$ and two commuting copies of the universal $2$-parameter VOA $\cW^{\gs\gp}_{\infty}$ of type $W(1^3, 2, 3^3, 4, \dots)$ constructed in \cite{CKL24}.

There are several more big $N=4$ SUSY vertex algebras which we also expect to arise as $1$-parameter quotients $\cW^{N=4}_{\infty}$:
\begin{enumerate} 
\item For $n\geq 1$ and $r \geq 1$, $W^k_{N=1}(\go\gs\gp(4n|2(2n-1)+2r) \oplus \mathbb{C}, f_{2n,2n|2n-1,2n-1})$ contains $V^{k'}_{N=1}(\gs\gp(2r))$ for some shifted level $k'$, and we consider the coset
$$\text{Com}(V^{k'}_{N=1}(\gs\gp(2r)), W^k_{N=1}(\go\gs\gp(4n|2(2n-1)+2r)\oplus \mathbb{C}, f_{2n,2n|2n-1,2n-1})).$$

\item For $n\geq 1$, consider the orbifold $W^k_{N=1}(\go\gs\gp(4n+1|2(2n-1))\oplus \mathbb{C}, f_{2n,2n|2n-1,2n-1})^{\mathbb{Z}_2}$.

\item For $n\geq 1$ and $r \geq 1$, $W^k_{N=1}(\go\gs\gp(4n+r|2(2n-1))\oplus \mathbb{C}, f_{2n,2n|2n-1,2n-1})$ contains $V^{k'}_{N=1}(\gs\go_r)$ for some shifted level $k'$, and consider
$$\text{Com}(V^{k'}_{N=1}(\gs\go_r), W^k_{N=1}(\go\gs\gp(4n+r|2(2n-1))\oplus \mathbb{C}, f_{2n,2n|2n-1,2n-1}))^{\mathbb{Z}_2}.$$ Here the action of $\mathbb{Z}_2$ comes from the fact that the action of $\gs\go_r$ on $W^k_{N=1}(\go\gs\gp(4n+r|2(2n-1))\oplus \mathbb{C}, f_{2n,2n|2n-1,2n-1})$ lifts to $O(r)$.

\item For $n\geq 0$ and $r \geq 1$, $W^k_{N=1}(\go\gs\gp(4n|2(2n+1)+2r)\oplus \mathbb{C}, f_{2n,2n|2n+1,2n+1})$ contains $V^{k'}_{N=1}(\gs\gp_{2r})$ for some shifted level $k'$, and consider
$$\text{Com}(V^{k'}_{N=1}(\gs\gp_{2r}), W^k_{N=1}(\go\gs\gp(4n|2(2n+1)+2r)\oplus \mathbb{C}, f_{2n,2n|2n-1,2n-1})).$$ Note that in the case $n=0$, we get
$$\text{Com}(V^{k}_{N=1}(\gs\gp_{2r}), V^k_{N=1}(\gs\gp_{2r+2}\oplus \mathbb{C})).$$

\item For $n\geq 1$, consider the orbifold $W^k_{N=1}(\go\gs\gp(4n+1|2(2n+1))\oplus \mathbb{C}, f_{2n,2n|2n+1,2n+1})^{\mathbb{Z}_2}$.

\item For $n\geq 0$ and $r \geq 2$, $W^k_{N=1}(\go\gs\gp(4n+r|2(2n+1))\oplus \mathbb{C}, f_{2n,2n|2n+1,2n+1})$ contains $V^{k'}_{N=1}(\gs\go_r)$ for some shifted level $k'$, and consider
$$\text{Com}(V^{k'}_{N=1}(\gs\go_r), W^k_{N=1}(\go\gs\gp(4n+r|2(2n+1))\oplus \mathbb{C}, f_{2n,2n|2n+1,2n+1}))^{\mathbb{Z}_2}.$$ As above, the action of $\mathbb{Z}_2$ comes from the fact that the action of $\gs\go_r$ on $W^k_{N=1}(\go\gs\gp(4n+r|2(2n+1))\oplus \mathbb{C}, f_{2n,2n|2n+1,2n+1})$ lifts to $O(r)$. When $n = 0$, we get
$$\text{Com}(V^{k}_{N=1}(\gs\go_r), V^k_{N=1}(\go\gs\gp(r|2)\oplus \mathbb{C}) )^{\mathbb{Z}_2}.$$
\end{enumerate}

Using the description of orbifolds and SUSY affine cosets of SUSY $W$-algebras given by Corollary \ref{cor:SFGproperty}, it is straightforward to check using classical invariant theory that all of these VOAs have strong generating type some truncation of \eqref{N=4:firstattemptuniversal}, and are conformal extensions of the big $N=4$ algebra. As such, they can also be regarded as extensions of either the small $N=4$ algebra or the $N=3$ algebra as well. It is more difficult to find examples of $1$-parameter VOAs that are extensions of either the small $N=4$ algebra or the $N=3$ algebra, which do {\it not} admit an action of the big $N=4$ algebra.

\subsection{Examples of small $N=4$ SUSY VOAs but not big $N=4$ SUSY VOAs}
In this subsection, we introduce a $1$-parameter vertex algebra which is a conformal extension of the small $N=4$ algebra but not the big $N=4$ algebra.
For $n\geq 1$, let $\mathcal{W}(n) =  \mathcal{H}(4n) \otimes \mathcal{F}(4n)$, where $\mathcal{H}(n)$ is the rank $4n$ Heisenberg algebra with generators $\{\alpha^{1,i}, \alpha^{2,i}, \beta^{1,i}, \beta^{2,i}|\ i = 1,\dots, n\}$ satisfying
$$[\alpha^{1,i}{}_\lambda \beta^{1,j}] = \delta_{i,j}\lambda,  \qquad [\alpha^{2,i}{}_{\lambda} \beta^{2,j}] = \delta_{i,j}\lambda,$$
and $\cF(4n)$ is the rank $4n$ free fermion algebra with generators $\{b^{1,i}, b^{2,i}, c^{1,i}, c^{2,i}|\ i = 1,\dots, n\}$ satisfying
$$[b^{1,i}{}_\lambda c^{1,j}] = \delta_{i,j} ,  \qquad [b^{2,i}{}_\lambda c^{2,j}] = \delta_{i,j}.$$
$\mathcal{W}(n)$ has an action of the symplectic group $\text{Sp}(2n)$ by automorphisms, where each of the sets $\{\alpha^{1,i}, \alpha^{2,i}\}$, $\{\beta^{1,i}, \beta^{2,i}\}$, $\{b^{1,i}, b^{2,i}\}$, and $\{ c^{1,i}, c^{2,i}\}$ transforms as the standard $\text{Sp}(2n)$-module. The orbifold $\mathcal{W}(n)^{\text{Sp}(2n)}$ is strongly generated by the following fields, which close linearly under OPE:
\begin{equation} \label{spnorbifold}
\begin{split} & \sum_{i=1}^n \big(:\partial^r b^{1,i} \partial^s c^{1,i}: \ +\  :\partial^r b^{2,i} \partial^s c^{2,i}:\big),
\\ & \sum_{i=1}^n :\partial^r b^{1,i} \partial^s b^{2,i}: \ +\  :\partial^s b^{1,i} \partial^r b^{2,i}:\big) , \qquad  \sum_{i=1}^n \big(:\partial^r c^{1,i} \partial^s c^{2,i}:\ +\ :\partial^s c^{1,i} \partial^r c^{2,i}:\big)
\\ & \sum_{i=1}^n \big(:\partial^r \alpha^{1,i} \partial^s \beta^{1,i}:\  + \ :\partial^r \alpha^{2,i} \partial^s \beta^{2,i}:\big),
\\ & \sum_{i=1}^n\big( :\partial^r \alpha^{1,i} \partial^s \alpha^{2,i}:  \ - \ : \partial^s \alpha^{1,i}\partial^r \alpha^{2,i}:\big),\qquad \sum_{i=1}^n\big( :\partial^r \beta^{1,i} \partial^s \beta^{2,i}:\  - : \partial^s \beta^{1,i} \partial^r \beta^{2,i}:\big), 
\\ & \sum_{i=1}^n\big( :\partial^r \alpha^{1,i} \partial^s b^{1,i}: +\ : \partial^r \alpha^{2,i} \partial^s b^{2,i}:\big),\qquad \sum_{i=1}^n  \big(:\partial^r \beta^{1,i}  \partial^s c^{1,i} : +\ : \partial^s \beta^{2,i} c^{2,i}:\big),
\\ & \sum_{i=1}^n\big( :\partial^r \alpha^{1,i} \partial^s c^{2,i} :\ - :\partial^r \alpha^{2,i} \partial^s c^{1,i}:\big), \qquad \sum_{i=1}^n \big( :\partial^r \beta^{1,i} \partial^s b^{2,i} :\ - :\partial^r \beta^{2,i} \partial^s b^{1,i}:\big).
\end{split}
\end{equation}
Note that $\mathcal{W}(n)^{\text{Sp}(2n)}$ is a conformal extension of the small $N=4$ algebra with central charge $6n$, which is generated by
\begin{equation}
\begin{split} & \sum_{i=1}^n \big(: b^{1,i} c^{1,i}: \ +\  : b^{2,i} c^{2,i}: \big),\qquad \sum_{i=1}^n : b^{1,i}  b^{2,i}:, \qquad  \sum_{i=1}^n  :c^{1,i}  c^{2,i}:,
\\ & \sum_{i=1}^n \big(: \alpha^{1,i}  \beta^{1,i}:\  + \ : \alpha^{2,i}  \beta^{2,i}:\big) -\frac{1}{2} \big(: b^{1,i} \partial c^{1,i}: - : \partial b^{1,i} c^{1,i}:  + : b^{2,i} \partial c^{2,i}: - : \partial b^{2,i} c^{2,i}:\big),
\\ & \sum_{i=1}^n : \alpha^{1,i} \big( b^{1,i}: +\ :  \alpha^{2,i}  b^{2,i}:\big) ,\qquad \sum_{i=1}^n  :\big(: \beta^{1,i}  c^{1,i} : +\ : \beta^{2,i} c^{2,i}:\big)
\\ & \sum_{i=1}^n : \alpha^{1,i}  \big( c^{2,i} :\ - : \alpha^{2,i}  c^{1,i}:\big), \qquad \sum_{i=1}^n  \big(: \beta^{1,i}  b^{2,i} :\ - : \beta^{2,i} b^{1,i}:\big)
\end{split}
\end{equation}
It is straightforward to check that the fields \eqref{spnorbifold} close linearly under OPE and hence generate a Lie conformal superalgebra. The OPE relations among these fields are independent of $n$ except for the vacuum terms, which are linear in $n$. It follows that the OPE algebra of these fields can be defined over the polynomial ring $\mathbb{C}[\nu]$, where $\nu$ is a formal parameter replacing the discrete parameter $n$. The universal enveloping algebra of this Lie conformal superalgebra is a $1$-parameter vertex superalgebra $\cA(\nu)$ with the following features:
\begin{enumerate}
\item $\mathcal{A}(\nu)$ is simple for generic values of $\nu$, and is freely generated of type
$$ W\bigg(1^3, 2^2, 3^6, 4^2, 5^6, 6^2, \dots; \bigg(\frac{3}{2}\bigg)^4, \bigg(\frac{5}{2}\bigg)^4,\bigg(\frac{5}{2}\bigg)^4,\dots\bigg).$$
\item For each $n\in \mathbb{N}$, the quotient $\cA(\nu) / (\nu - n) \mathcal{A}(\nu)$ is not simple, and its simple quotient is isomorphic to $\mathcal{W}(n)^{\text{Sp}(2n)}$.
\item $\mathcal{A}(\nu)$ is a conformal extension of the small $N=4$ algebra with central charge $6\nu$,
\item $\mathcal{A}(\nu)$ does not contain the big $N=4$ algebra.
\end{enumerate}

A more interesting $1$-parameter vertex algebra which is a conformal extension of the small $N=4$ algebra but not the big one was recently conjectured to exist in \cite{BM25}. It is denoted by $\cW^{s,s}_{\infty}$ and is expected to be a unifying algebra for the $4D$ $N=4$ super Yang-Mills algebra $\mathcal{V}(A_{N-1})$ for $\mathfrak{su}(N)$. In particular, when the central charge of $\cW^{s,s}_{\infty}$ is specialized to $-3(N^2-1)$, its simple quotient is conjectured to be isomorphic to the algebra $\mathcal{V}(A_{N-1})$. Note that $\mathcal{V}(A_{N-1})$ is a special case of the algebras $\cW_{\Gamma}$ introduced by Beem, Rastelli and Meneghelli in \cite{BMR19} for any reflection group $\Gamma$; $\mathcal{V}(A_{N-1})$ is just the case $\mathcal{W}_{S_N}$ when $\Gamma$ is the symmetric group $S_N$. A free field realization of $\cW_{\Gamma}$ was proposed in \cite{BMR19} and many properties were conjectured. In the case $\Gamma= S_N$, a rigorous construction of $\mathcal{W}_{S_N}$ was given in \cite{AKM23}.

\subsection{Examples of $N=3$ SUSY VOAs but not big $N=4$ SUSY VOAs}
In this subsection, we introduce a $2$-parameter vertex algebra which is a conformal extension of the $N=3$ algebra but not of the big $N=4$ algebra.

It can readily be checked from the OPE algebra of the big $N=4$ algebra $W^k_{N=1}(D(2,1;\alpha)\oplus \mathbb{C}, f)$ for $f$ an odd minimal nilpotent, that it admits an order $2$ automorphism defined as follows:
\begin{equation} \begin{split}
& k \mapsto k, \qquad \alpha \mapsto \frac{1}{\alpha},
\\ & \xi \mapsto -\xi, \qquad \tilde{L} \mapsto \tilde{L},\qquad  \tilde{J}^i \mapsto \tilde{J}'^{i}, \qquad  \tilde{J}'^i \mapsto \tilde{J}^{i} \qquad i = 0,\pm,
\\ & \sigma^{++} \mapsto -\alpha \sigma^{++}, \qquad \sigma^{--} \mapsto -\frac{1}{\alpha} \sigma^{--}, \qquad \sigma^{+-}  \mapsto - \sigma^{-+}, \qquad \sigma^{-+} \mapsto - \sigma^{+-},
\\ & \tilde{G}^{++} \mapsto \tilde{G}^{++}, \qquad \tilde{G}^{--} \mapsto \tilde{G}^{--}, \qquad \tilde{G}^{-+} \mapsto \tilde{G}^{+-}, \qquad \tilde{G}^{+-} \mapsto \tilde{G}^{-+}.
\end{split}
\end{equation}
Consider the orbifold $W^k_{N=1}(D(2,1;\alpha)\oplus \mathbb{C},f)^{\mathbb{Z}_2}$ under this automorphism. It is straightforward to check that:
\begin{enumerate}[(i)]
\item The $N=3$ algebra given by \eqref{eq: N=3 inside big N=4} is fixed by this automorphism, so we have a conformal embedding of the $N=3$ algebra in $W^k_{N=1}(D(2,1;\alpha)\oplus \mathbb{C},f)^{\mathbb{Z}_2}$ after changing the conformal vector to $\tilde{L}$.
\item The orbifold $W^k_{N=1}(D(2,1;\alpha)\oplus \mathbb{C},f)^{\mathbb{Z}_2}$ is bigger than the $N=3$ algebra; for example it contains the field $:\partial \sigma^{++} \sigma^{++}:$.
\end{enumerate}

\appendix
\section{Structure of $N\geq 2$ superconformal algebras}

\subsection{$N=3$ superconformal algebra} \label{appendix: N=3 superconformal}

In this section, we show that the minimal SUSY W-algebra $W^k_{N=1}(\g,f)$ for $\g=\mathfrak{spo}(2|3)$ is isomorphic to the $N=3$ superconformal algebra. Since $\g_0^f\simeq \mathfrak{so}(2)$ is an one-dimensional even vector space, this algebra is isomorphic to the usual W-algebra $W^k(\g,F)$ tensored with one free fermion. The generators and relations of this algebra is studied in \cite[Section 8.5]{KW04}, and we adopt the notations used therein. 

\subsubsection{Dictionary between SUSY and nonSUSY W-algebras}
As a SUSY vertex algebra, $W^k_{N=1}(\mathfrak{spo}(2|3),f)$ is freely generated by
\begin{equation} \label{eq: N=3 SUSY generators}
    P,\quad  X,\quad  Y,\quad G,
\end{equation}
where $P$ and $G$ are odd and $X$ and $Y$ are even. In particular, $G$ in \eqref{eq: N=3 SUSY generators} is an $N=1$ superconformal vector of central charge $c=-6k-3$. Using the $\Lambda$-bracket introduced in \eqref{eq: Lambda bracket}, the remaining relations between the fields are given by
\begin{equation} \label{eq: N=3 relations}
    \begin{gathered}
    [G{}_{\Lambda}X]=(2\partial+2\lambda+\chi D)X, \quad [G{}_{\Lambda}Y]=(2\partial+2\lambda+\chi D)Y, \quad[G{}_{\Lambda}P]=(2\partial+\lambda+\chi D)P,\\
    [X{}_{\Lambda}Y ]=G+(\partial+\lambda+\chi D) P+\frac{c}{3}\lambda\chi, \quad  [P{}_{\Lambda}X]=X, \quad[P{}_{\Lambda}Y]=-Y, \quad [P{}_{\Lambda}P]=\frac{c}{3}\chi.
    \end{gathered}
\end{equation}
Now, let
\begin{equation} \label{eq: N=3 KW generators}
    \begin{gathered}
        \tilde{L}:=\frac{1}{2}DG, \quad \tilde{G}^0:=-\frac{1}{\sqrt{2}}G, \quad  \tilde{G}^+:=\frac{1}{2}DX, \quad \tilde{G}^-:=-DY,\\
        J^+:=\sqrt{2}\,X, \quad  J^-:=\sqrt{2}\,Y, \quad  J^0:=2DP, \quad  \Phi:=\frac{1}{\sqrt{2}}P,
    \end{gathered}
\end{equation}
then they recover the relations between those with the same names introduced in \cite[Section 8.5]{KW04}.

\subsubsection{$N=3$ SUSY structure} \label{appendix: N=3 structure}
Recall the redefined generators of the $N=3$ superconformal algebra in \eqref{eq: N=3 KW generators}. Fix any constants $b_1,b_2,,b_3\in \CC$ with $b_1\neq 0$ and consider
\begin{equation*}
    \begin{aligned}
        G_1=&\sqrt{2} b_2 \tilde{G}^0 + \frac{1-b_2^2}{\sqrt{2} b_1} \tilde{G}^- + \sqrt{2} b_1 \tilde{G}^+,\\
        G_2=&\frac{\sqrt{2}\big(b_2 b_3+\sqrt{b_1^2+b_3^2}\big)}{b1}\tilde{G}^0-\frac{(1+b_2^2)b_3+2b_2 \sqrt{b_1^2+b_3^2}}{\sqrt{2} b_1^2}\tilde{G}^- + \sqrt{2} b_3 \tilde{G}^+,\\
        G_3=&-\frac{\sqrt{-1}\sqrt{2}\big(b_3+b_2\sqrt{b_1^2+b_3^2}\big)}{b_1}\tilde{G}^0+\frac{\sqrt{-1}\big(2b_2b_3+(1+b_2^2)\sqrt{b_1^2+b_3^2}\big)}{\sqrt{2} b_1^2}\tilde{G}^- -\sqrt{-1}\sqrt{2}\sqrt{b_1^2+b_3^2}\tilde{G}^+,\\
        J_{12}=&-\frac{b_3+b_2 \sqrt{b_1^2+b_3^2}}{2b_1}J^0-\frac{2b_2 b_3+(1+b_2^2)\sqrt{b_1^2+b_3^2}}{2b_1^2}J^-+\frac{1}{2}\sqrt{b_1^2+b_3^2}J^+,\\
        J_{13}=&\frac{\sqrt{-1}\big(b_2b_3+\sqrt{b_1^2+b_3^2}\big)}{2b_1}J^0+\frac{\sqrt{-1}\big((1+b_2^2)b_3+2b_2\sqrt{b_1^2+b_3^2}\big)}{2b_1^2}J^--\frac{\sqrt{-1}}{2}b_3J^+,\\
        J_{23}=&-\frac{\sqrt{-1}}{2}b_2 J^0+\frac{\sqrt{-1}(1-b_2^2)}{2b_1}J^-+\frac{\sqrt{-1}}{2}b_1 J^+,\quad   K=\sqrt{-2}\Phi.
    \end{aligned}
\end{equation*}
Moreover, let $J_{ij}:=-J_{ji}$ if $i>j$. Then for each choice of pair $(i,j)$ with $i\neq j$, the following four fields
\begin{equation} \label{eq: N=2 embedding in N=3}
    J_{ij},\quad G_i,\quad G_j,\quad L
\end{equation}
satisfy the $\lambda$-bracket relations in \eqref{eq: N=2 conditions}. In other words, the fields \eqref{eq: N=2 embedding in N=3} generate the $N=2$ superconformal algebra, while $J_{ij}$ is an $N=2$ superconformal vector associated with $D_i:=(G_i)_{(0)}$ and $D_j:=(G_j)_{(0)}$. It implies that the $N=3$ superconformal algebra has an $N=3$ structure using $D_i=(G_i)_{(0)}$'s for $i=1,2,3$. In particular, if we choose $b_1=1$, $b_2=0$, and $b_3=0$, we get the simplest forms
\begin{equation*}
    \begin{gathered}
    G_1=\sqrt{2}\tilde{G}^++\frac{1}{\sqrt{2}}\tilde{G}^-, \quad G_2= \sqrt{2}\tilde{G}^0, \quad G_3=-\sqrt{-1}\big(\sqrt{2}\tilde{G}^+-\frac{1}{\sqrt{2}}\tilde{G}^-\big),\\
    J_{12}=\frac{1}{2}(J^+-J^-), \quad J_{13}=\frac{\sqrt{-1}}{2}J^0, \quad J_{23}=\frac{\sqrt{-1}}{2}(J^++J^-) \quad  K=\sqrt{-2}\Phi.
    \end{gathered}
\end{equation*}
\subsection{Small $N=4$ superconformal algebra} \label{appendix: small N=4}
For $\g=\mathfrak{psl}(2|2)$ and its odd minimal nilpotent $f$, we call the corresponding SUSY W-algebra $W^k_{N=1}(\g,f)$ the small $N=4$ superconformal algebra. Since $\g_0^f$ is trivial, this algebra is isomorphic to the nonSUSY W-algebra, which was studied in \cite{KW04}. The small $N=4$ algebra is freely generated by
\begin{equation} \label{eq: small N=4 KW generators}
    J^0,\quad J^+, \quad, J^-,\quad G^+,\quad, G^-, \quad \bar{G}^+, \quad \bar{G}^-,\quad L,
\end{equation}
where $L$ is a Virasoro field of central charge $c=-6(k+1)$. The even fields $J^0, J^{\pm}$ form affine $\mathfrak{sl}(2)$ being primary of weight $1$ and under the affine $\mathfrak{sl}(2)$ action, the odd fields $G^{\pm}$ and $\bar{G}^{\pm}$ form standard representations. Moreover, $G^{\pm}$ and $\bar{G}^{\pm}$ are primary of weight $\frac{3}{2}$. For explicit relations between the generators, refer to \cite[Section 8.4]{KW04}. Fix any constants $b_1,b_2\in \CC$ and let
\begin{equation} \label{eq: modified generators of small N=4}
    \begin{aligned}
    J_{12}&:=\sqrt{-1}J^0, \quad J_{13}:=\sqrt{-1}(J^++J^-), \quad J_{14}:=-(J^+-J^-),\\
    G_1&:=G^++b_1G^-+b_2 \bar{G}^++(1-b_1b_2)\bar{G}^-,\\
    G_2&:=-\sqrt{-1}(G^+-b_1 G^-+b_2 \bar{G}^+-(1-b_1b_2)\bar{G}^-),\\
    G_3&:=-\sqrt{-1}(b_1 G^++G^-+(1-b_1b_2)\bar{G}^+-b_2\bar{G}^-),\\
    G_4&:=b_1 G^+-G^--(1-b_1b_2)\bar{G}^++b_2\bar{G}^-.
    \end{aligned}
\end{equation}
Then, each $G_i$ satisfies $[G_i {}_{\lambda}G_i]=2L+\frac{c}{3}\lambda^2$. In other words, each $G_i$ is an $N=1$ superconformal vectors. Now, let $D_i:=(G_i)_{(0)}$ and
\begin{equation} \label{eq: modified generators2 of small N=4}
    J_{34}:=-J_{12}, \quad J_{24}:=J_{13}, \quad J_{23}:=-J_{14}
\end{equation}
Then, the four fields $J_{ij}, G_i, G_j, L$ for any $i<j$ form the $N=2$ superconformal algebra.

\subsection{Big $N=4$ superconformal algebra} \label{appendix: big N=4}
Let $\g=D(2,1;\alpha)$ for $\alpha\in \CC \setminus\{-1,0\}$ and $f$ be its odd minimal nilpotent in \eqref{eq: D(2,1;a) minimal}. In this section, we show that the SUSY W-algebra $W^k_{N=1}(\g\oplus \CC,f)$ has an $N=4$ supersymmetries. Moreover, we study the relationship between this W-algebra and $N=3$ or small $N=4$ superconformal algebra. The computations in this section were carried out using the Mathematica package \texttt{OPEdefs} \cite{Thiel91}.

To denote the generators and relations of the W-algebra, we take the first choice of $f$ in \eqref{eq: D(2,1;a) minimal}. Note that the choice of $f$ does not affect the structures of the SUSY W-algebra, since we have the relationship $W^k_{N=1}(\g\oplus \CC,f)\simeq W^k(\g\oplus \CC,F)\oplus \mathcal{F}(\g_0^f \oplus \CC)$.

\subsubsection{Generators}
We use notations in \cite{Musson12} to denote the elements of $\g$. As a SUSY vertex algebra, $W^k_{N=1}(\g,f)$ is freely generated by the following elements:
\begin{equation} \label{eq: generator SUSY D(2,1;a)}
    \begin{aligned}
           K_1=&\, E_2-F_3,\quad K_2=E_3-F_2,\quad K_3=H_2-H_3,\\
           J_1=&\, u_{-1}\otimes u_1\otimes u_{-1}+k \sqrt{2(\alpha+1)}DF_3+\frac{1}{\sqrt{2(1+\alpha)}}:\!H_1 E_2\!:+\frac{\alpha}{\sqrt{2(1+\alpha)}}:\!H_1 F_3\!:\\
           &-\frac{1+k}{\sqrt{2(1+\alpha)}}:\!E_2 H_2\!:+\frac{k}{\sqrt{2(1+\alpha)}}:\!E_2 H_3\!:-\frac{k}{\sqrt{2(1+\alpha)}}:\!H_2 F_3\!:-\frac{k-\alpha}{\sqrt{2(1+\alpha)}}:\!H_3 F_3\!:\\
           J_2=&\, u_{-1}\otimes u_{-1}\otimes u_{1}+k \sqrt{2(1+\alpha)} DE_3+\frac{k-\alpha}{\sqrt{2(1+\alpha)}}:\!E_3 H_3\!-\frac{k}{\sqrt{2(1+\alpha)}}:\!F_2 H_3\!:\\
           &+\frac{\alpha}{\sqrt{2(1+\alpha)}}:\!H_1 E_3\!:+\frac{1}{\sqrt{2(1+\alpha)}}:\!H_1 F_2\!:+\frac{k}{\sqrt{2(1+\alpha)}}:\!H_2 E_3\!-\frac{1+k}{\sqrt{2(1+\alpha)}}:\!H_2 F_2\!:,\\
           J_3=&\, u_{-1}\otimes u_{1}\otimes u_{1}-u_{-1}\otimes u_{-1}\otimes u_{-1}+k \sqrt{2(1+\alpha)} DH_3-\frac{k\sqrt{2}}{\sqrt{1+\alpha}}:\!E_2 E_3\!:+\frac{(1+k)\sqrt{2}}{\sqrt{1+\alpha}}:\!E_2 F_2\!:\\
           &-\frac{(k-\alpha)\sqrt{2}}{\sqrt{1+\alpha}}:\!E_3 F_3\!:+\frac{k\sqrt{2}}{\sqrt{1+\alpha}}:F_2 F_3:+\frac{1}{\sqrt{2(1+\alpha)}}:\!H_1 H_2\!:+\frac{\alpha}{\sqrt{2(1+\alpha)}}:\!H_1 H_3\!:,\\
           G=&\frac{2}{k^2}F_1-\frac{1}{k\sqrt{2(1+\alpha)}}(D(u_{-1}\otimes u_1\otimes u_1)+D(u_{-1}\otimes u_{-1}\otimes u_{-1}))\\&-\frac{1}{k(1+\alpha)}\big(:\!(DE_2)F_2\!:+\alpha :\!(DE_3)F_3\!:+:\!(DF_2)E_2\!:+\alpha :\!(DF_3)E_3\!:\big)\\
           &+\frac{1}{2k(1+\alpha)}\big((1+\alpha):\!(DH_1)H_1\!:-:\!(DH_2)H_2\!:-\alpha :\!(DH_3)H_3\!:\big)\\
           &+\frac{2}{k^2(1+\alpha)^2}(:\!E_2 H_2 F_2\!:+\alpha^2 :\!E_3 H_3 F_3\!:)\\
           &-\frac{\sqrt{2}}{k^2(1+\alpha)^{3/2}}:\!(E_2+\alpha F_3) (u_{-1}\otimes u_{-1}\otimes u_1)\!:-\frac{\sqrt{2}}{k^2(1+\alpha)^{3/2}}:\!(F_2+\alpha E_3) (u_{-1}\otimes u_{1}\otimes u_{-1})\!:\\
           &-\frac{1}{k^2\sqrt{2}(1+\alpha)^{3/2}}:\!((1+\alpha)H_1-H_2-\alpha H_3)(u_{-1}\otimes u_{-1}\otimes u_{-1})\!:\\
           &-\frac{1}{k^2\sqrt{2}(1+\alpha)^{3/2}}:\!((1+\alpha)H_1+H_2+\alpha H_3)(u_{-1}\otimes u_{1}\otimes u_1)\!:+\frac{1+k}{k}\partial H_1.
       \end{aligned}
   \end{equation}
   In particular, the odd field $G$ is $N=1$ superconformal with central charge 
\begin{equation}
    c=\frac{3}{2}(-1-4k),
\end{equation}
and the fields $K_i$'s (resp. $J_i$'s) are primary of conformal weight $1/2$ (resp. $1$) with $[G{}_{\lambda}K_i]=DK_i$ (rep. $[G{}_{\lambda}J_i]=DJ_i$). In addition to the generators above, the big $N=4$ superconformal algebra $W^k_{N=1}(\g\oplus \CC,f)$ has two extra free generators $\sigma$ and $D\sigma$, which generate the SUSY Heisenberg vertex algebra. To be explicit, $\sigma$ and $D\sigma$ are odd and even field, respectively, and the nonzero $\lambda$-brackets between $\sigma$ and $D\sigma$ are
\begin{equation*}
    [D\sigma{}_{\lambda}D\sigma]=k\lambda, \quad [\sigma{}_{\lambda}\sigma]=k,
\end{equation*}
while they commute with all the other free generators.

\subsubsection{Dictionary between SUSY and nonSUSY W-algebras} \label{appendix: dictionary big N=4}
Recall the generators of $W^k_{N=1}(\g,f)$ in \eqref{eq: generator SUSY D(2,1;a)}. In this section, we provide modifications of the generators so that they commute with the weight $1/2$ fields $K_1, K_2$, and $K_3$. Moreover, the redefined set of generators
\begin{equation} \label{eq: generators nonSUSY D(2,1;a)}
    \{L,G^{++},G^{--},G^{+-},G^{-+},J^0,J^+,J^-,J'^0,J'^+,J'^-\}
\end{equation}
freely generates the nonSUSY W-algebra $W^k(\g,F)$. The generators in \eqref{eq: generators nonSUSY D(2,1;a)} are named in accordance with those in \cite[Section 8.6]{KW04}. Let
\begin{equation} \label{eq: nonSUSY D(2,1;a) virasoro}
    L=\frac{1}{2}DG-L_{\mathcal{F}},
\end{equation}
where $L_{\mathcal{F}}$ is the conformal vector in $\mathcal{F}(\g_0^f)$. To be explicit, $L_{\mathcal{F}}$ is written as
\begin{equation*}
    L_{\mathcal{F}}=\frac{\alpha}{2k(1+\alpha)^2}\big(:\!(\partial K_1)K_2\!:+:\!(\partial K_2)K_1\!:+\frac{1}{2}:\!(\partial K_3)K_3\!:\big)
\end{equation*}
and the element \eqref{eq: nonSUSY D(2,1;a) virasoro} $L$ commutes with weight $1/2$ fields, since $K_i$'s are primary of conformal weight $1/2$ for both $\frac{1}{2}DG$ and $L_{\mathcal{F}}$. The weight $3/2$ generators are defined as follows:
\begin{equation} \label{eq: generator nonSUSY D(2,1;a) wt 3/2}
    \begin{aligned}
        G^{++}=&-\frac{\sqrt{-1}\sqrt{k}}{2} G+\frac{\sqrt{-1}}{2\sqrt{2k}\sqrt{1+\alpha}}DJ_3+\frac{\alpha\sqrt{-1}}{\sqrt{2}k\sqrt{k}(1+\alpha)^2\sqrt{1+\alpha}}:\!J_1 K_2\!:\\
        &+\frac{\alpha\sqrt{-1}}{\sqrt{2}k\sqrt{k}(1+\alpha)^2\sqrt{1+\alpha}}:\!J_2 K_1\!:+\frac{\sqrt{-1}(k+2\alpha+k\alpha)}{2\sqrt{k}(1+\alpha)^2}:\!(DK_1) K_2\!:-\frac{\sqrt{k}\sqrt{-1}}{2(1+\alpha)}:\!(DK_2) K_1\!:\\
        &-\frac{\alpha\sqrt{-1}}{4\sqrt{k}(1+\alpha)^2}:\!(DK_3) K_3\!:+\frac{\alpha^2 \sqrt{-1}}{k\sqrt{k}(1+\alpha)^4}:\!K_1 K_2 K_3\!:,\\
        G^{--}=&\frac{\sqrt{-1}\sqrt{k}}{2} G+\frac{\sqrt{-1}}{2\sqrt{2k}\sqrt{1+\alpha}}DJ_3+\frac{\alpha\sqrt{-1}}{\sqrt{2}k\sqrt{k}(1+\alpha)^2\sqrt{1+\alpha}}:\!J_1 K_2\!:\\
        &+\frac{\alpha\sqrt{-1}}{\sqrt{2}k\sqrt{k}(1+\alpha)^2\sqrt{1+\alpha}}:\!J_2 K_1\!:+\frac{\sqrt{k}\sqrt{-1}}{2(1+\alpha)}:\!(DK_1) K_2\!:-\frac{\sqrt{-1}(k+2\alpha+k\alpha)}{2\sqrt{k}(1+\alpha)^2}:\!(DK_2) K_1\!:\\
        &+\frac{\alpha\sqrt{-1}}{4\sqrt{k}(1+\alpha)^2}:\!(DK_3) K_3\!:-\frac{\alpha^2 \sqrt{-1}}{k\sqrt{k}(1+\alpha)^4}:\!K_1 K_2 K_3\!:,\\
        G^{+-}=&\frac{\sqrt{-1}}{\sqrt{2k}\sqrt{1+\alpha}}DJ_1+\frac{\alpha\sqrt{-1}}{\sqrt{2}k\sqrt{k}(1+\alpha)^2 \sqrt{1+\alpha}}:\!J_1 K_3\!:-\frac{\alpha \sqrt{-1}}{\sqrt{2}k \sqrt{k}(1+\alpha)^2\sqrt{1+\alpha}}:\!J_3 K_1\!:\\
        &+\frac{\sqrt{-1}(k+\alpha+k\alpha)}{2\sqrt{k}(1+\alpha)^2}:\!(DK_1) K_3\!:-\frac{\sqrt{-1}(k+\alpha+k\alpha)}{2\sqrt{k}(1+\alpha)^2}:\!(DK_3) K_1\!:,\\
        G^{-+}=&-\frac{\sqrt{-1}}{\sqrt{2k}\sqrt{1+\alpha}}DJ_2+\frac{\alpha\sqrt{-1}}{\sqrt{2}k\sqrt{k}(1+\alpha)^2 \sqrt{1+\alpha}}:\!J_2 K_3\!:+\frac{\alpha \sqrt{-1}}{\sqrt{2}k \sqrt{k}(1+\alpha)^2\sqrt{1+\alpha}}:\!J_3 K_2\!:\\
        &-\frac{\sqrt{-1}(k+\alpha+k\alpha)}{2\sqrt{k}(1+\alpha)^2}:\!(DK_2) K_3\!:+\frac{\sqrt{-1}(k+\alpha+k\alpha)}{2\sqrt{k}(1+\alpha)^2}:\!(DK_3) K_2\!:,
    \end{aligned}
\end{equation}
Next, define the weight $1$ generators.
\begin{equation} \label{eq: commuting sl(2)1}
    \begin{gathered}
    J^0=\frac{1}{k\sqrt{2(1+\alpha)}}J_3+DK_3+\frac{k-\alpha+k\alpha}{k(1+\alpha)^2}:K_1 K_2\!\!:, \\
     J^+=\frac{1}{k\sqrt{2(1+\alpha)}}J_1+DK_1+\frac{k-\alpha+k\alpha}{2k(1+\alpha)^2}:\!K_1 K_3\!:,\\
    J^-=\frac{1}{k\sqrt{2(1+\alpha)}}J_2-DK_2+\frac{k-\alpha+k\alpha}{2k(1+\alpha)^2}:\!K_2 K_3\!:,
    \end{gathered}
\end{equation}
\begin{equation} \label{eq: commuting sl(2)2}
    \begin{gathered}
        J'^0=\frac{1}{k\sqrt{2(1+\alpha)}}J_3+\frac{k+\alpha+k\alpha}{k(1+\alpha)^2}:\!K_1 K_2\!:, \quad J'^+=\frac{1}{k \sqrt{2(1+\alpha)}}J_2+\frac{k+\alpha+k\alpha}{2k(1+\alpha)^2}:\!K_2 K_3\!:\\
        J'^-=\frac{1}{k \sqrt{2(1+\alpha)}}J_1+\frac{k+\alpha+k\alpha}{2k(1+\alpha)^2}:\!K_1 K_3\!:
    \end{gathered}
\end{equation}
Note that $\eqref{eq: commuting sl(2)1}$ and $\eqref{eq: commuting sl(2)2}$ form two commuting affine $\mathfrak{sl}(2)$ vertex algebras. The modified generators in \eqref{eq: nonSUSY D(2,1;a) virasoro} to $\eqref{eq: commuting sl(2)2}$ satisfy the $\lambda$-bracket relations, with a small correction for typographical error in \cite[Section 8.6]{KW04}. For completeness, we include below a corrected version of the part.
\begin{equation*}
    \begin{aligned}
        [G^{++}{}_{\lambda}G^{++}]&=\frac{2\alpha}{(1+\alpha)^2}:\!J^0 J'^0\!:,\quad [G^{--}{}_{\lambda}G^{--}]=\frac{2\alpha}{(1+\alpha)^2}:\!J^- J'^-\!:,\\
        [G^{++}{}_{\lambda}G^{-+}]&=\frac{\alpha}{(1+\alpha)^2}:\!J^0 J'^+\!:+\frac{\alpha}{1+\alpha}\left(\frac{\alpha}{1+\alpha}-k-1\right)(\partial+2\lambda)J'^+,\\
        [G^{++}{}_{\lambda}G^{+-}]&=-\frac{\alpha}{(1+\alpha)^2}:\!J^0 J'^+\!:-\frac{1}{1+\alpha}\left(\frac{1}{1+\alpha}-k-1\right)(\partial+2\lambda)J^+.
    \end{aligned}
\end{equation*}

We remark here that the relations remain unchanged even if the weight $3/2$ generators in \eqref{eq: generator nonSUSY D(2,1;a) wt 3/2} are defined with opposite signs. Now, rename the extra generators of the big $N=4$ superconformal algebra as follows.
\begin{equation} \label{eq: D(2,1;a) wt 1/2 rename}
    \begin{aligned}
    \sigma^{++}&=\frac{\sqrt{-1}}{\sqrt{2\alpha}}\sigma+\frac{\sqrt{-1}}{2(1+\alpha)} K_3, \quad \sigma^{--}=-\frac{\sqrt{-1}\sqrt{\alpha}}{\sqrt{2}}\sigma+\frac{\sqrt{-1}\alpha}{2(1+\alpha)}K_3,\\
    \sigma^{+-}&=-\frac{\sqrt{-1}}{1+\alpha}K_1, \quad \sigma^{-+}=\frac{\sqrt{-1}\alpha}{1+\alpha}K_2, \quad \xi=-D\sigma.
    \end{aligned}
\end{equation}
Then, these fields satisfy $[\sigma^{--}{}_{\lambda}\sigma^{++}]=k$, $[\sigma^{+-}{}_{\lambda} \sigma^{-+}]=k$, and $[\xi{}_{\lambda}\xi]=\lambda k$. In \eqref{eq: D(2,1;a) wt 1/2 rename}, $\sqrt{\alpha}$ is the complex number determined up to sign by the property $(\sqrt{\alpha})^2=\alpha$. Consider the new set of generators
\begin{equation} \label{eq: generators big N=4}
    \{\tilde{L},\tilde{G}^{++},\tilde{G}^{--},\tilde{G}^{+-},\tilde{G}^{-+},\tilde{J}^0,\tilde{J}^+,\tilde{J}^-,\tilde{J}'^0,\tilde{J}'^+,\tilde{J}'^-,\xi, \sigma^{++},\sigma^{--},\sigma^{+-},\sigma^{-+}\},
\end{equation}
where the tilde-labeled generators are defined as in \cite{KW04}. Note, however, $\tilde{J}'^0$ should be defined as
\begin{equation*}
    \tilde{J}'^0=J'^0-\frac{1}{k}:\!\sigma^{--}\sigma^{++}\!:-\frac{1}{k}:\!\sigma^{+-}\sigma^{-+}\!:.
\end{equation*}
In particular, the modified generator $\tilde{L}$ is a conformal vector with central charge $-6k$, and it is a super partner of the total $N=1$ superconformal in $W^k_{N=1}(\g\oplus\CC,f)$. To be explicit,
\begin{equation}\label{eq: big N=4 total Virasoro}
    \tilde{L}=\frac{1}{2}DG_{\textup{tot}}, \quad G_{\textup{tot}}=G+\frac{1}{k}:\!(D\sigma)\sigma\!:,
\end{equation}
where $\frac{1}{k}:\!(D\sigma)\sigma\!:$ is an $N=1$ superconformal vector inside the SUSY Heisenberg vertex algebra generated by $\sigma$. The generators in \eqref{eq: generators big N=4} satisfy the $\lambda$-bracket relations in \cite{KW04}, with the following corrections applied.
\begin{equation*}
    \begin{gathered}
        [\tilde{J}'^-{}_{\lambda}\tilde{G}^{--}]=0, \quad [\tilde{J}'^-{}_{\lambda}\tilde{G}^{-+}]=\tilde{G}^{--}+\frac{\lambda}{\alpha}\sigma^{--}, \quad[\tilde{J}^0{}_{\lambda}\sigma^{\pm \mp}]=\pm \sigma^{\pm \mp}, \quad [\tilde{J}'^0{}_{\lambda}\sigma^{\pm \mp}]=\mp \sigma^{\pm \mp},\\
        [\tilde{G}^{\scriptsize{\hspace{-5pt}\begin{array}{c}--\\[-1em]-+\end{array}}\hspace{-5pt}}{}_{\lambda}\sigma^{\scriptsize{\hspace{-5pt}\begin{array}{c}++\\[-1em]+-\end{array}}\hspace{-5pt}}]=\pm\frac{1}{2(1+\alpha)}(\tilde{J}'^0\mp \tilde{J}^0)+\left(\frac{1}{2\alpha}\right)^{1/2}\xi, \\
        [\tilde{G}^{\scriptsize{\hspace{-5pt}\begin{array}{c}--\\[-1em]-+\end{array}}\hspace{-5pt}}{}_{\lambda}\sigma^{\scriptsize{\hspace{-5pt}\begin{array}{c}+-\\[-1em]++\end{array}}\hspace{-5pt}}]=-\frac{1}{1+\alpha}\tilde{J}'^{\mp}, \quad
        [\tilde{G}^{\scriptsize{\hspace{-5pt}\begin{array}{c}+-\\[-1em]++\end{array}}\hspace{-5pt}}{}_{\lambda}\sigma^{\scriptsize{\hspace{-5pt}\begin{array}{c}++\\[-1em]+-\end{array}}\hspace{-5pt}}]=\pm \frac{1}{1+\alpha}\tilde{J}^+,\\
        [\tilde{G}^{\scriptsize{\hspace{-5pt}\begin{array}{c}++\\[-1em]-+\end{array}}\hspace{-5pt}}{}_{\lambda}\tilde{G}^{\scriptsize{\hspace{-5pt}\begin{array}{c}--\\[-1em]+-\end{array}}\hspace{-5pt}}]=\tilde{L}+\frac{1}{2(1+\alpha)}(\partial+2\lambda)(\pm \tilde{J}^0+\alpha\tilde{J}'^0)-\lambda^2 k,
    \end{gathered}
\end{equation*}
\subsubsection{$N=4$ SUSY structure} \label{appendix: N=4 structure of big}
In this section, we analyze the structure of the big $N=4$ superconformal algebra. In particular, we observe that this algebra has an $N=4$ SUSY structure.

Recall the total $N=1$ superconformal vector $G_{\textup{tot}}$ introduced in \eqref{eq: big N=4 total Virasoro}. In terms of the new generators in \eqref{eq: generators big N=4}, $G_{\textup{tot}}$ is written as
\begin{equation}
    G_1:=G_{\textup{tot}}=\sqrt{-1}\Big(\tilde{G}^{++}-\tilde{G}^{--}\Big),
\end{equation}
which implies that $D_1:=D=(G_1)_{(0)}$. Fix any nonzero complex number $\eta$ and define
\begin{equation*}
    \begin{gathered}
    G_2:=\sqrt{-1}\Big(\frac{\eta}{\sqrt{\alpha}} \tilde{G}^{+-}-\frac{\sqrt{\alpha}}{\eta}\tilde{G}^{-+}\Big), \quad G_3:=\tilde{G}^{++}+\tilde{G}^{--}, \quad G_4:=-\Big(\frac{\eta}{\sqrt{\alpha}} \tilde{G}^{+-}+\frac{\sqrt{\alpha}}{\eta}\tilde{G}^{-+}\Big),\\
    J_{12}:=\frac{1}{1+\alpha}\Big(\frac{\eta}{\sqrt{\alpha}}(\tilde{J}^+ +\alpha \tilde{J}'^-)-\frac{\sqrt{\alpha}}{\eta}(\tilde{J}^-+\alpha\tilde{J}'^+)\Big),\quad  J_{14}:=\frac{\sqrt{-1}}{1+\alpha}\Big(\frac{\eta}{\sqrt{\alpha}}(\tilde{J}^+ +\alpha \tilde{J}'^-)+\frac{\sqrt{\alpha}}{\eta}(\tilde{J}^-+\alpha\tilde{J}'^+)\Big),\\
    J_{34}:=\frac{1}{1+\alpha}\Big(\frac{\eta}{\sqrt{\alpha}}(\tilde{J}^+ -\alpha \tilde{J}'^-)-\frac{\sqrt{\alpha}}{\eta}(\tilde{J}^- -\alpha\tilde{J}'^+)\Big),\quad
    J_{23}:=\frac{\sqrt{-1}}{1+\alpha}\Big(\frac{\eta}{\sqrt{\alpha}}(\tilde{J}^+ -\alpha \tilde{J}'^-)+\frac{\sqrt{\alpha}}{\eta}(\tilde{J}^- -\alpha\tilde{J}'^+)\Big),\\
    J_{13}:=\frac{\sqrt{-1}}{1+\alpha}\Big(\tilde{J}^0+\alpha \tilde{J}'^0\Big), \quad J_{24}=-\frac{\sqrt{-1}}{1+\alpha}\Big(\tilde{J}^0-\alpha \tilde{J}'^0\Big).
    \end{gathered}
\end{equation*}
Then, each $G_i$ is an $N=1$ superconformal vector associated with the differential
\begin{equation}
    D_i:=(G_i)_{(0)}, \quad i=1,2,3,4.
\end{equation}
Moreover, these are odd derivations satisfying $[D_i,D_j]=2\delta_{i,j}\partial$, that is, the big $N=4$ superconformal algebra is an $N=4$ SUSY vertex algebra with supersymmetries $D_1, \cdots, D_4$. Furthermore, if we let $J_{ij}=-J_{ji}$ for $i>j$, then each $J_{ij}$ for $i\neq j$ is an $N=2$ superconformal vector associated with $D_i$ and $D_j$. Namely, we have $D_i J_{ij}=G_j$, whenever $i\neq j$. We further redefine the remaining free generators as
\begin{equation}
    \begin{gathered}
        \sigma_1:=-\frac{\sqrt{-1}\sqrt{2}}{\sqrt{\alpha}}\big(\sigma^{--}-\alpha \sigma^{++}\big), \quad \sigma_3:=\frac{2}{\sqrt{\alpha}}\big(\sigma^{--}+\alpha \sigma^{++}\big), \\
        \sigma_2:=2\sqrt{-1}\big(\eta \sigma^{+-}-\frac{1}{\eta}\sigma^{-+}\big), \quad \sigma_4:= -2\big(\eta \sigma^{+-}+\frac{1}{\eta}\sigma^{-+}\big), \quad \tilde{\xi}:=2\sqrt{2}\xi
    \end{gathered}
\end{equation}
so that $D_i \sigma_i=\tilde{\xi}$ for $i=1,2,3,4$. We remark here that one cannot find $N=4$ SUSY structure if we consider $W^k_{N=1}(D(2,1; \alpha),f)$ instead of $W^k_{N=1}(D(2,1;\alpha)\oplus \CC,f)$.

\subsubsection{Embedding of the $N=3$ superconformal algebra} \label{appendix: embedding of N=3}
Recall the generators \eqref{eq: generators big N=4} of the big $N=4$ superconformal algebra. Fix any nonzero constants $a_1$ and $a_2$, and let
\begin{equation} \label{eq: N=3 inside big N=4}
    \begin{aligned}
        \tilde{L}_{new} &:=\tilde{L}-\frac{1-\alpha}{2\sqrt{2}\sqrt{\alpha}}\partial \xi,\quad  \tilde{G}^+_{new}:=\frac{a_1}{\sqrt{2}} \Big(\tilde{G}^{++}-\frac{1-\alpha}{2}\partial \sigma^{++}\Big), \quad  \tilde{G}^-_{new}:=\frac{\sqrt{2}}{a_1}\Big(\tilde{G}^{--}-\frac{1-\alpha}{2\alpha}\partial \sigma^{--}\Big),\\
        \tilde{G}^0_{new}&:=\frac{1}{\sqrt{2}}\Big(a_2 \tilde{G}^{+-}+\frac{1}{a_2}\tilde{G}^{-+}\Big)-\frac{1-\alpha}{2\sqrt{2}}\Big(a_2 \partial \sigma^{+-}+\frac{1}{\alpha a_2}\partial \sigma^{-+}\Big), \quad  J^0_{new}:=\tilde{J}'^0+\tilde{J}^0,\\
        J^+_{new}&:=-a_1\Big(a_2 \tilde{J}^++\frac{1}{a_2}\tilde{J}'^+\Big), \quad J^-_{new}:=-\frac{1}{a_1}\Big(\frac{1}{a_2}\tilde{J}^-+a_2 \tilde{J}'^-\Big),\quad \Phi_{new}:=\frac{1+\alpha}{2\sqrt{2}}\Big(a_2 \sigma^{+-}-\frac{1}{\alpha a_2}\sigma^{-+}\Big).
    \end{aligned}
\end{equation}
Then, they satisfy the same relations with \eqref{eq: N=3 KW generators}. In \eqref{eq: N=3 inside big N=4}, one can find a different embedding by considering the following fields instead:
\begin{equation*}
    \begin{gathered}
    \tilde{G}^+_{new}:=\frac{a_1}{\sqrt{2}} \Big(\tilde{G}^{++}-\frac{1-\alpha}{2}\partial \sigma^{++}-\frac{\sqrt{2}\sqrt{\alpha}}{k}:\!\xi\sigma^{++}\!:\Big), \\
     \tilde{G}^-_{new}:=\frac{\sqrt{2}}{a_1}\Big(\tilde{G}^{--}-\frac{1-\alpha}{2\alpha}\partial \sigma^{--}-\frac{\sqrt{2}}{\sqrt{\alpha}k}:\!\xi \sigma^{--}\!:\Big),\\
     \tilde{G}^0_{new}:=\frac{1}{\sqrt{2}}\Big(a_2 \tilde{G}^{+-}+\frac{1}{a_2}\tilde{G}^{-+}\Big)-\frac{1-\alpha}{2\sqrt{2}}\Big(a_2 \partial \sigma^{+-}+\frac{1}{\alpha a_2}\partial \sigma^{-+}\Big)-\frac{1}{k}\Big(\sqrt{\alpha}a_2 :\!\xi \sigma^{+-}\!:+\frac{1}{\sqrt{\alpha}a_2}:\!\xi \sigma^{-+}\!:\Big).
    \end{gathered}
\end{equation*}

\subsubsection{Embedding of the small $N=4$ superconformal algebra} \label{appendix: embedding of small N=4}

Recall the generators \eqref{eq: generators big N=4} of the big $N=4$ superconformal algebra. Fix any constants $a_1,a_2$, and $a_3$ with $a_1\neq 0$. Let
\begin{equation} \label{eq: small N=4 embedding 1}
    \begin{gathered}
    J^0_{new}:=J^0, \quad J^+_{new}:=J^{+}, \quad J^-_{new}:=J^-, \quad  L:=\tilde{L}+\frac{\sqrt{\alpha}}{\sqrt{2}}\partial \xi, \\
    G^+_{new}:=a_1 \big(\tilde{G}^{++}+\alpha \partial \sigma^{++}\big)+a_2 \big(\tilde{G}^{+-}+\alpha \partial \sigma^{+-}\big), \quad G^-_{new}:=-a_1\big(\tilde{G}^{-+}+\partial \sigma^{-+}\big)+a_2 \big(\tilde{G}^{--}+\partial \sigma^{--}\big),\\
    \tilde{G}^+_{new}:=a_3 \big(\tilde{G}^{++}+\alpha \partial \sigma^{++}\big)-\frac{1-a_2a_3}{a_1}\big(\tilde{G}^{+-}+\alpha \partial \sigma^{+-}\big), \\
    \tilde{G}^-_{new}:=a_3\big(\tilde{G}^{-+}+\partial \sigma^{-+}\big)+\frac{1-a_2a_3}{a_1}\big(\tilde{G}^{--}+\partial \sigma^{--}\big).
    \end{gathered}
\end{equation}
Then, these fields generate the small $N=4$ superconformal algebra by satisfying the same relations with the fields \eqref{eq: small N=4 KW generators}. Instead of the fields $G^{\pm}_{new}$ and $\tilde{G}^{\pm}_{new}$ in \eqref{eq: small N=4 embedding 1}, one can make another embedding of the small $N=4$ superconformal algebra by considering
\begin{equation}\label{eq: small N=4 embedding 1 twisted}
    \begin{aligned}
    G^+_{new}&:=a_1 \big(\tilde{G}^{++}+\alpha \partial \sigma^{++}-\frac{\sqrt{2}\sqrt{\alpha}}{k}:\!\xi \sigma^{++}\!:\big)+a_2 \big(\tilde{G}^{+-}+\alpha \partial \sigma^{+-}-\frac{\sqrt{2}\sqrt{\alpha}}{k}:\!\xi \sigma^{+-}\!:\big), \\
    G^-_{new}&:=-a_1\big(\tilde{G}^{-+}+\partial \sigma^{-+}-\frac{\sqrt{2}}{\sqrt{\alpha}k}:\!\xi \sigma^{-+}\!:\big)+a_2 \big(\tilde{G}^{--}+\partial \sigma^{--}-\frac{\sqrt{2}}{\sqrt{\alpha}k}:\!\xi \sigma^{--}\!:\big),\\
    \tilde{G}^+_{new}&:=a_3 \big(\tilde{G}^{++}+\alpha \partial \sigma^{++}-\frac{\sqrt{2}\sqrt{\alpha}}{k}:\!\xi \sigma^{++}\!:\big)-\frac{1-a_2a_3}{a_1}\big(\tilde{G}^{+-}+\alpha \partial \sigma^{+-}-\frac{\sqrt{2}\sqrt{\alpha}}{k}:\!\xi \sigma^{+-}\!:\big), \\
    \tilde{G}^-_{new}&:=a_3\big(\tilde{G}^{-+}+\partial \sigma^{-+}-\frac{\sqrt{2}}{\sqrt{\alpha}k}:\!\xi \sigma^{-+}\!:\big)+\frac{1-a_2a_3}{a_1}\big(\tilde{G}^{--}+\partial \sigma^{--}-\frac{\sqrt{2}}{\sqrt{\alpha}k}:\!\xi \sigma^{--}\!:\big).
    \end{aligned}
\end{equation}
Moreover, one can find another embedding as follows by assigning the other $\mathfrak{sl}(2)$-triple to $(J^0_{new}, J^+_{new}, J^-_{new})$:
\begin{equation} \label{eq: small N=4 embedding 2}
    \begin{gathered}
    J^0_{new}:=J'^0, \quad J^+_{new}:=J'^{+}, \quad J^-_{new}:=J'^-, \quad L:=\tilde{L}-\frac{1}{\sqrt{2}\sqrt{\alpha}}\partial \xi, \\
    G^+_{new}:=a_1 \big(\tilde{G}^{++}- \partial \sigma^{++}\big)+a_2 \big(\tilde{G}^{-+}-\frac{1}{\alpha}\partial \sigma^{-+}\big), \quad G^-_{new}:=-a_1\big(\tilde{G}^{+-}-\partial \sigma^{+-}\big)-a_2 \big(\alpha\tilde{G}^{--}+\frac{1}{\alpha}\partial \sigma^{--}\big),\\
    \tilde{G}^+_{new}:=a_3 \big(\tilde{G}^{++}- \partial \sigma^{++}\big)-\frac{1-a_2 a_3}{a_1}\big(\tilde{G}^{-+}- \partial \sigma^{-+}\big), \\
    \tilde{G}^-_{new}:=a_3\big(\tilde{G}^{+-}-\partial \sigma^{+-}\big)+\frac{1- a_2a_3}{a_1}\big(\tilde{G}^{--}-\partial \sigma^{--}\big).
    \end{gathered}
\end{equation}
As in \eqref{eq: small N=4 embedding 1 twisted}, the embedding \eqref{eq: small N=4 embedding 2} can also be twisted as

\begin{equation} \label{eq: small N=4 embedding 2 twisted}
    \begin{aligned}
        G^+_{new}&:= a_1 \big(\tilde{G}^{++}- \partial \sigma^{++}-\frac{\sqrt{2}\sqrt{\alpha}}{k}:\!\xi \sigma^{++}\!:\big)+a_2 \big(\tilde{G}^{-+}-\frac{1}{\alpha}\partial \sigma^{-+}-\frac{\sqrt{2}}{\sqrt{\alpha}k}:\!\xi \sigma^{-+}\!:\big),\\
        G^-_{new}&:= -a_1\big(\tilde{G}^{+-}-\partial \sigma^{+-}-\frac{\sqrt{2}\sqrt{\alpha}}{k}:\!\xi \sigma^{+-}\!:\big)-a_2 \big(\alpha\tilde{G}^{--}+\frac{1}{\alpha}\partial \sigma^{--}-\frac{\sqrt{2}}{\sqrt{\alpha}k}:\!\xi \sigma^{--}\!:\big),\\
        \tilde{G}^+_{new}&:= a_3 \big(\tilde{G}^{++}- \partial \sigma^{++}-\frac{\sqrt{2}\sqrt{\alpha}}{k}:\!\xi \sigma^{++}\!:\big)-\frac{1-a_2 a_3}{a_1}\big(\tilde{G}^{-+}- \partial \sigma^{-+}-\frac{\sqrt{2}}{\sqrt{\alpha}k}:\!\xi \sigma^{-+}\!:\big), \\
        \tilde{G}^-_{new}&:= a_3\big(\tilde{G}^{+-}-\partial \sigma^{+-}-\frac{\sqrt{2}\sqrt{\alpha}}{k}:\!\xi\sigma^{+-}\!:\big)+\frac{1- a_2a_3}{a_1}\big(\tilde{G}^{--}-\partial \sigma^{--}-\frac{\sqrt{2}}{\sqrt{\alpha}k}:\!\xi \sigma^{--}\!:\big).\\
    \end{aligned}
\end{equation}

\bibliographystyle{abbrv}
\bibliography{refs.bib}

\begin{bibdiv}
\begin{biblist}

\bib{AM95}{article}{
      author={Adamovic, Drazen},
      author={Milas, Antun},
       title={Vertex operator algebras associated to modular invariant
  representations for $a^{(1)}_1$},
        date={1995},
     journal={Math. Res. Lett.},
      volume={2},
      number={5},
       pages={563\ndash 575},
      review={\MR{1359963}},
}

\bib{ACKL17}{article}{
      author={Arakawa, Tomoyuki},
      author={Creutzig, Thomas},
      author={Kawasetsu, Kazuya},
      author={Linshaw, Andrew~R.},
       title={Orbifolds and cosets of minimal {$\mathcal{W}$}-algebras},
        date={2017},
        ISSN={0010-3616,1432-0916},
     journal={Comm. Math. Phys.},
      volume={355},
      number={1},
       pages={339\ndash 372},
  url={https://doi-org-ssl.libproxy.snu.ac.kr/10.1007/s00220-017-2901-2},
      review={\MR{3670736}},
}

\bib{AKM23}{article}{
      author={Arakawa, Tomoyuki},
      author={Kuwabara, Toshiro},
      author={M\"oller, Sven},
       title={Hilbert schemes of points in the plane and quasi-lisse vertex
  algebras with n=4 symmetry},
        date={arXiv:2309.17308, 2023},
     journal={ArXiv eprints},
      eprint={arXiv:2309.17308},
}

\bib{BK03}{article}{
      author={Bakalov, Bojko},
      author={Kac, Victor~G.},
       title={Field algebras},
        date={2003},
        ISSN={1073-7928,1687-0247},
     journal={Int. Math. Res. Not.},
      number={3},
       pages={123\ndash 159},
  url={https://doi-org-ssl.libproxy.snu.ac.kr/10.1155/S1073792803204232},
      review={\MR{1932531}},
}

\bib{B1996}{article}{
      author={Barron, Katrina},
       title={A supergeometric interpretation of vertex operator
  superalgebras.},
        date={1996},
     journal={Internat. Math. Res. Notices},
      volume={1996},
      number={9},
       pages={409\ndash 430},
      review={\MR{1399408}},
}

\bib{B2000}{article}{
      author={Barron, Katrina},
       title={$n=1$ neveu-schwarz vertex operator superalgebras over grassmann
  algebras and with odd formal variables.},
        date={2000},
     journal={Representations and quantizations (Shanghai, 1998).},
       pages={9\ndash 35},
      review={\MR{1802167}},
}

\bib{B2003}{article}{
      author={Barron, Katrina},
       title={The notion of $n=1$ supergeometric vertex operator superalgebra
  and the isomorphism theorem},
        date={2003},
     journal={Commun. Contemp. Math.},
      volume={5},
      number={4},
       pages={481\ndash 567},
      review={\MR{2003210}},
}

\bib{B2004}{article}{
      author={Barron, Katrina},
       title={Superconformal change of variables for n=1 neveu-schwarz vertex
  operator superalgebras},
        date={2004},
     journal={J. Algebra},
      volume={277},
      number={2},
       pages={717–764},
      review={\MR{2067629}},
}

\bib{BM25}{article}{
      author={Bonetti, Federico},
      author={Meneghelli, Carlo},
       title={A universal w-algebra for $n=4$ super yang-mills},
        date={arXiv:2506.15678, 2025},
     journal={ArXiv eprints},
      eprint={arXiv:2506.15678},
}

\bib{BMR19}{article}{
      author={Bonetti, Federico},
      author={Meneghelli, Carlo},
      author={Rastelli, Leonardo},
       title={Voas labelled by complex reflection groups and 4d scfts},
        date={2019},
     journal={J. High Energy Phys.},
      volume={155},
         url={https://link.springer.com/article/10.1007/JHEP05(2019)155},
      review={\MR{MR3973380}},
}

\bib{CG2013}{article}{
      author={Candu, Constantin},
      author={Gaberdiel, Matthias},
       title={Duality in $n=2$ minimal model holography},
        date={2013},
     journal={J. High Energy Phys},
      volume={070},
      number={2},
}

\bib{CCS24+}{article}{
      author={Chen, Chih-Whi},
      author={Cheng, Shun-Jen},
      author={Suh, Uhi~Rinn},
       title={Whittaker modules of central extensions of {T}akiff superalgebras
  and finite supersymmetric {W}-algebras},
     journal={arXiv:2410.22819},
         url={https://doi.org/10.48550/arXiv.2410.22819},
}

\bib{CGN21}{article}{
      author={Creutzig, Thomas},
      author={Genra, Naoki},
      author={Nakatsuka, Shigenori},
       title={Duality of subregular w-algebras and principal w-superalgebras},
        date={2021},
        ISSN={0001-8708},
     journal={Advances in Mathematics},
      volume={383},
       pages={107685},
  url={https://www.sciencedirect.com/science/article/pii/S0001870821001237},
}

\bib{CKM24}{article}{
      author={Creutzig, Thomas},
      author={Kanade, Shashank},
      author={McRae, Robert},
       title={Tensor categories for vertex operator superalgebra extensions},
        date={2024},
     journal={Mem. Amer. Math.},
      volume={295},
      number={1472},
       pages={vi+181 pp.},
         url={https://doi.org/10.1090/memo/1472},
      review={\MR{4720880}},
}

\bib{CKL24}{article}{
      author={Creutzig, Thomas},
      author={Kovalchuk, Vladimir},
      author={Linshaw, Andrew},
       title={Building blocks for $\mathcal{W}$-algebras of classical types},
        date={2024},
     journal={arXiv:2409.03465},
}

\bib{CKLSS25}{article}{
      author={Creutzig, Thomas},
      author={Kovalchuk, Vladimir},
      author={Linshaw, Andrew},
      author={Song, Arim},
      author={Suh, Uhi~Rinn},
       title={Universal $n=2$ supersymmetric $\mathcal{W}_{\infty}$-algebra},
        date={2025},
     journal={In preparation},
}

\bib{CL19}{article}{
      author={Creutzig, Thomas},
      author={Linshaw, Andrew~R.},
       title={Cosets of affine vertex algebras inside larger structures},
        date={2019},
        ISSN={0021-8693,1090-266X},
     journal={J. Algebra},
      volume={517},
       pages={396\ndash 438},
         url={https://doi.org/10.1016/j.jalgebra.2018.10.007},
      review={\MR{3869280}},
}

\bib{CL22a}{article}{
      author={Creutzig, Thomas},
      author={Linshaw, Andrew~R.},
       title={Trialities of $\mathcal{W}$-algebras},
        date={2022},
     journal={Camb. J. Math.},
      volume={10},
       pages={69\ndash 194},
         url={https://intlpress.com/JDetail/1805782238384054273},
      review={\MR{4445343}},
}

\bib{CL22b}{article}{
      author={Creutzig, Thomas},
      author={Linshaw, Andrew~R.},
       title={Trialities of orthosymplectic $\mathcal{W}$-algebras},
        date={2022},
     journal={Adv. Math.},
      volume={409 Part B},
       pages={108678},
  url={https://www.sciencedirect.com/science/article/abs/pii/S0001870822004959},
      review={\MR{4481140}},
}

\bib{DK06}{article}{
      author={De~Sole, Alberto},
      author={Kac, Victor~G.},
       title={Finite vs affine {$W$}-algebras},
        date={2006},
        ISSN={0289-2316},
     journal={Jpn. J. Math.},
      volume={1},
      number={1},
       pages={137\ndash 261},
         url={https://doi.org/10.1007/s11537-006-0505-2},
      review={\MR{2261064}},
}

\bib{EH91}{article}{
      author={Evans, Jonathan},
      author={Hollowood, Tim},
       title={Supersymmetric {T}oda field theories},
        date={1991},
        ISSN={0550-3213,1873-1562},
     journal={Nuclear Phys. B},
      volume={352},
      number={3},
       pages={723\ndash 768},
         url={https://doi.org/10.1016/0550-3213(91)90105-7},
      review={\MR{1098247}},
}

\bib{FZ92}{article}{
      author={Frenkel, Igor~B.},
      author={Zhu, Yongchang},
       title={Vertex operator algebras associated to representations of affine
  and virasoro algebras},
        date={1992},
     journal={Duke Math. J.},
      volume={66},
      number={1},
       pages={123\ndash 168},
      review={\MR{1159433}},
}

\bib{Genra17}{article}{
      author={Genra, Naoki},
       title={Screening operators for {$\mathcal{W}$}-algebras},
        date={2017},
        ISSN={1022-1824,1420-9020},
     journal={Selecta Math. (N.S.)},
      volume={23},
      number={3},
       pages={2157\ndash 2202},
         url={https://doi.org/10.1007/s00029-017-0315-9},
      review={\MR{3663604}},
}

\bib{GSS25}{article}{
      author={Genra, Naoki},
      author={Song, Arim},
      author={Suh, Uhi~Rinn},
       title={Principal {SUSY} and non{SUSY} {W}-algebras and their {Z}hu
  algebras},
        date={arXiv:2502.05721, 2025},
     journal={ArXiv eprints},
      eprint={math.CO/2502.05721},
}

\bib{GKO85}{article}{
      author={Goddard, P.},
      author={Kent, A.},
      author={Olive, D.},
       title={Virasoro algebras and coset space models},
        date={1985},
     journal={Phys. Lett. B},
      volume={152},
      number={2},
       pages={88\ndash 92},
         url={https://doi.org/10.1016/0370-2693(85)91145-1},
      review={\MR{0778819}},
}

\bib{goodmanwallach}{book}{
      author={Goodman, Roe},
      author={Wallach, Nolan~R.},
       title={Symmetry, representations, and invariants},
   publisher={Springer},
        date={2009},
      volume={255},
}

\bib{H17}{article}{
      author={Heluani, Reimundo},
       title={Recent advances and open questions on the susy structure of the
  chiral de rham complex},
        date={2017},
     journal={J. Phys. A},
      volume={50},
      number={42},
       pages={423002},
      review={\MR{3717036}},
}

\bib{HK07}{article}{
      author={Heluani, Reimundo},
      author={Kac, Victor~G.},
       title={Supersymmetric vertex algebras},
        date={2007},
        ISSN={0010-3616},
     journal={Comm. Math. Phys.},
      volume={271},
      number={1},
       pages={103\ndash 178},
  url={http://lps3.doi.org.libproxy.snu.ac.kr/10.1007/s00220-006-0173-3},
      review={\MR{2283956}},
}

\bib{Ito91}{article}{
      author={Ito, Katsushi},
       title={Quantum {H}amiltonian reduction and {$N=2$} coset models},
        date={1991},
        ISSN={0370-2693,1873-2445},
     journal={Phys. Lett. B},
      volume={259},
      number={1-2},
       pages={73\ndash 78},
         url={https://doi.org/10.1016/0370-2693(91)90136-E},
      review={\MR{1104183}},
}

\bib{Ito92}{article}{
      author={Ito, Katsushi},
       title={{$N=2$} superconformal {${\bf C}{\rm P}_n$} model},
        date={1992},
        ISSN={0550-3213,1873-1562},
     journal={Nuclear Phys. B},
      volume={370},
      number={1},
       pages={123\ndash 142},
         url={https://doi.org/10.1016/0550-3213(92)90347-E},
      review={\MR{1148541}},
}

\bib{KPP23}{article}{
      author={Kac, Victor~G.},
      author={M\"oseneder~Frajria, Pierluigi},
      author={Papi, Paolo},
       title={Unitarity of minimal {$W$}-algebras and their representations
  {I}},
        date={2023},
        ISSN={0010-3616,1432-0916},
     journal={Comm. Math. Phys.},
      volume={401},
      number={1},
       pages={79\ndash 145},
  url={https://doi-org-ssl.libproxy.snu.ac.kr/10.1007/s00220-023-04637-5},
      review={\MR{4604893}},
}

\bib{KP81}{article}{
      author={Kac, Victor~G.},
      author={Peterson, Dale},
       title={Spin and wedge representations of infinite-dimensional lie
  algebras and groups},
        date={1981},
     journal={Proc. Natl. Acad. Sci. USA},
      volume={78},
      number={6},
       pages={3308\ndash 3312},
         url={https://doi.org/10.1073/pnas.78.6.3308},
      review={\MR{0619827}},
}

\bib{KRW03}{article}{
      author={Kac, Victor~G.},
      author={Roan, Shi-Shyr.},
      author={Wakimoto, Minoru},
       title={Quantum reduction for affine superalgebras},
        date={2003},
        ISSN={0010-3616},
     journal={Comm. Math. Phys.},
      volume={241},
      number={2-3},
       pages={307\ndash 342},
  url={http://lps3.doi.org.libproxy.snu.ac.kr/10.1007/s00220-003-0926-1},
      review={\MR{2013802}},
}

\bib{KT85}{article}{
      author={Kac, Victor~G.},
      author={Todorov, Ivan~T.},
       title={Superconformal current algebras and their unitary
  representations},
        date={1985},
        ISSN={0010-3616,1432-0916},
     journal={Comm. Math. Phys.},
      volume={102},
      number={2},
       pages={337\ndash 347},
         url={http://projecteuclid.org/euclid.cmp/1104114386},
      review={\MR{820579}},
}

\bib{KW04}{article}{
      author={Kac, Victor~G.},
      author={Wakimoto, Minoru},
       title={Quantum reduction and representation theory of superconformal
  algebras},
        date={2004},
        ISSN={0001-8708},
     journal={Adv. Math.},
      volume={185},
      number={2},
       pages={400\ndash 458},
         url={https://doi.org/10.1016/j.aim.2003.12.005},
      review={\MR{2060475}},
}

\bib{KW22}{article}{
      author={Kac, Victor~G.},
      author={Wakimoto, Minoru},
       title={On {F}ree {F}ield {R}ealization of {Q}uantum {A}ffine
  {W}-{A}lgebras},
        date={2022},
        ISSN={1432-0916},
     journal={Communications in Mathematical Physics},
      volume={395},
      number={2},
       pages={571\ndash 600},
}

\bib{LS23}{article}{
      author={Linshaw, Andrew~R.},
      author={Song, Bailin},
       title={The global sections of chiral de rham complexes on compact
  ricci-flat k\"ahler manifolds ii.},
        date={2023},
     journal={Comm. Math. Phys.},
      volume={399},
      number={1},
       pages={189\ndash 202},
      review={\MR{4567372}},
}

\bib{MadRag94}{article}{
      author={Madsen, Jens~Ole},
      author={Ragoucy, Eric},
       title={Quantum {H}amiltonian reduction in superspace formalism},
        date={1994},
        ISSN={0550-3213},
     journal={Nuclear Phys. B},
      volume={429},
      number={2},
       pages={277\ndash 290},
         url={https://doi.org/10.1016/0550-3213(94)00258-4},
      review={\MR{1299068}},
}

\bib{MSV99}{article}{
      author={Malikov, Fedor},
      author={Schechtman, Vadim},
      author={Vaintrob, Arkady},
       title={Chiral de rham complex},
        date={1999},
     journal={Comm. Math. Phys.},
      volume={204},
      number={2},
       pages={439\ndash 473},
      review={\MR{1704283}},
}

\bib{MRS21}{article}{
      author={Molev, Alexander},
      author={Ragoucy, Eric},
      author={Suh, Uhi~Rinn},
       title={Supersymmetric $w$-algebras},
        date={2021},
        ISSN={0377-9017},
     journal={Lett. Math. Phys.},
      volume={111},
      number={1},
       pages={Paper No. 6, 25},
  url={http://lps3.doi.org.libproxy.snu.ac.kr/10.1007/s11005-020-01346-1},
      review={\MR{4200967}},
}

\bib{Musson12}{book}{
      author={Musson, Ian~M.},
       title={Lie superalgebras and enveloping algebras},
      series={Graduate Studies in Mathematics},
   publisher={American Mathematical Society, Providence, RI},
        date={2012},
      volume={131},
        ISBN={978-0-8218-6867-6},
         url={https://doi-org-ssl.libproxy.snu.ac.kr/10.1090/gsm/131},
      review={\MR{2906817}},
}

\bib{Nak23}{article}{
      author={Nakatsuka, Shigenori},
       title={On {M}iura maps for {$\mathcal{W}$}-superalgebras},
        date={2023},
        ISSN={1083-4362,1531-586X},
     journal={Transform. Groups},
      volume={28},
      number={1},
       pages={375\ndash 399},
         url={https://doi.org/10.1007/s00031-021-09679-4},
      review={\MR{4552179}},
}

\bib{O86}{article}{
      author={Odake, Satoru},
       title={Extensions of $n=2$ superconformal algebra and calabi-yau
  compactification},
        date={1986},
     journal={Modern Phys. Lett. A},
      volume={4},
      number={6},
       pages={557\ndash 568},
      review={\MR{1016545}},
}

\bib{SY25}{article}{
      author={{Rinn Suh}, Uhi},
      author={Yoon, Sangwon},
       title={Supersymmetric extension of universal enveloping vertex
  algebras},
        date={2025},
        ISSN={0021-8693},
     journal={Journal of Algebra},
      volume={667},
       pages={35\ndash 74},
  url={https://www.sciencedirect.com/science/article/pii/S0021869324006835},
}

\bib{Songthesis}{thesis}{
      author={Song, Arim},
       title={Structural {A}nalysis of {S}upersymmetric {W}-algebras via
  {S}upersymmetric {V}ertex {A}lgebras},
        type={Ph.D. Thesis},
        date={2024},
}

\bib{Song24free}{article}{
      author={Song, Arim},
       title={{F}ree {F}ield {R}ealization of {S}upersymmetric {W}-algebras},
        date={arXiv:2404.05275, 2024},
     journal={ArXiv eprints},
      eprint={math.CO/2404.05275},
}

\bib{Thiel91}{article}{
      author={Thielemans, Kris},
       title={A {M}athematica™ {P}ackage for {C}omputing {O}perator {P}roduct
  {E}xpansions},
        date={1991},
     journal={International Journal of Modern Physics C},
      volume={02},
      number={03},
       pages={787\ndash 798},
      eprint={https://doi.org/10.1142/S0129183191001001},
         url={https://doi.org/10.1142/S0129183191001001},
}

\end{biblist}
\end{bibdiv}

\end{document}